\documentclass[a4paper,fleqn]{cas-dc}

\usepackage[numbers,square,sort]{natbib}
\usepackage{amsmath,amssymb,bm}
\usepackage{booktabs,tabularx,array,multirow}
\usepackage{graphicx}
\usepackage{subcaption}
\usepackage{float}
\usepackage{tikz}
\usetikzlibrary{arrows.meta,positioning}
\usepackage{orcidlink}
\usepackage{url}

\newcolumntype{Y}{>{\raggedright\arraybackslash}X}
\newcolumntype{P}[1]{>{\raggedright\arraybackslash}p{#1}}

\newif\ifjournalversion
\journalversionfalse % Set to \journalversiontrue for the journal submission.

\begin{document}

\shorttitle{RL for Spacecraft and Space-Mission Autonomy}
\shortauthors{Lu et al.}

\ExplSyntaxOn
\cs_set:Npn \__first_footerline:
  {
    \group_begin:
    \small
    \sffamily
    \ifnum\theblind>0\relax
    \else
      \__short_authors: :~
    \fi
    { \rmfamily \itshape Preprint }
    \group_end:
  }
\ExplSyntaxOff

\title[mode=title]{Reinforcement Learning for Spacecraft and Space-Mission Autonomy: Architectures, Applications, Deployment Challenges, and Future Perspectives}

\author[1]{Wenlong Lu\,\orcidlink{0000-0002-3237-1135}}
\author[1]{Ziyao Geng\,\orcidlink{0009-0002-5481-4416}}
\author[1]{Fanqing Tu\,\orcidlink{0009-0000-6805-6447}}
\author[1]{Qiang Shen\,\orcidlink{0000-0001-8428-2356}}\cormark[1]
\ead{qiangshen@sjtu.edu.cn}
\author[1]{Shufan Wu\,\orcidlink{0000-0002-8138-5533}}
\author[2,3]{Yongchun Xie}
\author[4]{Hyondong Oh\,\orcidlink{0000-0002-1051-9477}}

\affiliation[1]{organization={School of Aeronautics and Astronautics, Shanghai Jiao Tong University},city={Shanghai},postcode={200240},country={China}}
\affiliation[2]{organization={Beijing Institute of Control Engineering},city={Beijing},postcode={100190},country={China}}
\affiliation[3]{organization={Science and Technology on Space Intelligent Control Laboratory},city={Beijing},postcode={100190},country={China}}
\affiliation[4]{organization={Department of Mechanical Engineering, Korea Advanced Institute of Science and Technology},city={Daejeon},postcode={34141},country={South Korea}}
\cortext[cor1]{Corresponding author}

\begin{abstract}
Reinforcement learning (RL) has attracted growing interest in spacecraft autonomy because of its potential for adaptive decision-making, rapid onboard inference, and integration with conventional guidance, navigation, and control systems.
Drawing on 203 application papers, this review examines RL architectures, applications, and deployment challenges across six domains: attitude control; orbital guidance and trajectory optimization; rendezvous, proximity operations, and on-orbit servicing; planetary landing, small-body operations, and surface mobility; mission planning and scheduling; and multi-spacecraft coordination.
Beyond surveying individual algorithms, the review compares RL decision authority, system architectures, integration with conventional methods, constraint-handling mechanisms, and validation practices.
The analysis reveals that RL serves diverse roles, from direct feedback control and adaptive guidance to optimizer assistance and mission-level decision-making.
Hybrid architectures offer opportunities to combine learned policies with established controllers, planners, and safety mechanisms, although their reported performance depends on the complete system rather than the learned policy alone.
Safety, feasibility, and operational constraints vary across spacecraft tasks, while constraint-handling mechanisms differ in their enforcement scope; any safety claims depend on the underlying assumptions and implementation conditions.
Reported advantages over conventional approaches remain dependent on the selected baselines, operating conditions, and evaluation metrics, and validation evidence remains predominantly simulation-based.
Building on these findings, this review identifies research directions toward safe and robust learning, data-efficient adaptation, integrated spacecraft autonomy, and rigorous validation under realistic mission conditions.
\end{abstract}

\ifjournalversion
\begin{highlights}
\item Reviews reinforcement learning across six space-mission application domains.
\item Examines RL decision authority and contributions within hybrid architectures.
\item Distinguishes constraint handling from formal safety guarantees.
\item Reveals limited flight validation despite extensive simulation studies.
\item Identifies research gaps in adaptation, system integration, and validation.
\end{highlights}
\fi


\maketitle

\section{Introduction}
\label{sec:introduction}

Future space missions will increasingly operate where continuous ground supervision is impractical.
Communication delays, intermittent links, uncertain environments, long mission durations, and growing fleets make repeated onboard decisions important \citep{tipaldi2022reinforcementLearning,elkins2020adaptiveContinuous,stephenson2025learningPolicies}.
Classical guidance, navigation, and control (GNC) remains essential, while autonomous missions must also select objectives, adapt trajectories, allocate resources, and coordinate vehicles when models or contingency plans are incomplete \citep{tipaldi2022reinforcementLearning,elkins2020adaptiveContinuous}.

RL provides policies for sequential decisions, from attitude torque and rendezvous thrust to rover routes, observation schedules, and cooperative spacecraft actions.
Its appeal is not limited to replacing a controller: a learned policy may generate commands directly, adapt a conventional controller, supply a reference or optimizer warm start, or select tasks and search operators.
This review compares RL studies by the information available to each policy, the decisions it controls, the supporting modules retained, and the evaluation conditions reported.

For this review, spacecraft autonomy covers orbital vehicles, landers, spacecraft-mounted manipulators, and planetary mobility platforms when RL participates in a space-mission decision loop or its implementation and deployment are specifically investigated.
Planning, scheduling, teleoperation, orbital games, and servicing studies are included when they address RL-based decision-making, control, or deployment within these systems.
Because these tasks have different dynamics and risks, the review distinguishes application domain from agent organization and records the policy's control authority as the commands or mission decisions it can change.

Across the reviewed studies, RL is used for nonlinear feedback, adaptation to modeled changes, and repeated decisions for which an online optimization can be costly.
Elkins et al. report large-angle attitude slews and adaptation to disturbance torques unseen during training \citep{elkins2020adaptiveContinuous}.
After training, RL policies can replace repeated online optimization with policy evaluation: Tafanidis et al. report low computation for station-keeping policies, while Stephenson et al. derive RL policies that make scheduling decisions at variable intervals and describe them as a computationally inexpensive, closed-loop alternative to traditional preplanning \citep{tafanidis2025reinforcementLearning,stephenson2025learningPolicies}.
Hybrid RL designs retain useful structure: proportional--integral--derivative (PID) guidance speeds training of the twin delayed deep deterministic policy gradient (TD3) agent, and Hovell et al.'s RL-based deep-guidance policy supplies velocity commands to a conventional tracker \citep{zhang2020modelFree,hovell2020onDeep}.
RL safety architectures combine learned policies with mechanisms that screen or modify their actions: Dunlap et al. use runtime assurance (RTA) during RL training to enforce a dynamic velocity constraint, while Zhang et al. use online control barrier function (CBF) projection to filter RL capture commands in simulation \citep{dunlap2023runTime,zhang2026constrainedReinforcement}.
The growing interest in RL for spacecraft autonomy reflects its potential for adaptive feedback, computationally efficient policy evaluation, and integration with existing spacecraft control and planning modules.
The review evaluates these capabilities against the baselines and validation evidence reported for each application.

Prior reviews address spacecraft RL and related areas including spacecraft control, relative navigation, space robotics, and Earth-observation scheduling.
Tipaldi et al. survey RL for spacecraft control applications; adjacent reviews cover artificial intelligence (AI) trends in spacecraft guidance and control, machine learning for spacecraft control design, and deep learning for spacecraft dynamics, navigation, and control \citep{tipaldi2022reinforcementLearning,revIzzo2019,revShirobokov2021,revSilvestrini2022}.
Opromolla et al. review pose determination for close-proximity operations, Moghaddam et al. survey guidance, navigation, and control for in-orbit space robotics, and Song et al. survey deep learning for spacecraft relative navigation; Miralles et al. review machine learning across Earth-observation operations, while Wang et al. focus on deep reinforcement learning for Earth-observation satellite scheduling \citep{revOpromolla2017,revMoghaddam2021,revSong2022,revMiralles2023,revWang2026}.
Broader reviews cover safe learning, offline RL, and RL-assisted evolutionary optimization \citep{garcia2015comprehensiveSurvey,revBrunke2022,revGu2024,levine2020offlineReinforcement,revSongEA2024}.
Table~\ref{tab:prior-reviews} summarizes these scopes.

\begin{table*}[]
\caption{Representative prior reviews and what this review adds.}
\label{tab:prior-reviews}
\centering
\begin{tabularx}{\textwidth}{@{}lYY@{}}
\toprule
Review area & Earlier reviews and their focus & What this review adds \\
\midrule
Spacecraft RL & \citep{tipaldi2022reinforcementLearning}: control and planning applications & Assesses RL's added value over conventional methods through decision authority, system integration, and comparative evidence. \\
Spacecraft AI/GNC & \citep{revIzzo2019}; \citep{revShirobokov2021}; \citep{revSilvestrini2022}: broader learning and control & Examines how learned decisions interact with conventional GNC modules and where performance gains originate. \\
Robotics and navigation & \citep{revOpromolla2017}; \citep{revMoghaddam2021}; \citep{revSong2022}: operations and sensing & Evaluates sensing assumptions, safety mechanisms, and validation limitations in spacecraft operations. \\
Earth-observation & \citep{revMiralles2023}: machine learning across Earth-observation operations; \citep{revWang2026}: deep reinforcement learning satellite scheduling & Compares RL-assisted scheduling with conventional optimization in solution quality, computational cost, and adaptability. \\
General RL, safety, and optimization & \citep{garcia2015comprehensiveSurvey}; \citep{revBrunke2022}; \citep{revGu2024}; \citep{levine2020offlineReinforcement}; \citep{revSongEA2024}: safe RL, offline RL, and RL-assisted optimization & Connects constraint-handling mechanisms with spacecraft-specific requirements and deployment evidence. \\
\bottomrule
\end{tabularx}
\end{table*}

Literature records were identified through Google Scholar and supplemented by searches of IEEE, Elsevier, Springer, AIAA, and relevant conference proceedings pages.
Search terms combined reinforcement-learning descriptors, including ``reinforcement learning,'' ``deep reinforcement learning,'' and ``multi-agent reinforcement learning,'' with platform terms such as ``spacecraft,'' ``satellite,'' ``space robot,'' ``lander,'' and ``rover,'' and task terms covering attitude control, orbital guidance and trajectory optimization, rendezvous and proximity operations, planetary landing and surface mobility, mission planning and scheduling, and multi-spacecraft coordination.
The search covered literature available through October 2026.
Studies were screened by title, abstract, and full text; application papers entered the 203-paper corpus if an RL policy or learning component directly influenced a space-mission control action or decision, or if the paper specifically evaluated the implementation or deployment of such a component in a spacecraft system.
General RL methods, reviews, and adjacent non-RL studies served as methodological context or comparators and were excluded from the count.
Peer-reviewed conference and journal papers were eligible.
Preprints were considered when relevant technical details were available; preprint, conference, and journal versions describing the same study were consolidated as one application paper, whereas separate papers reporting a shared experimental campaign remained separate papers but counted as one demonstration when reporting the number of distinct demonstrations.
Each application paper is assigned to a primary domain based on its primary mission objective and the decision variables addressed by the RL component---attitude control; orbital guidance and trajectory optimization; rendezvous, proximity operations, and servicing; planetary landing and surface mobility; mission planning and scheduling; or multi-spacecraft coordination---and cross-domain papers may also be discussed in other chapters.
Section~\ref{sec:testing-readiness} assigns each application paper one primary validation category, even when a paper reports multiple settings; the categories describe test settings, not a maturity ranking.
Study evidence was assessed qualitatively by examining baseline comparability, modeled constraints and uncertainties, validation settings, and reporting of performance and computational requirements.

The review is guided by four questions:
\begin{enumerate}
  \item Which spacecraft decisions have been studied with RL, and what can each policy control?
  \item How are RL policies integrated with conventional controllers, optimizers, estimators, and safety mechanisms?
  \item What evidence supports the adaptability, safety, computational efficiency, and mission-level applicability of RL-based approaches?
  \item Which decision models and validation practices could help establish when RL adds value and how to assess dependable autonomy?
\end{enumerate}

Across six application domains, this review critically assesses reported RL advantages over conventional control and optimization methods, considering decision authority, system integration, constraint handling, baseline selection, and validation evidence.
It identifies research directions toward safe, robust, and dependable spacecraft autonomy.
After the foundations, six technical chapters examine attitude control, orbital guidance, proximity operations and servicing, planetary landing and mobility, mission planning and scheduling, and multi-spacecraft autonomy; the final chapter develops evidence-linked directions for decision models, adaptation, integrated autonomy, and testing.

\begin{figure}[]
\centering
\includegraphics[width=\columnwidth]{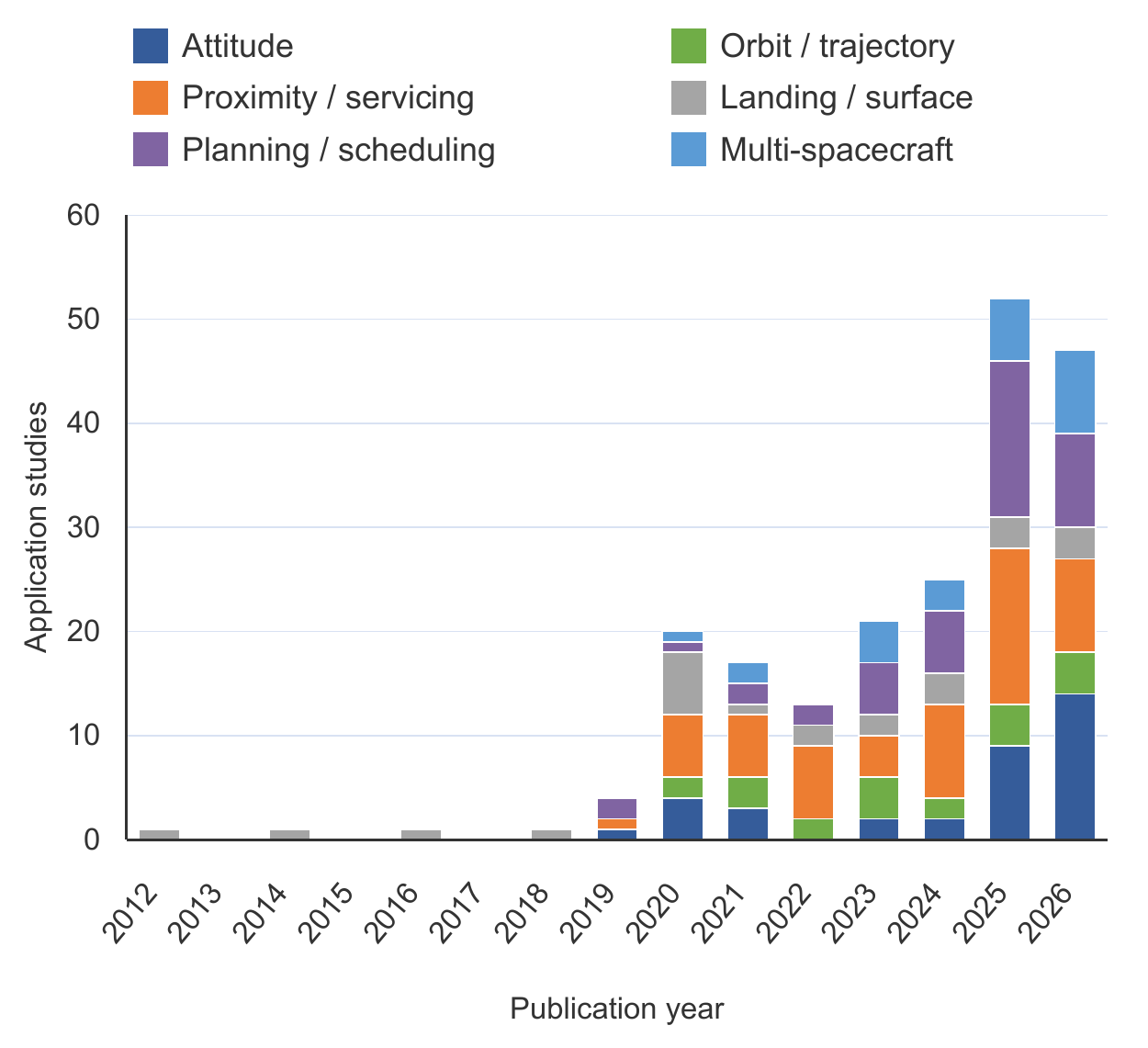}
\caption{Publication-year distribution of 203 papers on reinforcement learning for spacecraft and related mission systems, grouped across six technical chapters.}
\label{fig:annual-domain-trend}
\end{figure}

Figure~\ref{fig:annual-domain-trend} shows the publication-year distribution of these 203 papers across the six technical chapters; each paper is counted once under its primary chapter, using publication-year information from Google Scholar and publisher records.

\section{Foundations of Reinforcement Learning for Spacecraft Autonomy}
\label{sec:foundations}

Spacecraft RL studies begin by defining what the policy observes, which actions it may take, how the system evolves, and what outcomes the reward values.
Partial observations, safety constraints, uncertain dynamics, variable-duration decisions, and interactions among agents motivate extensions beyond the basic Markov decision process (MDP).
This section introduces the common formulation, its spacecraft-relevant variants, and the distinction between single-agent and multi-agent learning used throughout the application chapters.

\subsection{MDP Formulation and Spacecraft-Specific Design Choices}

An MDP specifies the states, actions, transition dynamics, and rewards used to evaluate a policy.
In spacecraft problems, the formulation depends on state and action representations, decision horizon, and reward design.
The basic discrete-time MDP assumes a fully observed Markov state and fixed-duration decision steps; partial observations, variable-duration decisions, and explicit cost or risk criteria require extensions or additional modeling choices.
A finite-horizon or discounted infinite-horizon MDP can be written as
\begin{equation}
  \mathcal{M}=\langle\mathcal{S},\mathcal{A},P,r,\gamma,H\rangle,
\end{equation}
where $s\in\mathcal{S}$ is the decision state, $a\in\mathcal{A}$ is the action, $P(s'\mid s,a)$ is the transition law, $r(s,a,s')$ is the reward, $\gamma\in[0,1]$ is the discount factor, and $H$ is the number of decision steps or $\infty$ for a continuing task \citep{sutton2018reinforcementLearning}.
For an initial-state distribution $\rho_0$, a policy $\pi_\theta$ parameterized by $\theta$ is evaluated by its expected return and may be optimized to maximize
\begin{equation}
  J(\pi_\theta)=\mathbb{E}_{s_0\sim\rho_0,\pi_\theta,P}\!\left[\sum_{t=0}^{H-1}\gamma^t r_t\right].
  \label{eq:return}
\end{equation}
Here, $r_t=r(s_t,a_t,s_{t+1})$ denotes the immediate reward at decision step $t$.
For a stationary infinite-horizon discounted MDP with bounded rewards, $0\le\gamma<1$, and an attained optimal action, the state-value function, action-value function, and Bellman optimality equation are given by \citep{sutton2018reinforcementLearning}:
\begin{equation}
  V^\pi(s)=\mathbb{E}_{\pi,P}\!\left[\left.\sum_{k=0}^{\infty}\gamma^k r_{t+k}\,\right|s_t=s\right].
  \label{eq:state-value}
\end{equation}
\begin{equation}
  Q^\pi(s,a)=\mathbb{E}_{\pi,P}\!\left[\left.\sum_{k=0}^{\infty}\gamma^k r_{t+k}\,\right|s_t=s,a_t=a\right].
  \label{eq:action-value}
\end{equation}
\begin{equation}
  Q^*(s,a)=\mathbb{E}_{P}\!\left[\left.r_t+\gamma\max_{a'\in\mathcal{A}}Q^*(s_{t+1},a')\,\right|s_t=s,a_t=a\right].
  \label{eq:bellman-optimality}
\end{equation}
For continuing tasks, $H=\infty$ when the discounted return is well defined, for example with bounded rewards and $\gamma<1$; using $\gamma=1$ requires additional termination or convergence conditions.
For finite-horizon transfers, docking, landing, or schedules, time remaining should be part of the decision state when it changes the optimal action, or the policy should be time dependent, $\pi_t(a\mid s)$.

\begin{figure*}[]
\centering
\includegraphics[width=\textwidth]{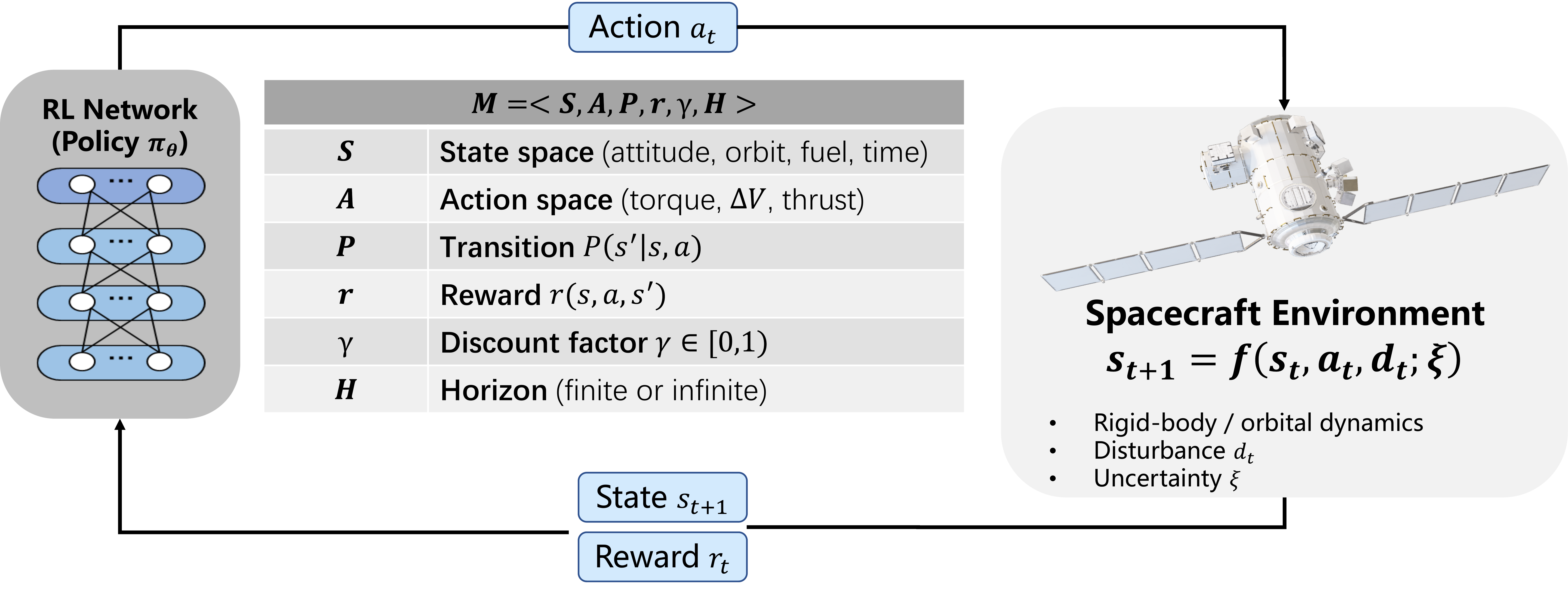}
\caption{MDP decision loop and spacecraft examples.}
\label{fig:spacecraft-mdp-framework}
\end{figure*}

The spacecraft-specific design choice is which information the deployed policy receives and what it controls.
Inputs may include vehicle or relative states, resource queues, and state estimates; actions may command torque, thrust, target selection, or hybrid maneuvers.
Quaternion sign ambiguity, variable-size target sets, actuator saturation, and discrete event logic all affect state or action representation.
Reward functions may combine accuracy, effort, elapsed time, and constraint margins; their scale and timing shape the learned trade-offs.
Reference frames and symmetry-aware representations can improve learning efficiency and generalization, while uncertainty-aware training and evaluation provide a basis for assessing policy robustness under changing spacecraft dynamics.
Figure~\ref{fig:spacecraft-mdp-framework} summarizes how the policy, spacecraft environment, state, action, transition, and reward fit into a decision loop, alongside spacecraft-specific examples of these MDP elements.

\subsection{Extended MDP Formulations for Spacecraft Autonomy}

Several extensions of the standard MDP are relevant to spacecraft autonomy.
These formulations address partial observations, uncertain dynamics, operational constraints, variable decision intervals, and decentralized coordination.

Partially observable Markov decision processes (POMDPs) represent incomplete observations \citep{kaelbling1998planningAnd}.
Given the transition and observation model, a belief over the hidden state is a sufficient statistic; an estimated or recurrent state is not automatically a calibrated belief.
Constrained Markov decision processes (CMDPs) impose bounds on expected cumulative costs under a policy \citep{achiam2017constrainedPolicy}, providing a principled way to manage mission-level constraint budgets but not by itself precluding constraint violations on individual trajectories.
Conditional value at risk (CVaR) instead places explicit weight on the tail of the cumulative cost distribution rather than only its expectation \citep{chow2015riskSensitive}.
Iyengar develops robust dynamic programming for transition uncertainty under rectangular uncertainty sets \citep{iyengar2005robustDynamic}.
Nilim and El Ghaoui study robust control when transition matrices are uncertain \citep{nilim2005robustControl}.
Distributionally robust MDPs optimize against an ambiguity set of probability distributions over uncertain parameters, rather than a single nominal distribution \citep{xu2010distributionallyRobust}.
A semi-Markov decision process (semi-MDP) represents decisions with variable durations; elapsed time affects the transition, accumulated reward, and discounting, as formalized for temporal abstraction by Sutton et al. \citep{sutton1999betweenMdps}.
A decentralized partially observable Markov decision process (Dec-POMDP) models cooperative agents whose joint actions affect shared dynamics and team reward while each agent acts from local observation histories; Bernstein et al. analyze the complexity of decentralized control in this setting \citep{bernstein2002complexity}.
Duff formulates a Bayes-adaptive MDP by augmenting the physical state with a Bayesian information state over unknown transition probabilities; observed transitions update that state, allowing the policy to weigh immediate reward against information that can improve later decisions \citep{duff2002optimal}.
Table~\ref{tab:decision-models} summarizes these distinctions.

These are modeling choices rather than application categories.
Section~\ref{sec:mdp-variants} connects them to spacecraft tasks while distinguishing explicit formulations from similar problem structures, such as randomized tests that do not define a robust MDP.
A recurrent state can summarize observation history but does not itself guarantee a calibrated belief, and reward penalties can encourage constraint compliance without enforcing a hard bound.

\begin{table*}[]
\caption{Decision formulations and illustrative spacecraft uses.}
\label{tab:decision-models}
\centering
\setlength{\tabcolsep}{3pt}
\renewcommand{\arraystretch}{1.0}
\begin{tabularx}{\textwidth}{@{}lYYY@{}}
\toprule
Formulation & Core distinction & Spacecraft applications and potential uses & Key evaluation check \\
\midrule
MDP & Fully observed Markov state; fixed-duration decisions; transition may be unknown to the learner & Illustrative: attitude feedback, powered descent, trajectory planning & Check state sufficiency and performance over the modeled range \\
POMDP & Partial observations; policy uses observation history or a model-based belief & Reviewed: explicit POMDP formulations in active imaging and reconstruction, visual rendezvous/docking and proximity operations, and mission planning; implementations include belief-based approaches, image-conditioned policies, and frame-stacked observations \citep{brandonisio2021reinforcementLearning,brandonisio2024deepReinforcement,supreeth2026visionGuided,zhang2026sampleEfficient,harris2019spacecraftDecision,harris2022generationOf} & Validate observation/estimation under sensing limits or degradation \\
CMDP & Bounds expected cumulative cost; distinct from per-trajectory guarantees and tail-risk objectives such as CVaR & Reviewed: debris avoidance \citep{mu2024autonomousSpacecraft}, constraint-aware attitude control \citep{lu2025safeReinforcement}, and lunar-rover planning with action masks, Lagrangian constraint handling, and reward penalties \citep{lu2026constrainedReinforcement}.
Potential: propellant, thermal, actuator, and pointing budgets & Report expected costs and per-run violations; add tail-risk metrics when relevant \\
Robust MDP & Optimizes worst-case expected return over a set of transition models & Potential: bounded uncertainty in inertia, aerodynamics, gravity, or disturbances & Justify the uncertainty set; compare nominal and worst-case performance \\
Distributionally robust MDP & Optimizes worst-case expected return over an ambiguity set of probability distributions & Potential: uncertain transition distributions or model-parameter distributions & Justify the ambiguity set; report worst-case performance and conservatism \\
Bayes-adaptive MDP & Adds a Bayesian information state over unknown model parameters; observations update it \citep{duff2002optimal} & Potential: decisions that learn unknown target or vehicle parameters & Assess the prior and value of information-seeking actions \\
Semi-MDP & Models variable-duration decisions, including elapsed time in rewards and discounting & Reviewed: Earth-observation scheduling as a partially observable semi-Markov scheduling formulation \citep{stephenson2025learningPolicies} & Account for duration-dependent reward and decision timing \\
Dec-POMDP & Cooperative decentralized control with local observation histories; joint actions determine transitions and shared team reward & Related: multi-agent POMDP studies; inspection and formation studies use local observations without explicitly specifying complete Dec-POMDP models \citep{luo2026researchOn,zheng2025deepReinforcement,dunlap2024deepReinforcement,jin2026deepReinforcement} & Specify local policies, communication, and team-size assumptions \\
\bottomrule
\end{tabularx}
\end{table*}

\subsection{RL Learning Paradigms and System Architectures}

The distinction between single-agent and multi-agent RL is about who observes, decides, and receives reward.
Foerster et al. study agents that maximize shared utility and note that one agent's learning can make the environment non-stationary for the others \citep{foerster2016learningTo}.
In a multi-spacecraft task, a maneuver can also change another vehicle's opportunity or risk, making it harder to attribute mission-level outcomes to individual decisions.
Centralized training with decentralized execution (CTDE) can use global information during learning, but it does not remove the execution-time information gap.
Fully decentralized learning lacks a centralized critic, while communication delay, packet loss, and changing team size further affect coordination.

Algorithm choice is a separate question.
The model-based Dreamer method learns behavior through latent imagination \citep{hafner2019dreamTo}.
Conservative Q-learning addresses offline policy learning from a fixed dataset \citep{kumar2020conservativeQ}.
Model-agnostic meta-learning trains an initialization for rapid adaptation \citep{finn2017modelAgnostic}.
Rakelly et al. use probabilistic context variables for off-policy meta-reinforcement learning (meta-RL) \citep{rakelly2019efficientOff}.
Levy et al. learn multi-level temporal hierarchies with hindsight \citep{levy2017learningMulti}.
Nachum et al. introduce an off-policy hierarchical reinforcement learning agent that trains higher- and lower-level policies from off-policy experience and demonstrates high sample efficiency in simulated robot tasks \citep{nachum2018dataEfficient}.
These learning paradigms offer complementary capabilities for spacecraft autonomy, from model-based planning and offline policy learning to rapid adaptation and hierarchical decision-making.
Across spacecraft tasks, RL may issue actuator commands, adapt guidance parameters, generate references, assist optimizers, or make mission-level decisions; these roles differ in decision variables, time scales, and constraints.
Validation records distinguish model fidelity, test coverage, and execution platform, including closed-loop hardware-in-the-loop (HIL), physical testbeds, and flight tests.
Table~\ref{tab:unified-taxonomy} organizes these dimensions for consistent comparison across applications, recording model fidelity, test coverage, and execution platform separately.
\begin{table*}[]
\caption{Comparison dimensions for spacecraft-RL studies.}
\label{tab:unified-taxonomy}
\centering
\begin{tabularx}{\textwidth}{@{}P{0.19\textwidth}YY@{}}
\toprule
Comparison dimension & What is recorded & Cross-study question \\
\midrule
Mission task and decision level & Domain/objective; feedback control, guidance/reference generation, trajectory/maneuver planning, or mission planning/scheduling; decision variables, targets/requests/resources, and cadence & Which mission function and decision level are addressed? \\
RL role & Direct policy; residual/adaptive augmentation; reference or maneuver generation; heuristic/operator selection; optimizer assistance or warm start; estimator/context adaptation; supervisory/safety decision & What does RL decide or modify? \\
Decision model & Formulation; state/observation/history/belief; action; transition model/uncertainty; reward/cost; decision horizon & Is the model stated explicitly or inferred from the task? \\
Agent organization and information & Agent number (single or multi-agent); training/execution structure (centralized, CTDE, or fully decentralized); objective (cooperative or adversarial); communication, team size, and training--execution information gap & Who acts, with what information, and when? \\
System architecture & Stand-alone, hierarchical, or hybrid RL; retained controller, optimizer/planner, estimator/perception module, and safety monitor/filter & Which modules remain alongside RL? \\
Constraint treatment & Reward/cost; expected-cost constraint; action mask; projection/barrier; shield; runtime filter; feasibility repair & How are constraints handled, and what does the mechanism cover? \\
Baseline and comparison & Baseline method; observation and model information; constraint consistency; training, tuning, and online-computation budgets & Are information and resources comparable? \\
Model fidelity & Analytical/simplified dynamics, higher-fidelity simulation, and model validation & How realistic is the tested model? \\
Test coverage & Nominal, randomized/Monte Carlo, stress, and out-of-distribution cases; training--test relationship & Which conditions were varied or held out? \\
Execution platform and reporting & Simulation; embedded timing; processor-in-the-loop; closed-loop controller HIL; optical/sensor bench; air-bearing/free-floating test; flight; outcomes, runtime, failures, statistics, and code/data availability & Which policies and system components ran on each platform, and what evidence was reported? \\
\bottomrule
\end{tabularx}
\end{table*}

\begin{figure*}[]
\centering
\begin{tikzpicture}[>=Latex,
  task/.style={draw=black!55,rounded corners=2pt,fill=black!5,align=center,text width=2.05cm,minimum height=8mm,font=\scriptsize},
  role/.style={draw=black!55,rounded corners=2pt,fill=blue!6,align=center,text width=3.40cm,minimum height=8mm,font=\scriptsize},
  module/.style={draw=black!55,rounded corners=2pt,fill=green!7,align=center,text width=3.05cm,minimum height=8mm,font=\scriptsize},
  bound/.style={draw=black!55,rounded corners=2pt,fill=black!3,align=center,text width=3.05cm,minimum height=8mm,font=\scriptsize},
  head/.style={font=\small\bfseries,align=center}]
\node[head] at (0,0.7) {Mission domain};
\node[head] at (3.6,0.7) {What RL decides};
\node[head] at (7.5,0.7) {Conventional modules retained};
\node[head] at (11.4,0.7) {What was tested};
\node[task] (a1) at (0,0) {Attitude};
\node[role] (b1) at (3.6,0) {Direct torque / gain tuning / gimbal steering};
\node[module] (c1) at (7.5,0) {Controller / observer / safety filter};
\node[bound] (d1) at (11.4,0) {Faults, saturation, timing};
\node[task] (a2) at (0,-1.05) {Orbital guidance};
\node[role] (b2) at (3.6,-1.05) {Direct thrust / optimizer tuning / warm start / uncertainty adaptation};
\node[module] (c2) at (7.5,-1.05) {$Q$-law / optimizer / targeter};
\node[bound] (d2) at (11.4,-1.05) {Off-nominal dynamics};
\node[task] (a3) at (0,-2.10) {Proximity};
\node[role] (b3) at (3.6,-2.10) {Approach / guidance tuning / sensing / arm};
\node[module] (c3) at (7.5,-2.10) {Navigation / tracker / safety filter};
\node[bound] (d3) at (11.4,-2.10) {Sensing and contact};
\node[task] (a4) at (0,-3.15) {Landing};
\node[role] (b4) at (3.6,-3.15) {Thrust / guidance tuning / route or hop};
\node[module] (c4) at (7.5,-3.15) {Guidance / estimator / planner};
\node[bound] (d4) at (11.4,-3.15) {Image loop versus dynamics};
\node[task] (a5) at (0,-4.20) {Scheduling};
\node[role] (b5) at (3.6,-4.20) {Schedule / request / search / repair};
\node[module] (c5) at (7.5,-4.20) {Feasibility checks / optimizer};
\node[bound] (d5) at (11.4,-4.20) {Requests absent from training};
\node[task] (a6) at (0,-5.25) {Multi-spacecraft};
\node[role] (b6) at (3.6,-5.25) {Control / reference / sensing / reconfiguration / game};
\node[module] (c6) at (7.5,-5.25) {Tracker / safety layer / coordinator};
\node[bound] (d6) at (11.4,-5.25) {Communication and team shift};
\foreach \i in {1,...,6}{\draw[->,black!65] (a\i.east)--(b\i.west);\draw[->,black!65] (b\i.east)--(c\i.west);\draw[->,black!65] (c\i.east)--(d\i.west);}
\end{tikzpicture}
\caption{RL roles, retained modules, and test issues across six mission domains.}
\label{fig:review-framework}

\end{figure*}
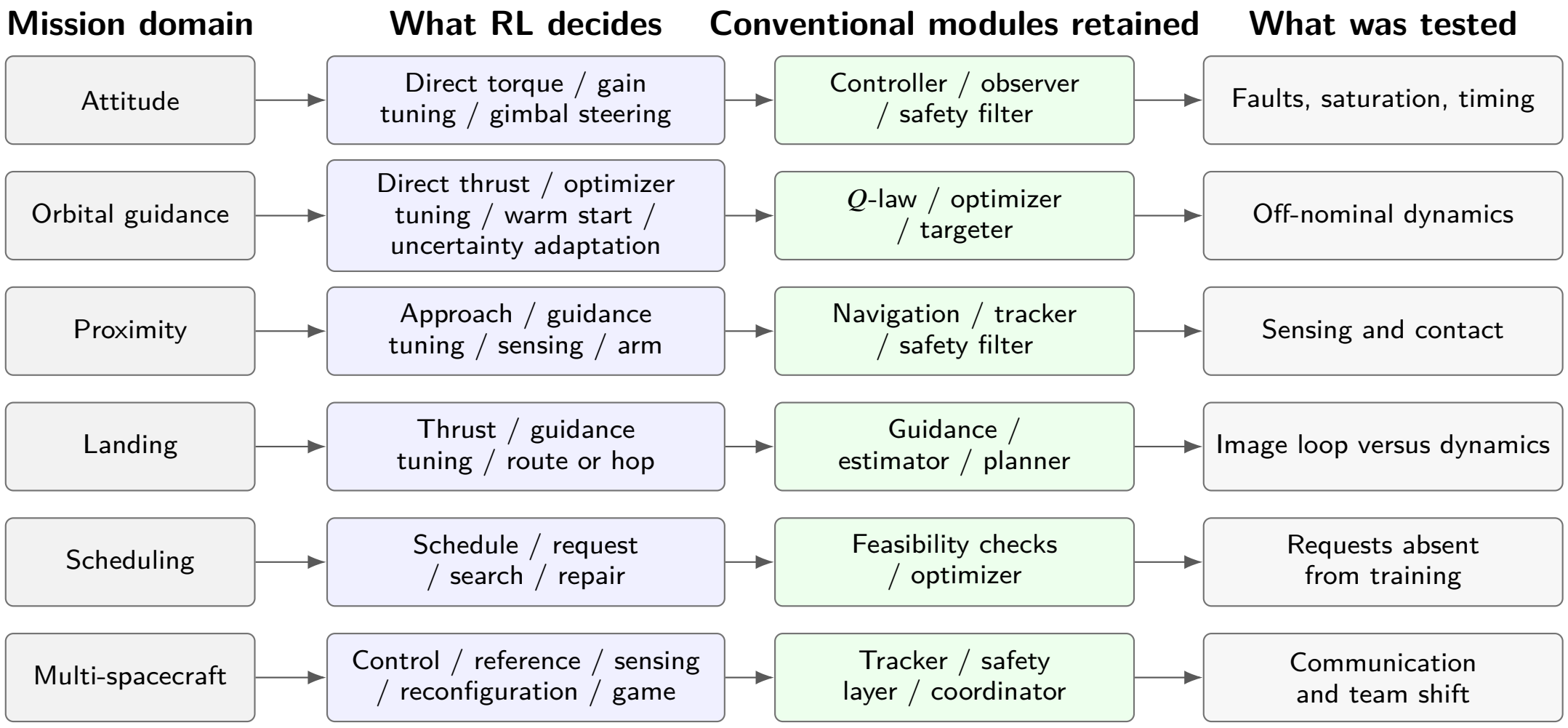

Across the six mission domains, RL takes on these roles while conventional controllers, estimators, planners, and safety mechanisms may remain part of the system.
Figure~\ref{fig:review-framework} summarizes how the policy decision, retained modules, and test conditions vary across those domains.

\begin{figure}[]
\centering
\includegraphics[width=0.94\columnwidth]{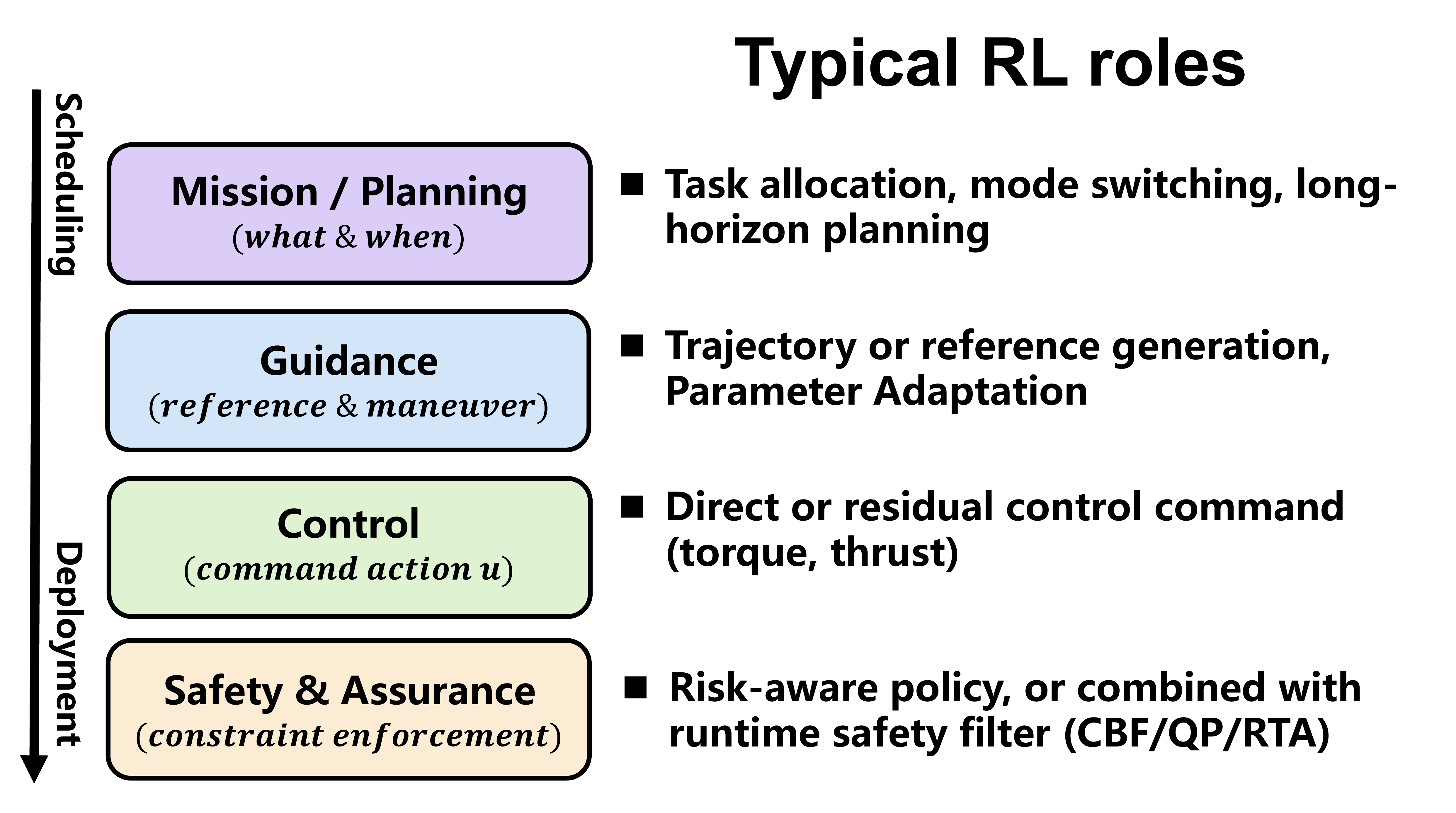}
\caption{Mission and control decisions influenced by RL.}
\label{fig:rl-decision-authority}
\end{figure}

The same policy role can act at different levels, from actuator control and guidance to planning and safety-related decisions.
Figure~\ref{fig:rl-decision-authority} groups these roles by the mission decision they affect.

\begin{figure}[]
\centering
\begin{subfigure}[t]{0.48\columnwidth}\centering
  \includegraphics[width=\linewidth]{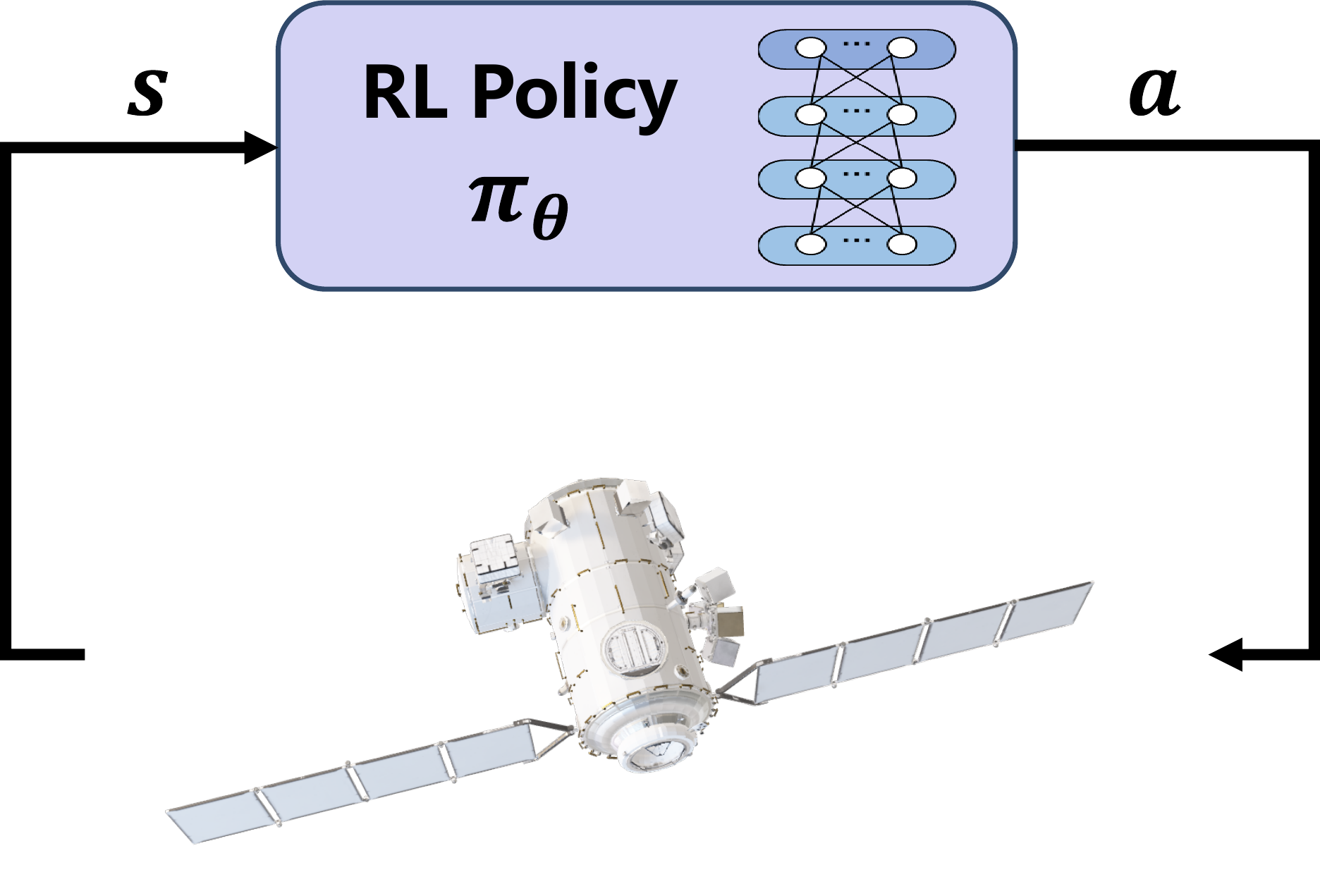}
  \caption{Single-agent RL}
\end{subfigure}\hfill
\begin{subfigure}[t]{0.48\columnwidth}\centering
  \includegraphics[width=\linewidth]{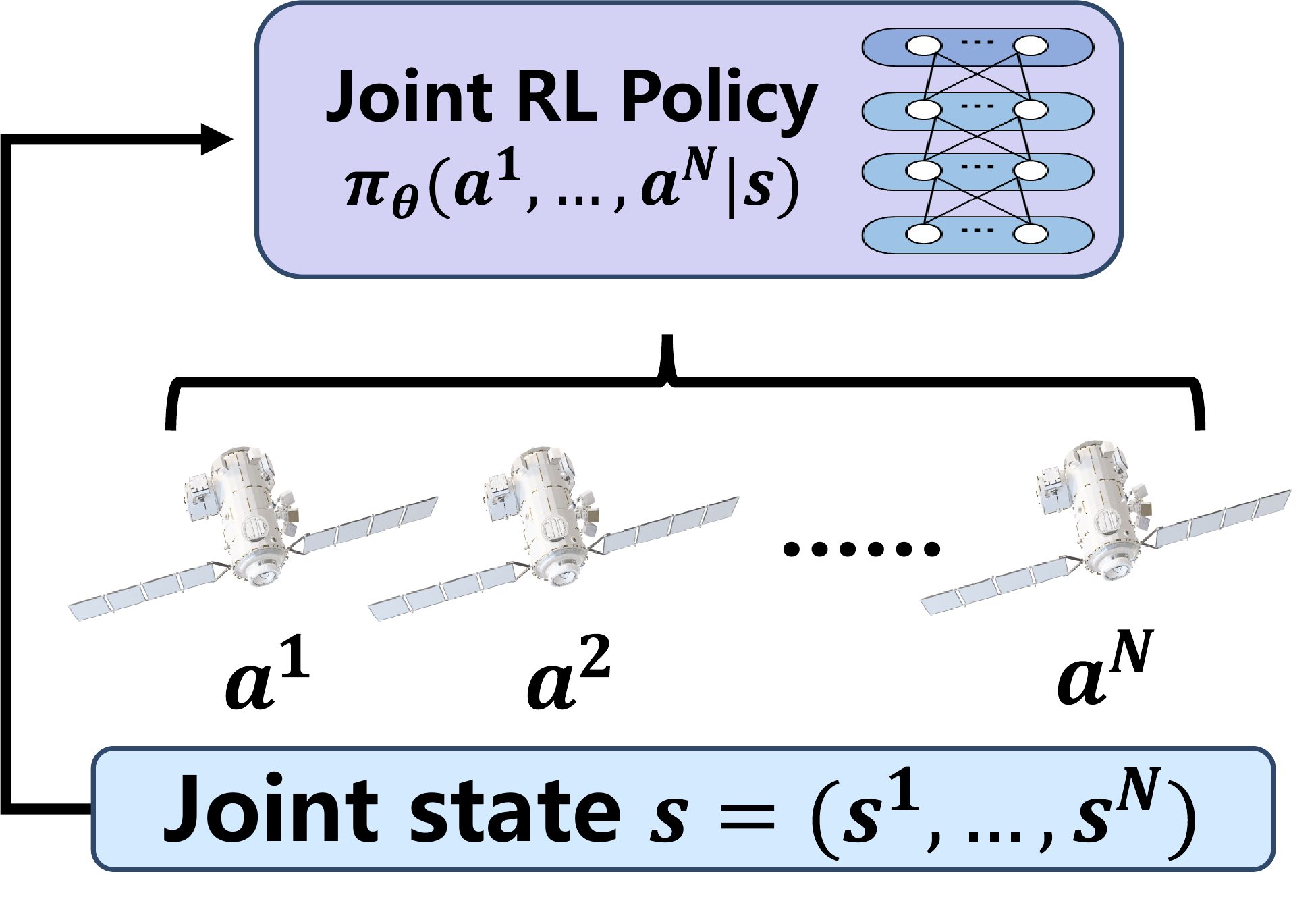}
  \caption{Centralized execution}
\end{subfigure}\par\smallskip
\begin{subfigure}[t]{0.48\columnwidth}\centering
  \includegraphics[width=\linewidth]{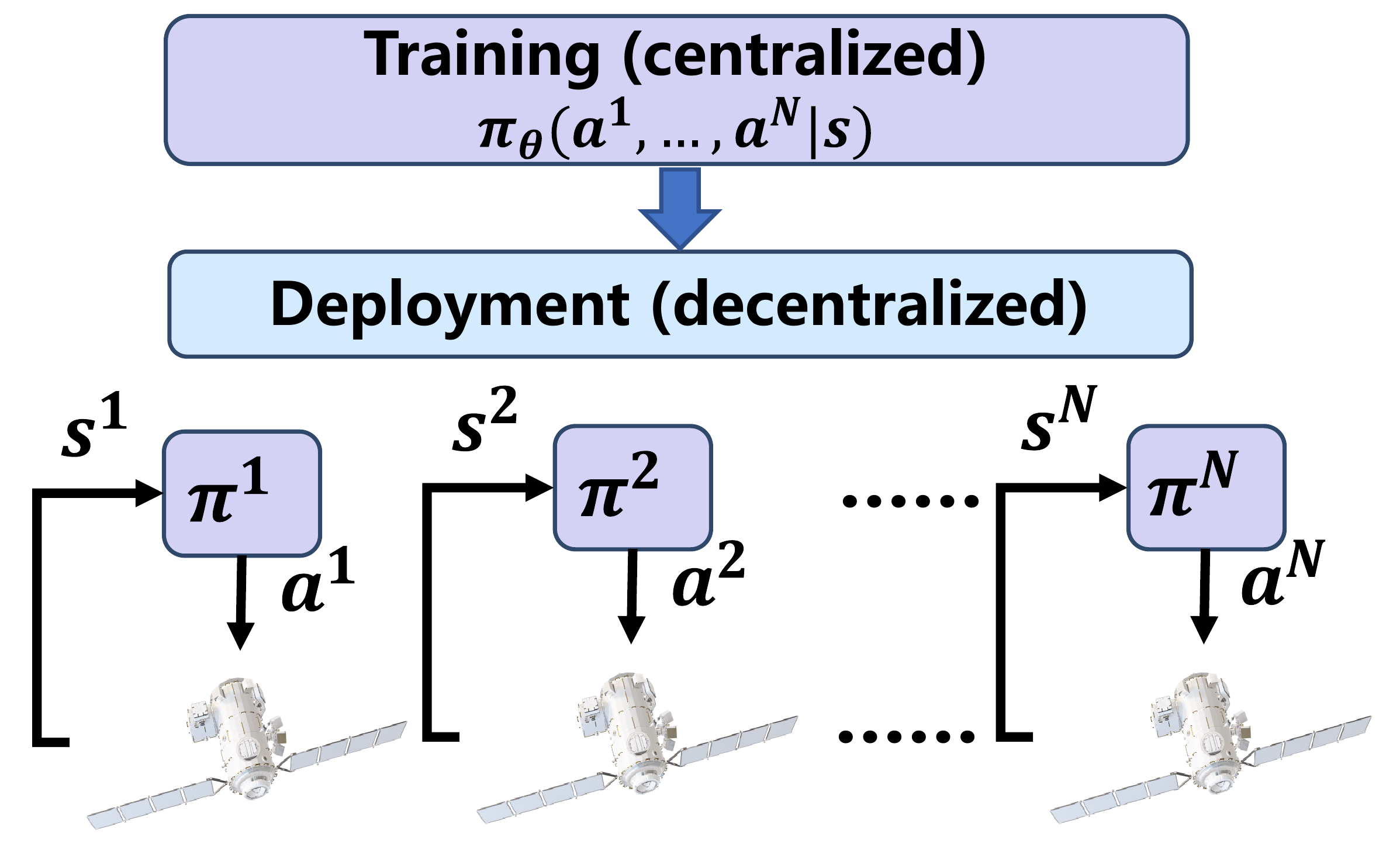}
  \caption{CTDE}
\end{subfigure}\hfill
\begin{subfigure}[t]{0.48\columnwidth}\centering
  \includegraphics[width=\linewidth]{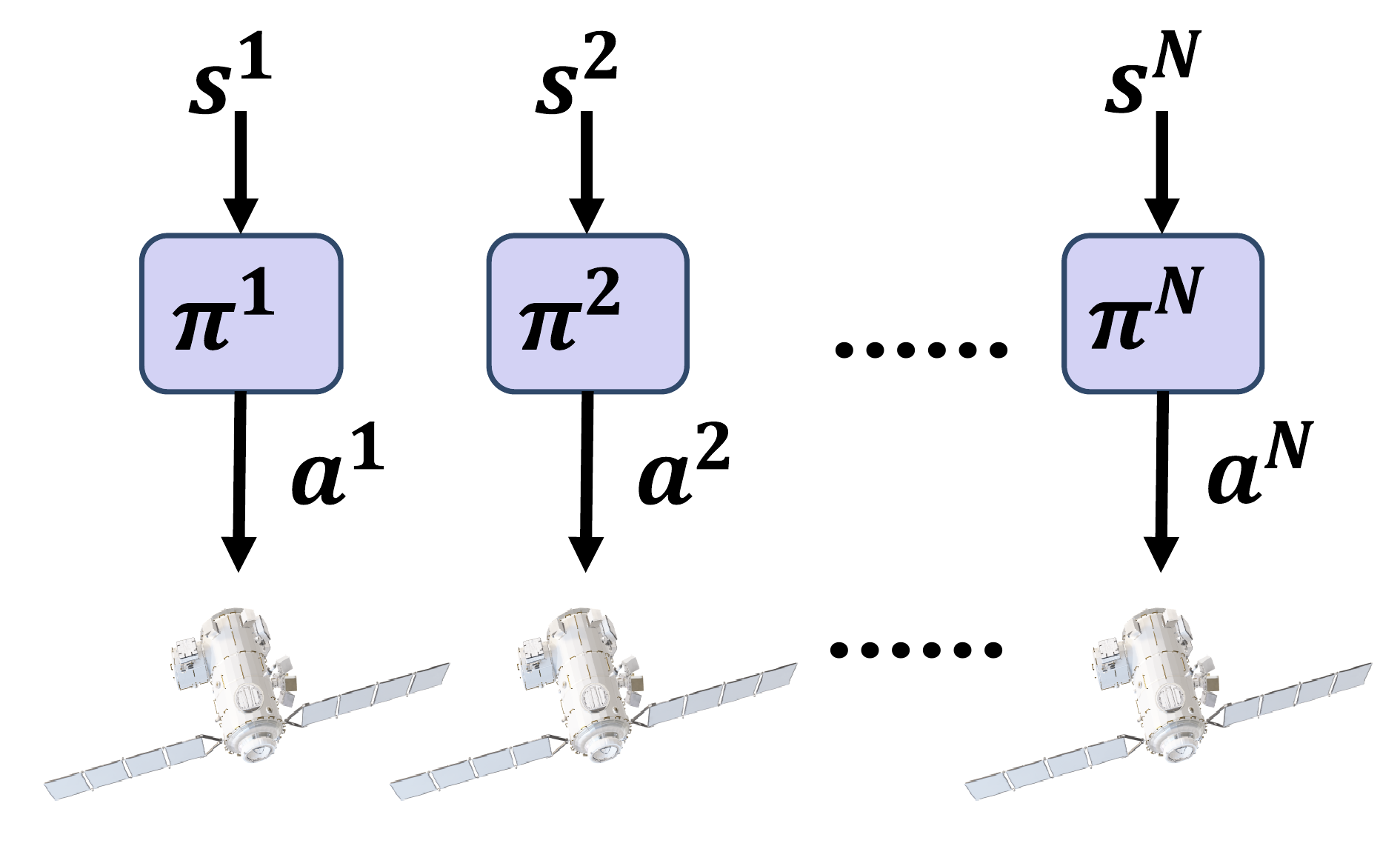}
  \caption{Decentralized agents}
\end{subfigure}\par\smallskip
\begin{subfigure}[t]{0.96\columnwidth}\centering
  \includegraphics[width=\linewidth]{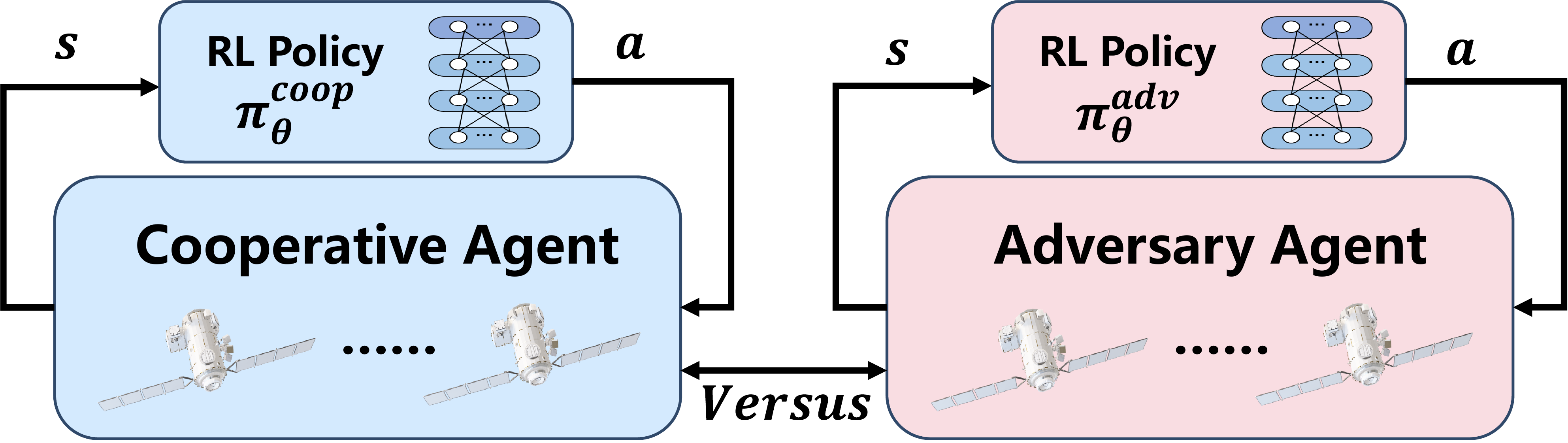}
  \caption{Adversarial RL}
\end{subfigure}
\caption{Single-agent and multi-agent organizations for spacecraft RL.}
\label{fig:agent-organization}
\end{figure}

Application domain and agent organization are separate: a multi-spacecraft mission need not use multi-agent reinforcement learning (MARL), and MARL also applies beyond spacecraft swarms.
Agent organizations can differ in how policies share information and divide training and execution, and these design patterns may overlap.
Figure~\ref{fig:agent-organization} contrasts single-agent, centralized, CTDE, fully decentralized, and adversarial organizations.

\begin{figure*}[]
\centering
\begin{subfigure}[t]{0.19\textwidth}\centering
  \includegraphics[height=1.45in,keepaspectratio]{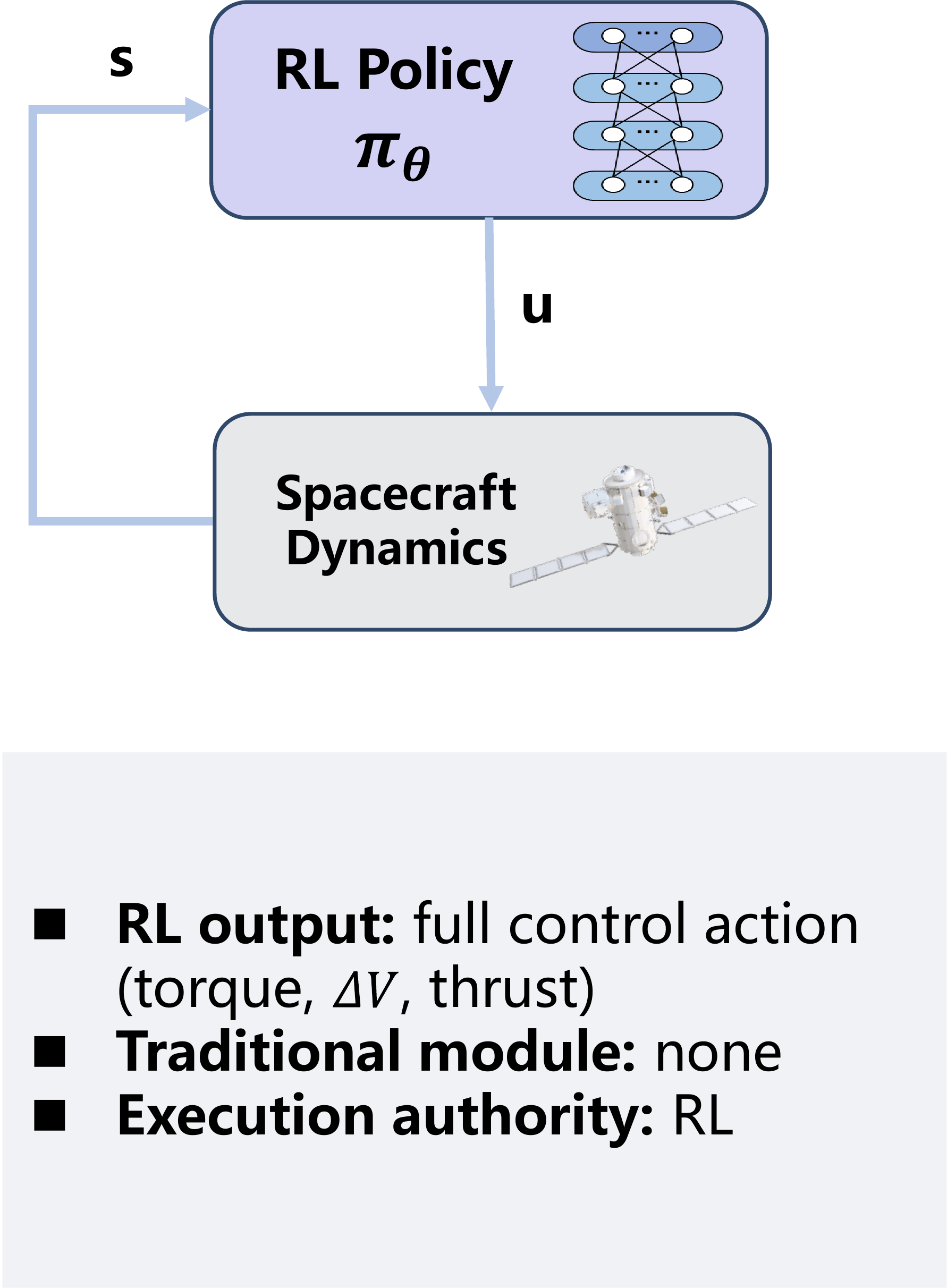}
  \caption{Direct RL}
\end{subfigure}\hfill
\begin{subfigure}[t]{0.19\textwidth}\centering
  \includegraphics[height=1.45in,keepaspectratio]{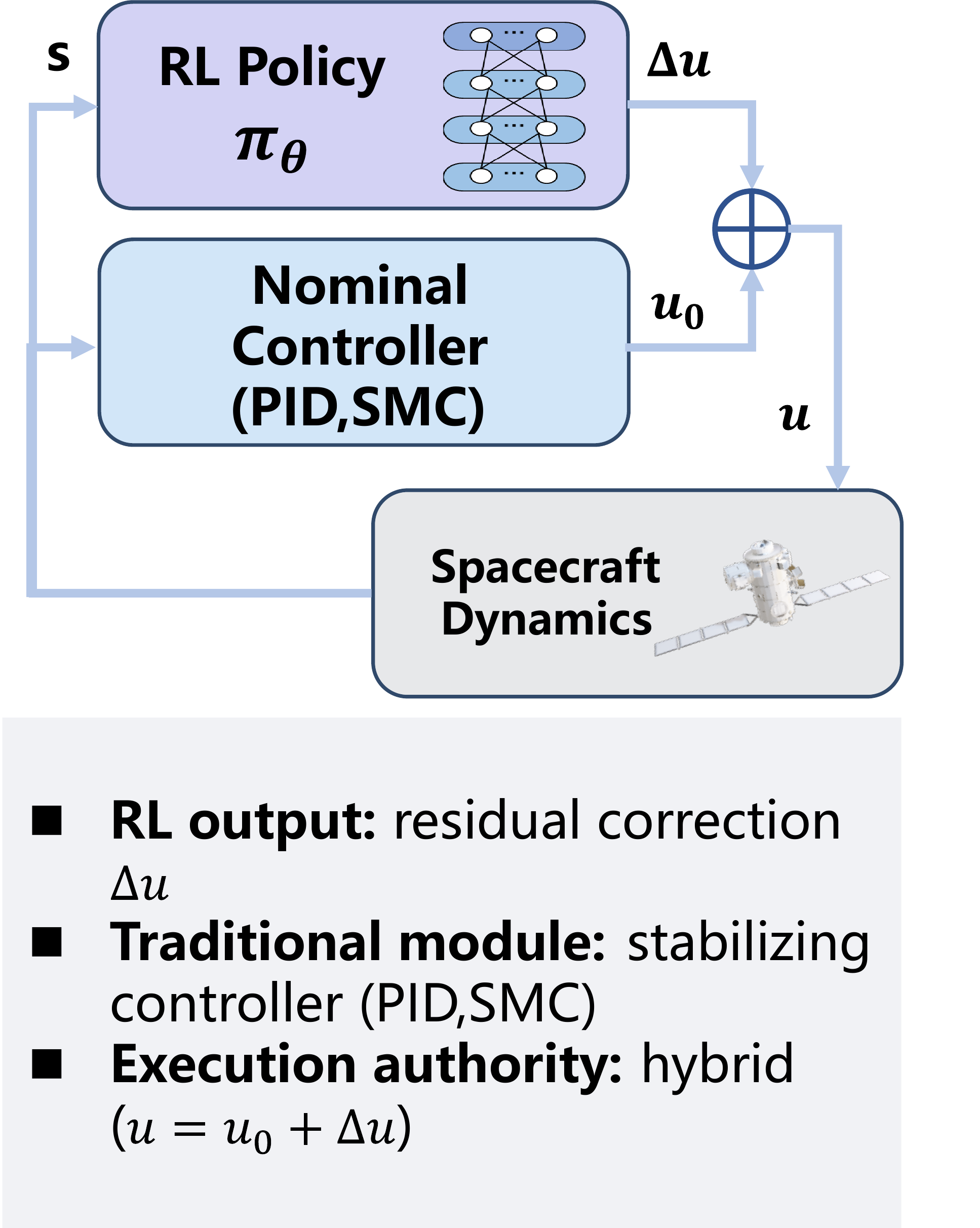}
  \caption{Residual RL}
\end{subfigure}\hfill
\begin{subfigure}[t]{0.19\textwidth}\centering
  \includegraphics[height=1.45in,keepaspectratio]{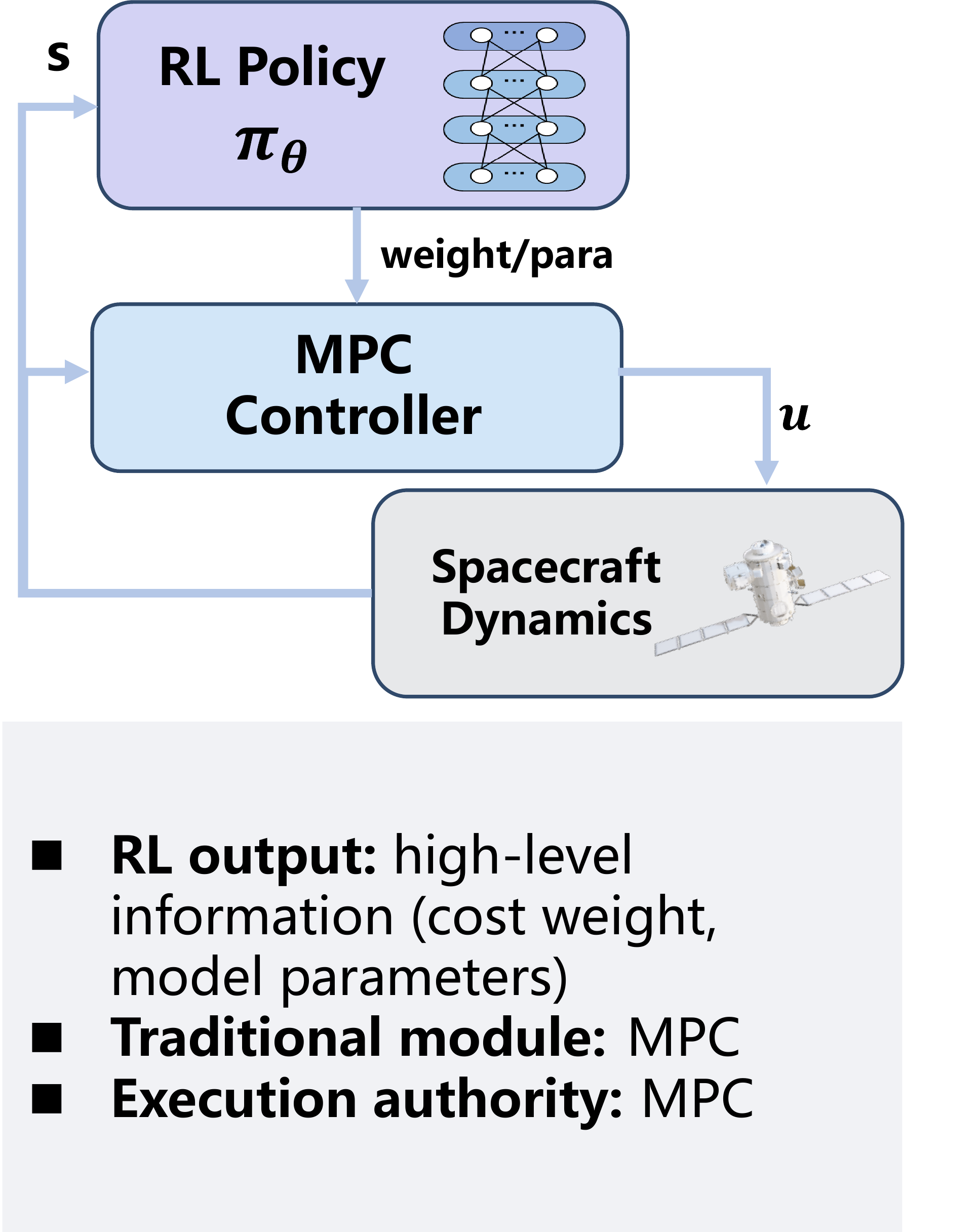}
  \caption{RL-assisted MPC}
\end{subfigure}\hfill
\begin{subfigure}[t]{0.19\textwidth}\centering
  \includegraphics[height=1.45in,keepaspectratio]{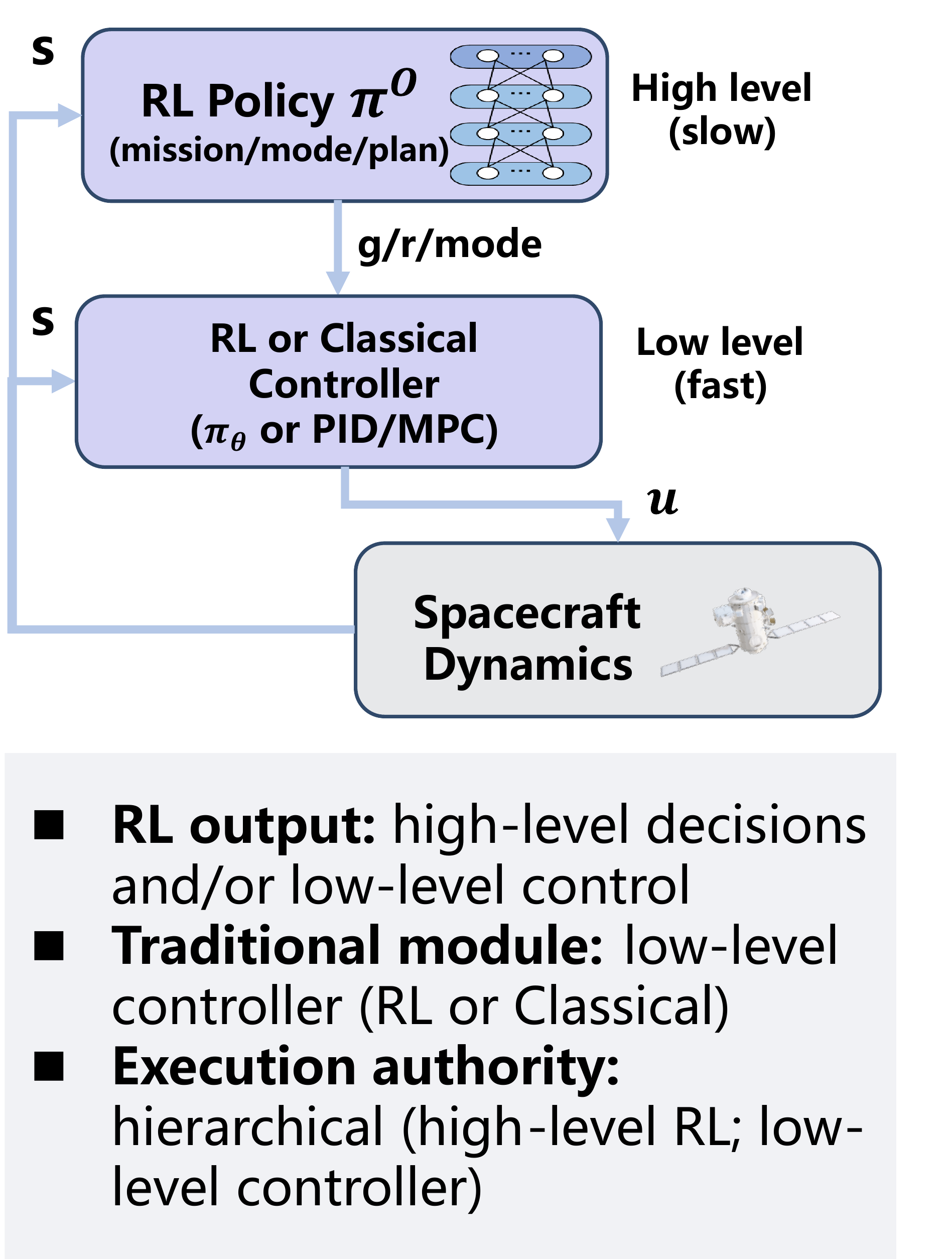}
  \caption{Hierarchical control with RL}
\end{subfigure}\hfill
\begin{subfigure}[t]{0.19\textwidth}\centering
  \includegraphics[height=1.45in,keepaspectratio]{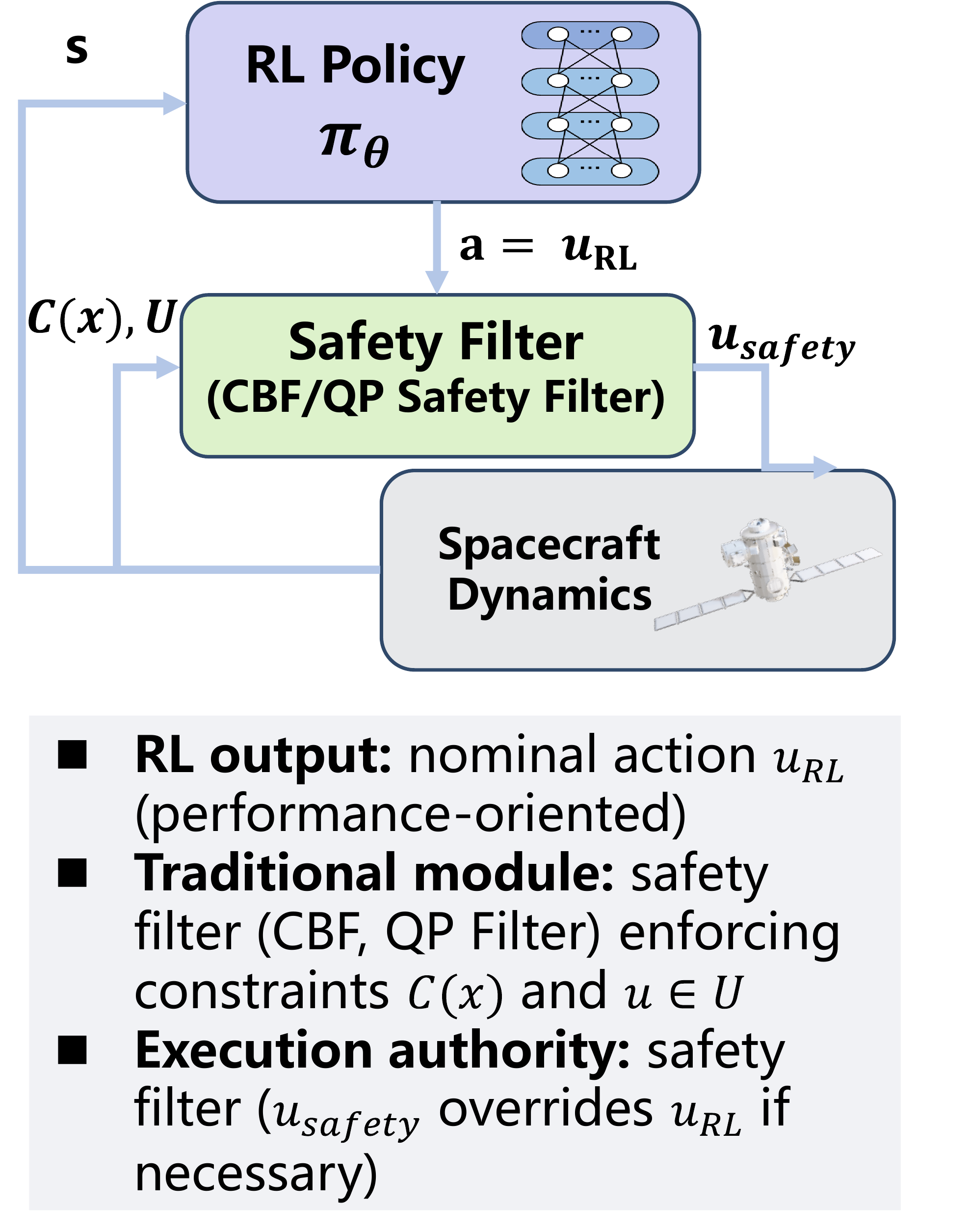}
  \caption{Safety-filtered RL (CBF/QP)}
\end{subfigure}
\caption{Common ways to integrate RL with spacecraft control and planning.}
\label{fig:rl-architectures}
\end{figure*}

Agent organization concerns how policies are arranged across agents, information, training, and execution; control architecture concerns how a learned policy connects to conventional modules.
Direct RL issues actions, residual RL augments a baseline controller, RL-assisted model predictive control (MPC) guides optimization, hierarchical control with RL assigns decisions across levels, and CBF constraints in a quadratic program (QP) safety filter modify candidate policy actions.
These architectures can overlap, and Figure~\ref{fig:rl-architectures} contrasts how they assign decision authority between the learned policy and retained modules.

The architectures in Figure~\ref{fig:rl-architectures} can be summarized by the following schematic input--output relations.
\paragraph{Direct RL.}
\begin{equation}
u_t=\pi_\theta(s_t).
\end{equation}
\paragraph{Residual RL.}
\begin{equation}
\left\{
\begin{aligned}
u_t&=u_{\mathrm{nom}}(s_t)+\Delta u_t,\\
\Delta u_t&=\pi_\theta(s_t).
\end{aligned}
\right.
\end{equation}
\paragraph{RL-assisted MPC with parameter adaptation.}
\begin{equation}
\left\{
\begin{aligned}
p_t&=\pi_\theta(s_t),\\
\mathbf{U}_t^*&\in\arg\min_{\mathbf{U}\in\mathcal{F}(s_t;p_t)}\\
&\qquad J_{\mathrm{MPC}}(s_t,\mathbf{U};p_t),\\
u_t&=[\mathbf{U}_t^*]_0.
\end{aligned}
\right.
\end{equation}
\paragraph{RL-assisted MPC with a warm start.}
\begin{equation}
\left\{
\begin{aligned}
\mathbf{U}_t^{(0)}&=\pi_\theta(s_t),\\
\mathbf{U}_t^*&=\operatorname{SolveMPC}(s_t;\mathbf{U}_t^{(0)}),\\
u_t&=[\mathbf{U}_t^*]_0.
\end{aligned}
\right.
\end{equation}
\paragraph{Hierarchical control with RL.}
\begin{equation}
\left\{
\begin{aligned}
\eta_k&=\pi_\theta^{\mathrm{high}}(s_{t_k}),\\
u_t&=\kappa(s_t,\eta_k),\quad t_k\leq t<t_{k+1}.
\end{aligned}
\right.
\end{equation}
\paragraph{Safety-filtered RL.}
\begin{equation}
\left\{
\begin{aligned}
u_{\mathrm{RL},t}&=\pi_\theta(s_t),\\
u_t&=\mathcal{F}_{\mathrm{safe}}(s_t,u_{\mathrm{RL},t}).
\end{aligned}
\right.
\end{equation}
Here, $\theta$ denotes learned policy parameters, $s_t$ is the policy input, $u_{\mathrm{nom}}$ is the baseline-controller command, and $\Delta u_t$ is the learned residual.
$p_t$ denotes MPC parameters, and $\mathcal{F}(s_t;p_t)$ is the feasible set determined by the prediction model and applicable state, input, and other constraints.
$\mathbf{U}_t^{(0)}$ is the solver's initial guess, $\operatorname{SolveMPC}$ returns the optimized sequence $\mathbf{U}_t^*$, and $[\mathbf{U}_t^*]_0$ is its first command.
$\eta_k$ is the high-level goal or reference held from update time $t_k$ until $t_{k+1}$, and $\kappa$ denotes the retained low-level controller or policy.
$u_{\mathrm{RL},t}$ is the candidate RL command and $\mathcal{F}_{\mathrm{safe}}$ is the safety filter that returns the applied command.
The policy expressions denote deployed action outputs; for a stochastic policy, the output is sampled from $\pi_\theta(\cdot\mid s_t)$.
The warm-start relation applies when RL initializes the solver rather than changing the MPC objective or model.
The safety-filter relation refers to action filtering or projection, with the CBF/QP formulation given in Section~\ref{sec:constraint-safety}.

The value of each design depends on both the decisions assigned to RL and the conditions under which it is evaluated.
The abbreviation DoF denotes degrees of freedom; numeric prefixes indicate the count, as in 3-DoF and 6-DoF.
To keep the comparison concise, Table~\ref{tab:domain-evidence-overview} presents two selected studies per domain to illustrate differing RL decision roles, system configurations, and validation settings.
\begin{table*}[]
\caption{Selected studies illustrating RL decision roles and validation evidence across six application domains.}
\label{tab:domain-evidence-overview}
\centering
\begingroup
\footnotesize
\setlength{\tabcolsep}{3pt}
\renewcommand{\arraystretch}{1.0}
\begin{tabularx}{\textwidth}{@{}P{0.13\textwidth}P{0.14\textwidth}P{0.12\textwidth}P{0.18\textwidth}P{0.12\textwidth}Y@{}}
\toprule
Study and task & Policy input & RL decision & Modules and constraints & Comparator & Tests and limits \\
\midrule
Attitude: hybrid tracking \citep{sarjoughian2026hardwareValidated} & Tracking error and derivative & Incremental torque & Super-twisting/fuzzy control; wheel-torque bounds & Super-twisting, fuzzy, RL & High-fidelity simulation; closed-loop HIL on a reaction-wheel platform.
HIL rankings vary with noise and unmodeled dynamics; stability is retained \\
Attitude: magnetic control \citep{munoz2025feasibilityOf} & Magnetometer derivatives/averages; angular-velocity error & Three-axis magnetorquer duty-cycle command & Magnetic attitude-control model; pulse-width modulation actuation & UPMSat-2 rate requirement & STM32F407 processor-in-the-loop; satellite environment simulated in MATLAB/Simulink; embedded response similar to simulation \\
Orbit: off-nominal recovery \citep{lafarge2023adaptiveClosed} & Off-nominal state; reference orbit & Targeter initial conditions and timing & Neural-network initialization; multiple-shooting/differential-correction targeter & Crossing-control orbit maintenance & Three-body Monte Carlo and an $N$-body ephemeris case; targeter refines the learned initial guess \\
Orbit: station-keeping \citep{tafanidis2025reinforcementLearning} & Relative orbital elements & Thrust direction, start time, burn duration & High-fidelity dynamics; low-thrust limits & Optimal-control station-keeping & Monte Carlo: trained policies run faster after offline training.
Successful optimizer trials use less fuel, but the optimal-control program has a lower success rate in the reported along-track comparison \\
Proximity: rendezvous \citep{yang2022modelBased} & Relative position and velocity & One of 27 discrete three-axis thrust actions via a neural-network-compressed dynamic-programming policy & Clohessy--Wiltshire model; terminal conditions & PID; model-free RL & Relative-motion simulation; policy size/runtime assessed on a workstation, not onboard \\
Proximity: docking \citep{dunlap2023runTime} & Planar chief-relative state & Desired planar force filtered by runtime assurance & Dynamic velocity constraint & No filter; penalty-only; four assurance designs & Planar linearized simulation; includes test points outside training; no hardware execution \\
Landing: powered descent \citep{furfaro2020adaptiveGeneralized} & Position and velocity & Guidance gains; time-to-go & Thrust and glide-slope constraints & Fuel-optimal solution & Sampled two- and three-dimensional descent simulations \\
Landing: image-based GNC \citep{federici2026opticalBench} & Camera images; recurrent history & Three-axis thrust and torque & Recurrent image pipeline; scenario-specific approach/glide-slope and camera field-of-view constraints & Simulated versus optical-bench images & 100-trajectory simulation/optical-bench comparison using camera/Jetson hardware; moderate error increase; scene and spacecraft motion remain simulated \\
Scheduling: solver assistance \citep{li2026solvingThe} & Solution quality; operator weights & Destroy, repair, and position-operator weights & Roulette-wheel selection; position heuristics; tabu and simulated-annealing search & ALNS/TPF; ablations & Repeated computational scheduling instances \\
Scheduling: variable-duration tasking \citep{stephenson2025learningPolicies} & Spacecraft state/resources; request opportunities & Imaging request or charging action & Basilisk simulation; optional power shield & One-hour open-loop schedule; reward upper bound & High-fidelity closed-loop simulation; no hardware execution \\
Multi-spacecraft: formation \citep{jin2026deepReinforcement} & Local state; formation curve & Desired state/velocity references & Rigid-body dynamics; Lloyd refinement is part of the proposed method & Formation-configuration comparisons & Numerical simulations; eight air-bearing simulators track precomputed, not live RL, trajectories \\
Multi-spacecraft: inspection \citep{dunlap2024deepReinforcement} & Chief-relative state; neighbor ranges & Deputy control actions & Runtime assurance filters for translation and rotation constraints & Policies without inter-agent information & 6-DoF multi-agent simulation \\
\bottomrule
\end{tabularx}
\endgroup
\end{table*}

Table~\ref{tab:cross-domain-baselines} summarizes representative conventional baselines and task-dependent RL contributions that warrant comparative evaluation, rather than universally established advantages.

\begin{table}[]
\centering
\caption{Conventional baselines and potential RL contributions across spacecraft application domains.}
\label{tab:cross-domain-baselines}
\footnotesize
\setlength{\tabcolsep}{3pt}
\renewcommand{\arraystretch}{1.0}
\begin{tabularx}{\columnwidth}{@{}P{0.19\columnwidth}P{0.30\columnwidth}Y@{}}
\toprule
Application domain & Conventional baselines & Potential RL contributions to evaluate \\
\midrule
Attitude control & PD, SMC, LQR, MPC & Adaptation to model variations; cross-condition performance \\
Orbital guidance & $Q$-law, indirect methods, direct collocation, MPC & Fast repeated decisions; optimizer warm starts; off-nominal recovery \\
Rendezvous and docking & CW/TH-based guidance, MPC, optimal control & Closed-loop adaptation under uncertainty; online efficiency; perception-guidance coupling \\
Planetary landing & ZEM--ZEV, optimal guidance, trajectory optimization & Adaptation to off-nominal descent; perception-informed guidance \\
Mission scheduling & MILP, ALNS, heuristic optimization & Fast repeated decisions; cross-instance reuse; event-driven replanning \\
Multi-spacecraft coordination & Consensus control, distributed optimization, game-theoretic methods & Decisions with limited communication; changing interactions; coordination performance \\
\bottomrule
\end{tabularx}
\end{table}

\subsection{Constraint-Handling and Safety Mechanisms}
\label{sec:constraint-safety}

Spacecraft RL must account for constraints such as pointing exclusions, actuator limits, keep-out geometry, resource budgets, and operational rules.

Table~\ref{tab:domain-safety-constraints} summarizes common task-level safety and feasibility constraints and operational/resource constraints across the six application domains; the listed constraints are not necessarily enforced simultaneously or explicitly in every reviewed study.

\begin{table*}[]
\centering
\caption{Representative safety, feasibility, and operational constraints across spacecraft RL application domains.}
\label{tab:domain-safety-constraints}
\small
\setlength{\tabcolsep}{5pt}
\renewcommand{\arraystretch}{1.05}
\begin{tabularx}{\textwidth}{@{}P{0.20\textwidth}YY@{}}
\toprule
Application domain & Task-level safety and feasibility constraints & Operational and resource constraints \\
\midrule
Attitude control & Forbidden pointing regions; angular-rate limits; pointing-error limits & Actuator saturation and availability; control-effort limits; maneuver time \\
Orbital guidance & Collision avoidance; exclusion regions; terminal-state and orbital-deviation limits & Propellant budgets; thrust magnitude and duration; maneuver timing \\
Proximity operations & Keep-out zones; collision avoidance; safe approach conditions; docking alignment & Thrust and control limits; fuel and maneuver time \\
Planetary landing and surface mobility & Hazard avoidance; touchdown safety; obstacle clearance; terrain traversability & Fuel and thrust limits; energy and mobility resources \\
Mission planning and scheduling & Time-window compliance; scheduling conflicts; task-precedence requirements & Power, onboard memory, communication, and observation-time budgets \\
Multi-spacecraft coordination & Inter-spacecraft collision avoidance; formation geometry and separation requirements & Communication availability; coordination and maneuver budgets \\
\bottomrule
\end{tabularx}
\end{table*}

Constraint-handling mechanisms address these requirements at different stages of learning and execution.
The following five approaches have distinct scopes and should not be interpreted as a universal hierarchy of safety guarantees.

\paragraph{Reward penalties.}
A penalty incorporates constraint cost $c_t$ into task reward $r_t$, allowing the policy to trade task performance against violations.
For penalty weight $\lambda$ and a finite horizon of $T$ decision steps, a common objective is
\begin{equation}
\max_{\pi}\;\mathbb{E}_{\pi}\!\left[\sum_{t=0}^{T-1}\gamma^t(r_t-\lambda c_t)\right].
\end{equation}
The result depends on penalty scaling and training, and does not by itself guarantee constraint satisfaction on every trajectory.
\paragraph{CMDPs.}
CMDPs specify a budget on expected cumulative constraint cost while maximizing task return \citep{achiam2017constrainedPolicy}.
For discount factor $\gamma$, horizon $T$, and cost budget $d$, the objective is
\begin{equation}
\left\{
\begin{aligned}
\max_{\pi}\;\mathbb{E}_{\pi}\!\left[\sum_{t=0}^{T-1}\gamma^t r_t\right],\\
\text{s.t.}\quad
\mathbb{E}_{\pi}\!\left[\sum_{t=0}^{T-1}\gamma^t c_t\right]\le d.
\end{aligned}
\right.
\end{equation}
This expresses a policy-level cost limit, but individual trajectories may still exceed the budget.
\paragraph{Action masks.}
Action masks remove encoded invalid choices before a discrete action is sampled \citep{lu2026constrainedReinforcement}.
Let $z_{\theta}(s,a)$ denote the policy logit and $\mathcal{A}_{\mathrm{valid}}(s)$ the encoded valid-action set.
The masked logits and policy can be written as
\begin{equation}
\left\{
\begin{aligned}
\tilde z_{\theta}(s,a)=
\begin{cases}
z_{\theta}(s,a), & a\in\mathcal{A}_{\mathrm{valid}}(s),\\
-\infty, & a\notin\mathcal{A}_{\mathrm{valid}}(s),
\end{cases}\\
\pi_{\theta}^{\mathrm{mask}}(a\mid s)
=\operatorname{softmax}_{a}\!\left(\tilde z_{\theta}(s,a)\right),
\quad\mathcal{A}_{\mathrm{valid}}(s)\ne\emptyset.
\end{aligned}
\right.
\end{equation}
It cannot enforce constraints that are absent from the mask or directly handle continuous control values, and an empty valid-action set requires a defined fallback.
\paragraph{CBF/QP filters.}
CBF/QP filters define a safe set and can minimally modify an RL command to satisfy the encoded barrier condition \citep{wijk2024safeSpacecraft,wijayatunga2026metaReinforcement}.
For control-affine dynamics $\dot{x}=f(x)+g(x)u$, safe set $\mathcal{C}=\{x:h(x)\ge0\}$, a relative-degree-one safety function $h(x)$ (so $u$ appears in its first derivative), candidate command $u_{\mathrm{RL}}$, and actuator set $\mathcal{U}$, the filter is
\begin{equation}
\left\{
\begin{aligned}
u_{\mathrm{safe}}=\arg\min_{u\in\mathcal{U}}
\frac{1}{2}\lVert u-u_{\mathrm{RL}}\rVert_W^2,\\
\text{s.t.}\quad
\nabla h(x)^{\mathsf T}\!\left(f(x)+g(x)u\right)
+\alpha\!\left(h(x)\right)\ge0.
\end{aligned}
\right.
\end{equation}
Here, $W\succ0$ is a weighting matrix and $\alpha$ is an appropriate extended class-$\mathcal{K}$ function.
For higher-relative-degree constraints, this first-order condition is insufficient; a high-order CBF or another suitable reformulation is required.
Forward invariance additionally requires an initially safe state $x(0)\in\mathcal{C}$, feasibility of the barrier-constrained control input, and the regularity assumptions underlying the CBF condition.
In practice, the safety property also depends on model accuracy, sampling, and state estimation.
\paragraph{Runtime assurance.}
Runtime assurance uses a monitor to accept an RL command when its assurance test passes and otherwise switch to a backup controller \citep{dunlap2023runTime}.
With monitor $\mathcal{M}_{\mathrm{RTA}}$, the applied control is
\begin{equation}
u_t=\begin{cases}
u_{\mathrm{RL},t}, & \mathcal{M}_{\mathrm{RTA}}(x_t,u_{\mathrm{RL},t})=\mathrm{pass},\\
u_{\mathrm{backup},t}, & \mathcal{M}_{\mathrm{RTA}}(x_t,u_{\mathrm{RL},t})=\mathrm{fail}.
\end{cases}
\end{equation}
Here, $x_t$ is the monitored state, $u_{\mathrm{RL},t}$ is the candidate learned command, and $u_{\mathrm{backup},t}$ is the backup-controller command.
The assurance claim depends on monitor correctness, backup-controller safety, and the conditions under which switching preserves safety.

\begin{figure}[]
\centering
\begin{subfigure}[t]{0.92\columnwidth}\centering
  \includegraphics[width=\linewidth]{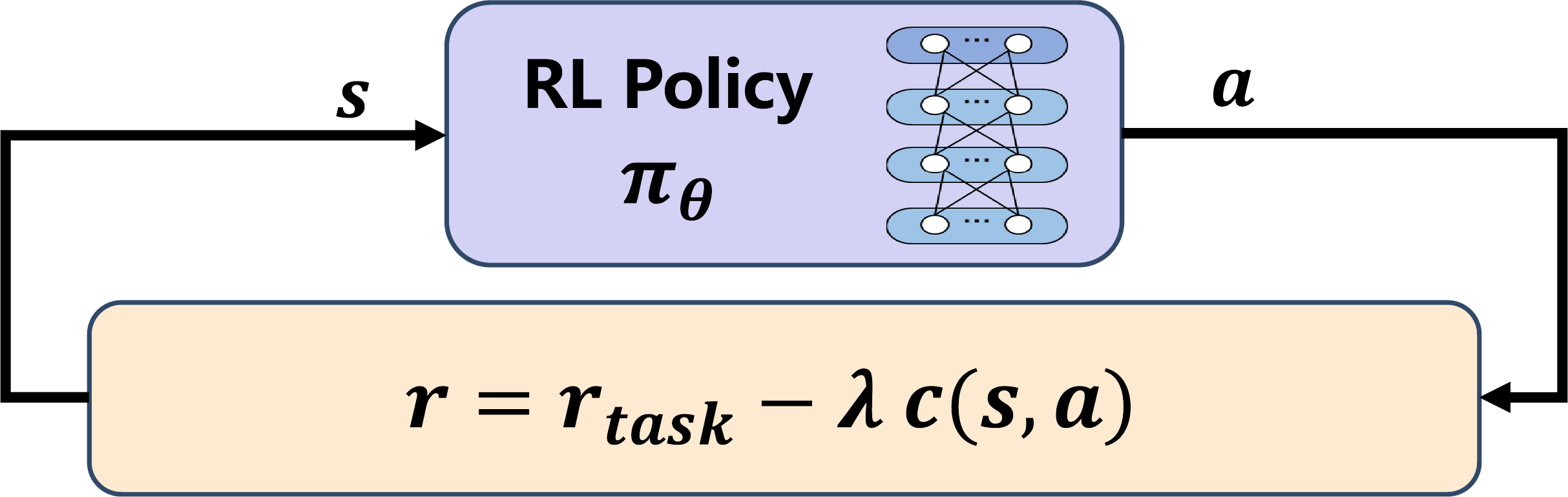}
  \caption{Reward penalty}
\end{subfigure}\par\smallskip
\begin{subfigure}[t]{0.92\columnwidth}\centering
  \includegraphics[width=\linewidth]{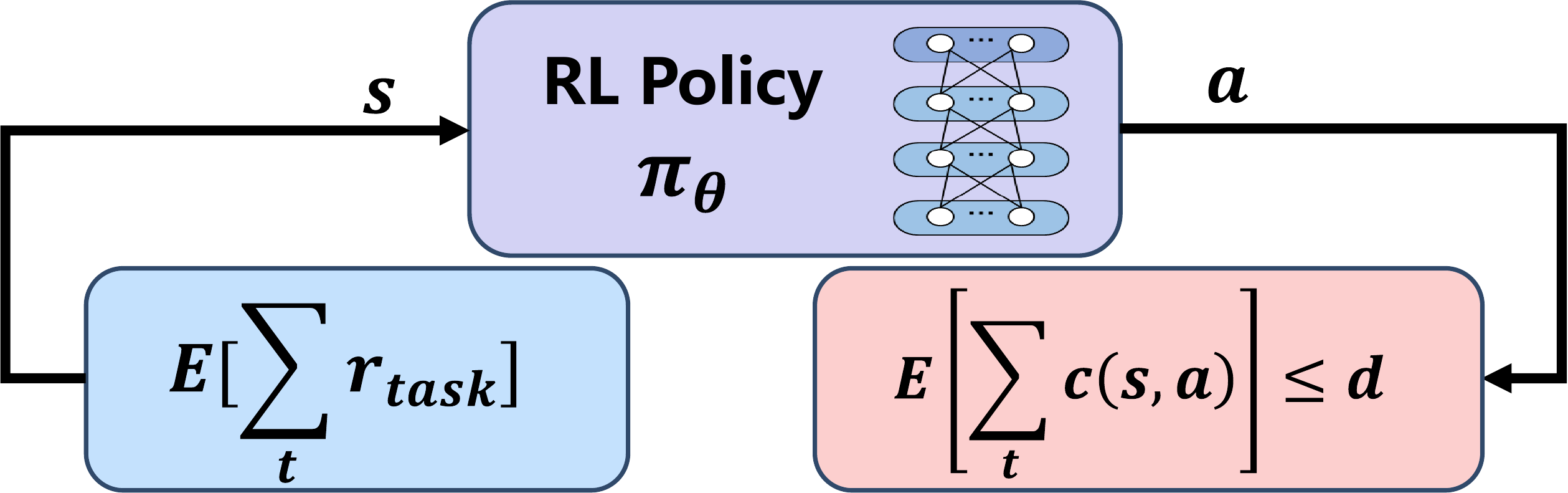}
  \caption{CMDP}
\end{subfigure}\par\smallskip
\begin{subfigure}[t]{0.92\columnwidth}\centering
  \includegraphics[width=\linewidth]{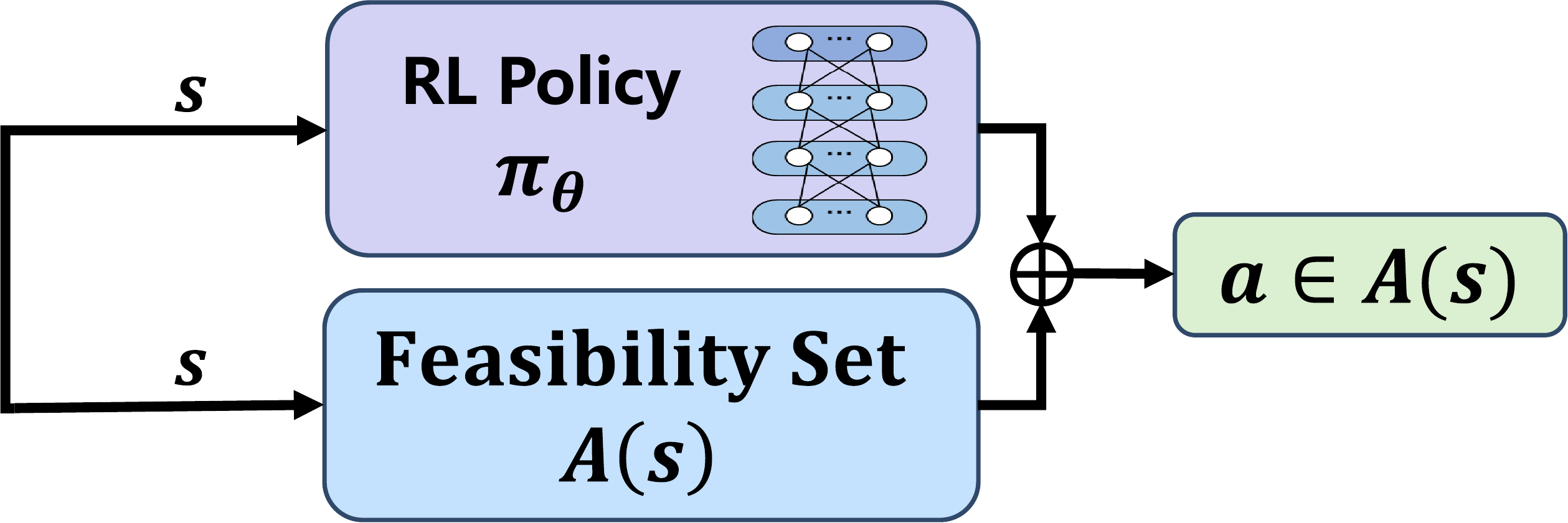}
  \caption{Action mask}
\end{subfigure}\par\smallskip
\begin{subfigure}[t]{0.92\columnwidth}\centering
  \includegraphics[width=\linewidth]{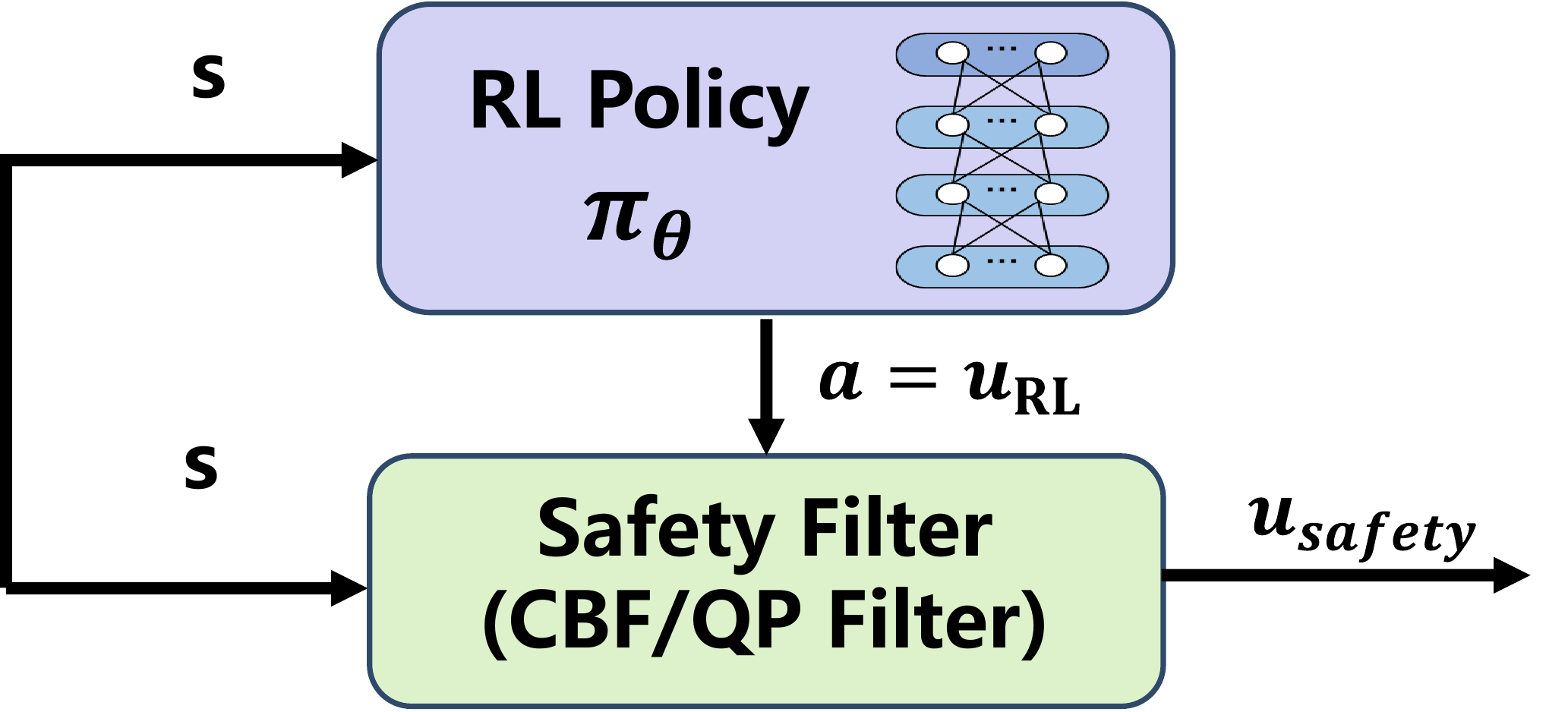}
  \caption{CBF/QP safety filter}
\end{subfigure}\par\smallskip
\begin{subfigure}[t]{0.92\columnwidth}\centering
  \includegraphics[width=\linewidth]{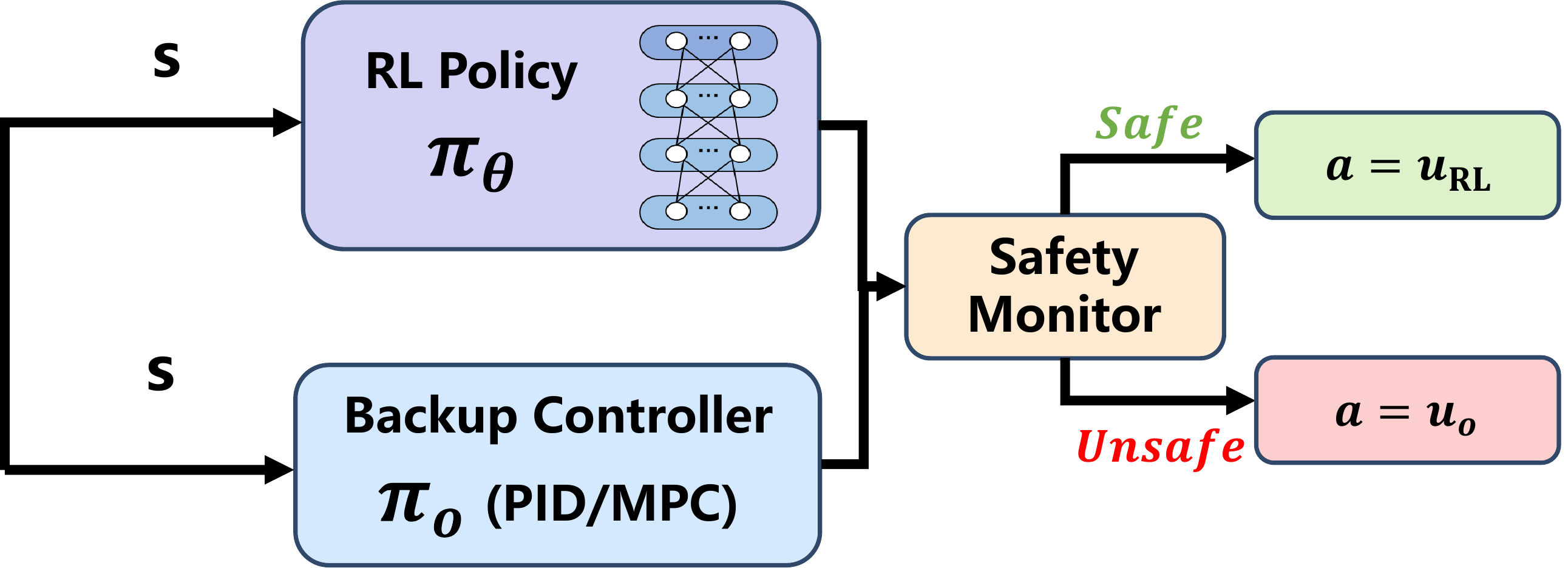}
  \caption{Runtime assurance}
\end{subfigure}
\caption{Five approaches to constraint handling in reinforcement learning.}
\label{fig:safety-mechanisms}
\end{figure}

Figure~\ref{fig:safety-mechanisms} provides a schematic overview of the five mechanisms.

Table~\ref{tab:assurance-scope} compares where each mechanism enters the decision pipeline and the assumptions that delimit its constraint handling.
\begin{table*}[]
\centering
\caption{Comparison of constraint-handling mechanisms by intervention point and scope.}
\label{tab:assurance-scope}
\small
\setlength{\tabcolsep}{4pt}
\renewcommand{\arraystretch}{1.05}
\begin{tabularx}{\textwidth}{@{}P{0.14\textwidth}P{0.16\textwidth}YY@{}}
\toprule
Mechanism & Where it enters & What it directly constrains or checks & Main scope or limitation \\
\midrule
Reward penalty & Learning objective & Task reward and weighted constraint costs & Soft trade-off; penalty tuning does not guarantee trajectory-wise constraint satisfaction \\
CMDP & Policy objective & Expected cumulative constraint cost & Constrains expected cost when the policy-level constraint is satisfied; individual trajectories may still violate constraints \\
Action mask & Discrete action selection & State-dependent admissibility of encoded actions & Enforces only encoded restrictions; requires a fallback when no valid action is available \\
CBF/QP filter & Control execution & Candidate control actions subject to barrier-derived safety conditions & Properties depend on model accuracy, QP feasibility, sampling, and state estimation \\
RTA & Runtime supervision & Monitoring of learned commands and switching to a backup controller & Depends on monitor correctness, backup safety, and safe switching conditions \\
\bottomrule
\end{tabularx}
\end{table*}

Across spacecraft applications, constraint-handling mechanisms differ in their intervention points, enforcement scope, and underlying assumptions.
Their effectiveness and safety guarantees depend on the specific mission constraints, system dynamics, and implementation conditions.
The following sections examine how these mechanisms are integrated into RL-based spacecraft systems and evaluated in representative applications.

\section{Spacecraft Attitude Control}
\label{sec:attitude}

Spacecraft attitude control spans nominal regulation and agile slews, fault recovery, constraint-aware maneuvers, and flexible or dynamically coupled systems.
PID, optimal, and sliding-mode control, together with observers and safety filters, remain important references.
Across these tasks, RL may issue direct torque or actuator commands, adapt a conventional controller, generate a maneuver checked by a retained safety layer, or coordinate actions in a flexible formation; this chapter compares these roles and the conditions under which they have been tested.

\paragraph{Rigid-body attitude model and RL interface.}
To provide a common mathematical reference for the attitude-control studies, consider a rigid spacecraft whose attitude is represented by the scalar-last unit quaternion
\begin{equation}
\boldsymbol q =
[\boldsymbol q_v^{\mathrm T},q_4]^{\mathrm T},
\qquad
\|\boldsymbol q\|_2=1.
\end{equation}
Here, $\boldsymbol q$ maps the body frame to the inertial frame, $\boldsymbol\omega$ is the body angular velocity expressed in body coordinates, and the Hamilton quaternion convention is used.
For this rigid-body model, $\mathbf J\in\mathbb{R}^{3\times3}$ is the constant, symmetric positive-definite inertia matrix about the spacecraft center of mass, expressed in body-fixed coordinates.
$I_3$ is the three-dimensional identity matrix, and $[\boldsymbol q_v]_\times$ is the skew-symmetric matrix satisfying $[\boldsymbol q_v]_\times\boldsymbol v=\boldsymbol q_v\times\boldsymbol v$.
The corresponding kinematics and rigid-body dynamics are
\begin{equation}
\left\{
\begin{aligned}
\dot{\boldsymbol q}_v
&=
\frac{1}{2}
\left(q_4 I_3+[\boldsymbol q_v]_\times\right)
\boldsymbol\omega,\\
\dot q_4
&=
-\frac{1}{2}\boldsymbol q_v^{\mathrm T}\boldsymbol\omega,\\
\mathbf J\dot{\boldsymbol\omega}
&=
-\boldsymbol\omega\times
(\mathbf J\boldsymbol\omega)
+\boldsymbol u_{\mathrm{act}}+\boldsymbol d,
\end{aligned}
\right.
\end{equation}
Here, $\boldsymbol u_{\mathrm{act}}$ denotes the equivalent actuator-generated torque applied to the rigid-body model and abstracts actuator internal dynamics.
Explicit reaction-wheel or CMG models require additional states, such as wheel momentum or speed, or gimbal angle and rate, together with the corresponding actuator-to-body torque mapping.
The disturbance term $\boldsymbol d$ collects external disturbances and unmodeled torques.

For a desired unit quaternion $\boldsymbol q_d$, a sign-invariant scalar attitude error is
\begin{equation}
e_R =
2\arccos\!\left(
\left|\boldsymbol q_d^{\mathrm T}\boldsymbol q\right|
\right),
\qquad e_R\in[0,\pi].
\end{equation}
This quantity is the principal rotation angle between the desired and actual attitudes and is invariant to the quaternion double covering.

For a representative direct-torque RL formulation, the state and action may be written as
\begin{equation}
\left\{
\begin{aligned}
\boldsymbol q_{e,t}
&=\boldsymbol q_{d,t}^{-1}\otimes\boldsymbol q_t,\\
\boldsymbol\omega_{e,t}
&=\boldsymbol\omega_t-\mathbf R(\boldsymbol q_{e,t})^{\mathrm T}\boldsymbol\omega_{d,t},\\
s_t
&=[\boldsymbol q_{e,t}^{\mathrm T},\boldsymbol\omega_{e,t}^{\mathrm T}]^{\mathrm T},\\
a_t
&=\boldsymbol u_{c,t},
\end{aligned}
\right.
\end{equation}
Here, $\boldsymbol q_d$ is the desired attitude, $\boldsymbol\omega_d$ is the desired angular velocity expressed in the desired body frame, and $\mathbf R(\boldsymbol q_e)$ is the rotation matrix associated with the relative quaternion.
The operator $\otimes$ denotes Hamilton quaternion multiplication.
$\boldsymbol\omega$ and the resulting angular-velocity error are expressed in the current body frame.
For nominal direct-torque control, $\boldsymbol u_{\mathrm{act},t}=\boldsymbol u_{c,t}$.
In actuator-level formulations, the action may instead represent reaction-wheel torque, CMG gimbal rate, or magnetic dipole commands; under actuator faults, the commanded action and applied torque may differ.

\subsection{Nominal Attitude Regulation, Tracking, and Agile Maneuvering}

Nominal regulation and tracking combine attitude determination, reference generation, feedback computation, actuator allocation, and, when needed, momentum management or terminal-rate regulation.
PID, sliding-mode, and optimal control remain strong references.
RL appears both as direct nonlinear feedback and as an adaptive layer that tunes or coordinates a retained controller, offering a way to respond to changing maneuvers without repeatedly redesigning a fixed gain schedule.

Proximal policy optimization (PPO) is tested on torque-saturated and reaction-wheel systems with randomized inertia; it improves on a single quaternion-rate-feedback tuning across the tested range and approaches case-specific tuning \citep{vedant2019reinforcementLearning}.
Gao et al. compare improved deep deterministic policy gradient (DDPG) and TD3 with proportional--derivative (PD) control for unconstrained, state-constrained, and online-optimization tasks \citep{gao2020satelliteAttitude}.
Elkins et al. test direct RL feedback for large-angle slews, disturbance rejection, and transfer to an untrained spacecraft configuration \citep{elkins2020adaptiveContinuous}.
PID guidance accelerates TD3 exploration; compared with end-to-end TD3, PID-Guide TD3 trains faster and reaches higher convergence accuracy, while the paper reports lower control performance for PID alone \citep{zhang2020modelFree}.
Zheng et al. retain a sliding-mode observer/controller for disturbance rejection and use RL to compensate residual uncertainty \citep{zheng2021reinforcementLearning}.
For control-moment-gyro (CMG) maneuvers, soft actor--critic (SAC) commands gimbal rates during the slew; conventional control then handles terminal pointing, and the study considers momentum-vector recovery \citep{oghim2025deepReinforcement}.
Xue et al. use actor--critic learning to optimize energy within a fixed-time controller, with an observer supplying convergence structure and a command filter handling saturation \citep{xue2026reinforcementLearning}.
Barrenechea et al. port the hybrid online policy adaptation strategy (HOPAS) to the Satellite Experimental Processing Platform (SEPP) on ESA's OPS-SAT, testing the software and telemetry interfaces on the Engineering and Flight Models \citep{barrenechea2023enabling}.
In-orbit tests exercised the Algorithm Driver without the HOPAS AI algorithm; the paper reports that closed-loop platform control and in-orbit execution of the AI algorithm were not achieved.
LeLaR deploys a simulation-trained RL attitude controller on the InnoCube 3U nanosatellite, where it commands reaction-wheel torques for inertial-pointing maneuvers; in seven repeated trials, it meets the $1^\circ$ per-axis criterion in all seven, compared with three of seven for the onboard PD controller, reported as a contextual reference \citep{11457578}.
Lu et al. use an RL-based attitude controller with redundant actuators for multisatellite moving-target tracking; simulations report accurate and stable tracking under the tested scenarios \citep{lu2025multisatelliteScheduling}.
These studies highlight the flexibility of RL in spacecraft attitude control.
Direct policies provide nonlinear feedback for large-angle maneuvers and respond to the tested disturbance or configuration changes, while hybrid approaches can accelerate policy learning, compensate residual model uncertainty, and retain conventional control for terminal pointing \citep{elkins2020adaptiveContinuous,zhang2020modelFree,zheng2021reinforcementLearning,oghim2025deepReinforcement}.
Together, these roles support agile maneuvering and tracking across the tasks studied.

Dong et al. test critic-only adaptive dynamic programming (ADP) with a constrained estimator in an attitude HIL loop \citep{dong2021learningBased}.
Muñoz et al. run the RL agent on an embedded board in a processor-in-the-loop setup, with the UPMSat-2 attitude-control model running in Simulink on a host computer \citep{munoz2025feasibilityOf}.
Against gain-scheduled PID in re-entry control, domain randomization improves off-policy RL across sampled dynamics, while a hybrid controller performs better in the tested out-of-distribution cases \citep{fabisch2026deep}.
Trained policies can generate commands at runtime, while only some architectures also update policy or controller parameters during operation; conventional feedback offers a more established stability and tuning basis.

\subsection{Robust and Fault-Tolerant Attitude Control}

Robust and fault-tolerant attitude control must maintain pointing under disturbances or changing dynamics and recover when sensors or actuators degrade.
Robust control and disturbance observers address model variation; fault-detection systems separate diagnosis from compensation.
RL can adapt its control law to faults or use health estimates supplied by another module, complementing observers and robust feedback when the spacecraft response changes after a failure.

\begin{figure}[]
\centering
\includegraphics[width=\columnwidth]{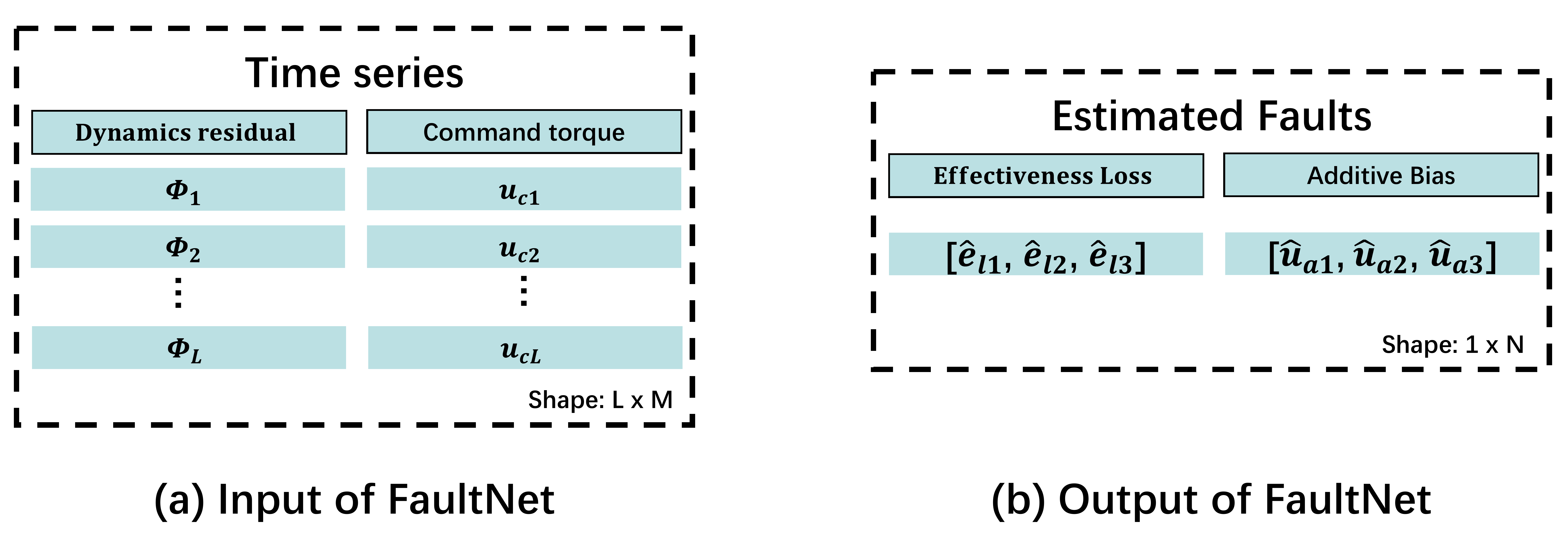}
\caption{FaultNet inputs and estimated actuator faults \citep{lu2025neuralNetwork}.}
\label{fig:faultnet-input-output}
\end{figure}

\paragraph{Fault diagnosis and policy conditioning.}
Following the loss-of-effectiveness convention used by Lu et al. \citep{lu2025neuralNetwork}, a representative three-axis actuator-fault model is
\begin{equation}
\left\{
\begin{aligned}
\boldsymbol u_{\mathrm{act},t}
&=
(I_3-\mathbf E_{\ell,t})\boldsymbol u_{c,t}
+\boldsymbol u_{a,t},\\
\mathbf E_{\ell,t}
&=
\operatorname{diag}
\!\left(e_{\ell 1,t},e_{\ell 2,t},e_{\ell 3,t}\right),
\qquad e_{\ell i,t}\in[0,1].
\end{aligned}
\right.
\end{equation}
Here, $\boldsymbol u_{c,t}$ is the commanded torque, $\boldsymbol u_{\mathrm{act},t}$ is the applied torque, $\mathbf E_{\ell,t}$ is the loss-of-effectiveness matrix, and $\boldsymbol u_{a,t}$ is the additive bias fault.
Both fault terms may vary with time; $e_{\ell i,t}=0$ denotes a healthy channel and $e_{\ell i,t}=1$ denotes complete loss of effectiveness \citep{lu2025neuralNetwork}.
Figure~\ref{fig:faultnet-input-output} illustrates the time-series inputs and estimated faults used by FaultNet.

Two representative RL information structures distinguish implicit accommodation from estimator-conditioned control:
\begin{equation}
\left\{
\begin{aligned}
\boldsymbol u_{c,t}
&=
\pi_\theta(s_t),\\
\boldsymbol u_{c,t}
&=
\pi_\theta\!\left(
s_t,
\widehat{\mathbf E}_{\ell,t},
\widehat{\boldsymbol u}_{a,t}
\right).
\end{aligned}
\right.
\end{equation}
The first policy receives no explicit fault estimate and must accommodate changes through the observed state or history and the variations represented during training.
The second conditions its decision on estimates supplied by fault-diagnosis or health-management modules; Lu et al. instantiate this form with estimated effectiveness loss and additive bias \citep{lu2025neuralNetwork}.

Xiao et al. combine RL-based fixed-time optimal control with prescribed-performance tracking and fault/disturbance estimation \citep{xiao2023fixedTime}.
Rafiee et al. pair model-free Q-learning with an observer that detects actuator faults and estimates their aggregate effect, so the policy needs no explicit fault parameters \citep{rafiee2024activeFault}.
Jin and Yang use actor--critic learning to generate feedforward compensation for inertia uncertainty within a prescribed-performance controller, while an adaptive law compensates actuator faults and external disturbances; simulations report stable attitude control under actuator faults \citep{jin2024faultTolerant}.
FaultNet uses a long short-term memory (LSTM) network to diagnose actuator-effectiveness loss and additive bias as input to an RL fault-tolerant controller \citep{lu2025neuralNetwork}.
Lu et al. use parallel differential evolution to estimate reaction-wheel effectiveness loss and additive torque bias, then provide these estimates to an RL fault-tolerant controller; the integrated response to time-varying faults is evaluated in simulation \citep{lu2026gpu}.
Yang et al. adapt an offline RL controller online with a neural observer and critic update, without an accurate spacecraft model \citep{yang2025reinforcementLearning}.
For high-order fully actuated dynamics, Wu et al. combine identifier--actor--critic learning with sliding-mode tracking under actuator faults \citep{wu2026adaptiveFault}.
Mamba--Kolmogorov--Arnold network (Mamba-KAN) estimates actuator health from temporal data for physics-regularized DDPG \citep{he2026mambaKan}.
A related closed-loop design couples long-sequence health estimation with meta-Q-learning \citep{he2026closedLoop}.
Dalahmeh et al. introduce TD3-HD, combining Hindsight Experience Replay (HER) with Dimension-Wise Clipping (DWC) to learn attitude recovery from simulated actuator degradation \citep{dalahmeh2026intelligentSpacecraft}.
StAR-RL embeds a Lyapunov-decrease penalty in the TD3 critic loss, uses sliding-mode reference actions in reward shaping, and activates a sliding-mode control (SMC)-based safety fallback when stability criteria are violated; its continuous policy is discretized for magnetorquer commands \citep{ramezani2026starRl}.
In 3-DoF reaction-wheel HIL tests, RL, super-twisting, and fuzzy controllers show different rankings under sensor noise and unmodeled dynamics \citep{sarjoughian2026hardwareValidated}.
Across these studies, fault information reaches RL through observers, learned estimators, or explicit policy inputs, while some policies adapt to modeled degradation without a separate fault estimate \citep{rafiee2024activeFault,lu2025neuralNetwork,lu2026gpu,he2026mambaKan,he2026closedLoop,yang2025reinforcementLearning,dalahmeh2026intelligentSpacecraft}.
Robust feedback and disturbance observers remain useful partners in these designs \citep{xiao2023fixedTime,jin2024faultTolerant,wu2026adaptiveFault,ramezani2026starRl}.

\subsection{Safety- and Constraint-Aware Attitude Control}

\begin{figure}[]
\centering
\includegraphics[width=\columnwidth]{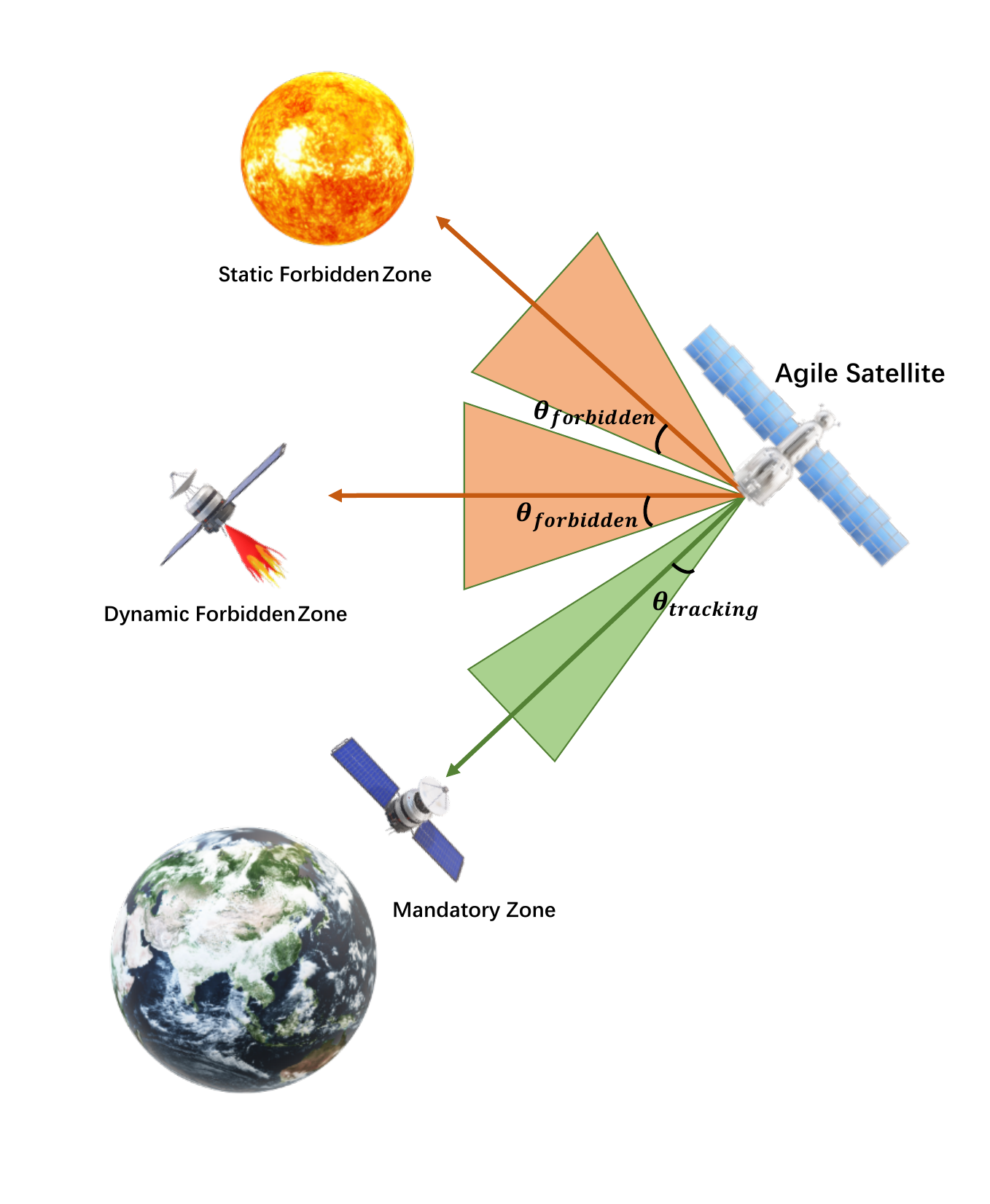}
\caption{Representative pointing constraints for spacecraft attitude maneuvers, including forbidden regions and tracking requirements \citep{dong2020reinforcementLearning,yang2021adpBased,yang2025deepReinforcement}.}
\label{fig:attitude-constraints}
\end{figure}

Attitude maneuvers during on-orbit operations must respect exclusion cones, angular-rate limits, actuator saturation, and pointing restrictions along the commanded maneuver.
Barrier functions, reference governors, and constrained control provide established references; RL safety claims depend on whether limits are penalized, optimized in expectation, or enforced by a runtime filter.
Figure~\ref{fig:attitude-constraints} illustrates representative pointing geometry with static and dynamic forbidden zones and a mandatory tracking zone.

The general mechanisms are summarized in Section~\ref{sec:constraint-safety}.
The studies below illustrate barrier-based costs, CMDP constraints, and constrained policy training in attitude control.

Dong et al. add barrier terms to the cost function in their RL framework for forbidden attitudes and angular-rate limits, and test it in simulation and HIL \citep{dong2020reinforcementLearning}.
Yang et al. use ADP for reorientation with actuator misalignment and forbidden pointing constraints, and evaluate the controller in numerical simulations and HIL experiments \citep{yang2021adpBased}.
SAC with a compact representation of one keep-out zone and curriculum training learns reorientation around the excluded region \citep{yang2025deepReinforcement}.
Its Monte Carlo tests include constraint violations, motivating a shielded extension.
Lu et al. combine CMDP-based safe RL with data-driven actuator-fault estimates for agile tracking \citep{lu2025safeReinforcement}.
Li et al. progressively train RL tracking policies on the special orthogonal group (SO(3)) Gym under pointing and torque-amplitude constraints, and evaluate them with internal disturbances \citep{li2026spacecraftAttitude}.
Across these attitude-control studies, RL supports reorientation and tracking under pointing restrictions, actuator limits, and modeled faults, while CMDP costs and barrier-based terms provide complementary ways to incorporate selected operational constraints into learning \citep{dong2020reinforcementLearning,yang2021adpBased,yang2025deepReinforcement,lu2025safeReinforcement,li2026spacecraftAttitude}.

\subsection{Flexible and Dynamically Coupled Spacecraft Control}

Flexible appendages, captured objects, and close formations add structural or orbital modes that are excited by ordinary attitude actions.
Operations therefore couple manoeuvre planning, vibration suppression, actuator limits, and sometimes relative-motion control.
Modal control, robust control, and disturbance-rejection schemes give useful baselines; RL can tune or adapt these controllers across changing dynamics, with its value best tested on flexible modes and parameters outside training.

Hu et al. use DDPG to tune a linear active-disturbance-rejection controller for a piezoelectric flexible spacecraft, combining model-independent adaptation with vibration suppression \citep{hu2025reinforcementLearning}.
Ameli et al. use temporal-difference actor--critic learning and a radial basis function (RBF) network to adapt boundary-control gains for attitude tracking and vibration suppression under actuator limits \citep{ameli2025reinforcementLearning}.
For a rigid--flexible satellite with a large space net, Sun et al. combine an RL parameter regulator with observer-based compensation and saturation-aware feedback, testing disturbance rejection and vibration suppression in simulation \citep{sun2026adaptiveAttitude}.
After target capture, Yan et al. use SAC to tune sliding-mode parameters and add residual compensation in a 9-DoF attitude--orbit model; their ablation attributes most of the response to the sliding-mode backbone \citep{yan2026intelligentAttitude}.
Shi et al. apply TD3 to leader--follower control of two ultra-close flexible spacecraft using coupled translation, attitude, and vibration dynamics \citep{shi2026reinforcementLearning}.
These studies extend RL-based attitude control to flexible spacecraft and coupled rotational--translational dynamics.
Learned parameter adaptation, residual compensation, and cooperative policies provide complementary approaches to vibration suppression, post-capture stabilization, and close-formation control \citep{hu2025reinforcementLearning,ameli2025reinforcementLearning,sun2026adaptiveAttitude,yan2026intelligentAttitude,shi2026reinforcementLearning}.
Further experimental validation under representative structural dynamics and actuator limitations will help assess their practical applicability.

\subsection{Cross-Subdomain Findings}

One demonstrated use case for direct RL is operation across large-angle maneuvers, inertia variation, disturbances, or altered spacecraft configurations without case-specific retuning.
Across randomized inertia, PPO performs slightly better than a quaternion-rate-feedback controller tuned over the same mass range and similarly to case-specific tuning; direct RL has also been tested in simulation on spacecraft configurations not represented during training \citep{vedant2019reinforcementLearning,elkins2020adaptiveContinuous}.
These comparisons support adaptability relative to fixed tuning; they do not establish general superiority over well-retuned PD, SMC, or model-predictive control.

In hybrid designs, the reported performance belongs to the complete architecture.
PID guidance accelerates TD3 exploration, retained feedback rejects selected disturbances or handles terminal pointing, and observers, command filters, and sliding-mode or fixed-time synchronized controllers provide estimation, convergence structure, saturation handling, or baseline stabilization \citep{zhang2020modelFree,zheng2021reinforcementLearning,oghim2025deepReinforcement,xue2026reinforcementLearning}.
In fault-tolerant designs, RL may use estimated actuator faults or health states, generate feedforward or learned compensation torques, or adapt directly to modeled degradation; retained estimators, adaptive laws, and safety fallbacks may also contribute to fault identification, robustness, and stability \citep{rafiee2024activeFault,lu2025neuralNetwork,he2026mambaKan,dalahmeh2026intelligentSpacecraft,jin2024faultTolerant,wu2026adaptiveFault,ramezani2026starRl}.
The same attribution issue appears in flexible-spacecraft control: ablation in the attitude--orbit study assigns most of the response to the sliding-mode backbone rather than the learned residual \citep{yan2026intelligentAttitude}.

These architectures differ in tracking accuracy, computational cost, and safety assurance.
Direct policies can provide low-latency embedded inference, but evaluation should report inference time separately from offline training and hyperparameter-search costs; hybrid policies retain a more established control structure, but their incremental contribution requires ablation \citep{munoz2025feasibilityOf,yan2026intelligentAttitude}.
Reward penalties, barrier-augmented costs, and CMDP formulations address selected pointing, torque, angular-rate, and forbidden-attitude constraints through different optimization mechanisms, without necessarily guaranteeing trajectory-wise satisfaction \citep{dong2020reinforcementLearning,yang2025deepReinforcement,lu2025safeReinforcement,li2026spacecraftAttitude}.
HIL results further show that the relative performance of RL, super-twisting, and fuzzy-control combinations depends on sensor noise and unmodeled dynamics rather than yielding one uniformly dominant controller \citep{sarjoughian2026hardwareValidated}.
Comparisons should therefore report tracking error, overshoot or settling time, control torque, fault-recovery time, constraint violations, inference latency, offline training cost, and baseline-tuning effort under matched conditions.

The reviewed evidence remains dominated by numerical simulation, including tests under varying inertia, modeled actuator degradation, and coupled spacecraft dynamics. Further physical validation under representative structural modes and actuator limitations is particularly important for flexible and dynamically coupled spacecraft control.
HIL, processor-in-the-loop, flight-software integration, and an in-orbit attitude-control demonstration strengthen selected claims but do not remove these cross-study comparability limits; the flight-software integration case does not demonstrate in-orbit closed-loop execution of its RL algorithm \citep{dong2021learningBased,munoz2025feasibilityOf,sarjoughian2026hardwareValidated,barrenechea2023enabling,11457578}.
Overall, the evidence supports RL as a means of extending adaptation across operating conditions and augmenting established control structures; its contribution to complete-system performance depends on baseline tuning, retained modules, evaluation conditions, and the costs assigned to training, computation, control effort, and safety.

\section{Orbital Guidance and Trajectory Optimization}
\label{sec:orbit}

Orbital guidance and trajectory optimization cover low-thrust transfers, cislunar and multibody trajectories, deep-space guidance, and long-duration station-keeping.
RL may generate closed-loop guidance directly, assist an optimizer, or approximate a family of solutions for repeated correction; its central trade-off with conventional trajectory optimization is fast reusable decisions versus explicit feasibility and propellant control.

\paragraph{Orbital dynamics and RL interface.}
A common reference model for continuous low-thrust guidance represents the spacecraft by its inertial position $\boldsymbol r$, velocity $\boldsymbol v$, and mass $m$:
\begin{equation}
\left\{
\begin{aligned}
\dot{\boldsymbol r}
&=\boldsymbol v,\\
\dot{\boldsymbol v}
&=-\frac{\mu_c}{\lVert\boldsymbol r\rVert^3}\boldsymbol r
  +\frac{\boldsymbol T}{m}
  +\boldsymbol a_d,\\
\dot m
&=-\frac{\lVert\boldsymbol T\rVert}{I_{\mathrm{sp}}g_0},
\qquad
\lVert\boldsymbol T\rVert\leq T_{\max}.
\end{aligned}
\right.
\end{equation}
Here, $\mu_c$ is the central body's gravitational parameter, $\boldsymbol T$ is the thrust vector, $T_{\max}$ is the maximum thrust magnitude, $\boldsymbol a_d$ collects additional modeled perturbations, $I_{\mathrm{sp}}$ is the specific impulse, and $g_0$ is standard gravity.
The associated physical state is
\begin{equation}
\boldsymbol x_{\mathrm{orb}}
=
[\boldsymbol r^\top,\boldsymbol v^\top,m]^\top.
\end{equation}
In a minimal fully observed formulation, the RL variables may be written as
\begin{equation}
s_t=\boldsymbol x_{\mathrm{orb}}(t),
\qquad
a_t=\boldsymbol T_t.
\end{equation}
The policy observation may additionally include time or time-to-go, target or terminal errors, reference-orbit information, navigation measurements, or estimated latent dynamics.
Other studies instead use classical orbital elements, modified equinoctial elements, relative orbital elements, or image-based observations.
For impulsive guidance and station-keeping, the continuous thrust action is replaced by maneuver timing and discrete $\Delta\boldsymbol v$ commands with the corresponding propellant update.

\paragraph{Trajectory-optimization problem.}
Many comparisons in this section can be interpreted through a representative finite-horizon optimal-control problem:
\begin{equation}
\left\{
\begin{aligned}
\min_{\boldsymbol T(\cdot),\,t_f}\quad
& w_m\left[m_0-m(t_f)\right]+w_t t_f,\\
\mathrm{s.t.}\quad
& \dot{\boldsymbol x}_{\mathrm{orb}}
  =
  f_{\mathrm{orb}}
  \left(\boldsymbol x_{\mathrm{orb}},\boldsymbol T,t\right),\\
& \boldsymbol x_{\mathrm{orb}}(0)
  =\boldsymbol x_{\mathrm{orb},0},\\
& \boldsymbol x_{\mathrm{orb}}(t_f)\in\mathcal X_f,\\
& \boldsymbol x_{\mathrm{orb}}(t)\in\mathcal X_{\mathrm{adm}},\\
& \lVert\boldsymbol T(t)\rVert\leq T_{\max}.
\end{aligned}
\right.
\end{equation}
Here, $t_f$ is the final time, $m_0$ is initial spacecraft mass, $\boldsymbol x_{\mathrm{orb},0}$ is the initial state, and $T_{\max}$ is the maximum thrust magnitude.
$\mathcal X_f$ represents terminal requirements, and $\mathcal X_{\mathrm{adm}}$ can encode path and operational constraints.
The nonnegative weights $w_m$ and $w_t$ include any normalization needed to combine propellant use and transfer time.
This formulation is representative rather than universal: some reviewed studies fix the transfer time, optimize only propellant or time, impose different terminal tolerances, or include tracking, control-effort, safety, and robustness terms.

Conventional direct and indirect optimal-control methods solve for a trajectory and its associated control history for a specified initial condition.
A direct RL policy instead learns a state-dependent mapping $a_t=\pi_\theta(s_t)$, with $a_t$ corresponding here to a thrust or maneuver command, typically over a distribution of initial conditions or modeled uncertainties.
RL-assisted approaches retain an optimizer or guidance law and use learning to tune its parameters, construct an initial guess, or provide a startup trajectory.
Consequently, faster policy evaluation does not by itself imply lower propellant consumption, higher terminal accuracy, or lower total computational cost once offline training is included.

\subsection{Low-Thrust Orbital Transfer and Guidance}

Low-thrust guidance selects thrust direction, magnitude, and timing while balancing propellant, transfer time, eclipses, and terminal constraints.
Optimal control, $Q$-law guidance, and nonlinear programming remain key comparators; RL roles range from tuning these methods to issuing thrust commands directly.

The main distinction is whether RL adjusts a retained method, initializes a solver, or generates thrust decisions.
Arora and Dutta use RL to adapt the objective weights of sequential low-level optimizations, which continue to construct each trajectory \citep{arora2020reinforcementLearning}.
Holt et al. make Lyapunov $Q$-law parameters state-dependent to improve optimality while keeping closed-loop guidance \citep{holt2020lowThrust}.
Their follow-up modifies the law to preserve asymptotic stability and tests transfers from geostationary transfer orbit (GTO) to geostationary Earth orbit (GEO), and from low Earth orbit (LEO) to GEO with eclipses and Monte Carlo navigation and thruster errors \citep{holt2021optimalQ}.
The effect on the stability--optimality balance differs between the two transfer families, so RL's gain is tied to how the learned law interacts with the orbit and mission objective.
For off-nominal near-rectilinear halo orbit (NRHO) recovery, LaFarge et al. use an RL-trained network to provide startup solutions for multiple-shooting targeting; differential correction refines the maneuver, and Monte Carlo convergence falls as departure from the reference orbit grows \citep{lafarge2023adaptiveClosed}.

Direct policies instead choose low-thrust actions.
Casas et al. learn model-free pericenter-raising guidance without prior dynamics or a reference trajectory, reporting quasi-optimal control under modeled uncertainty \citep{casas2022lowThrust}.
Zaidi et al. cascade policies to select thrust directions using orbital-element rewards for multi-revolution transfers \citep{zaidi2023cascadedDeep}.
They report shorter transfer times than the tested automated optimization solvers across the reported GTO-to-GEO and Super-GTO-to-GEO scenarios, while comparisons with traditional numerical optimizers are favorable in some cases and comparable in others.
Their attention actor--critic extends low-thrust optimization to geocentric and cislunar cases, with improved reported sample efficiency and tests of modest initial-orbit perturbations after offline training \citep{zaidi2025singleAgent}.
A later cascade applies deep reinforcement learning (DRL) to transfers toward GEO and NRHO; in four two-body cases, it reports shorter transfer times than the tested automated baselines and results comparable to traditional solvers, while its NRHO transfer is slightly longer than the traditional reference and uses a mildly relaxed terminal tolerance \citep{zaidi2025automatedTrajectory}.
Direct policies provide reusable closed-loop thrust decisions after training, while learned objective weights, state-dependent $Q$-law parameters, and targeter initializations retain optimization, guidance, or targeting structure around RL \citep{arora2020reinforcementLearning,holt2021optimalQ,lafarge2023adaptiveClosed,zaidi2023cascadedDeep}.

\subsection{Cislunar, Multibody, and Deep-Space Trajectory Guidance}

Cislunar and deep-space guidance faces sensitive multi-body dynamics, navigation error, and long correction horizons.
RL studies use direct feedback, optimizer initialization, image-conditioned navigation, and adaptation to uncertain dynamics; multiple shooting, direct collocation, and nonlinear programming remain important comparators.

\paragraph{CR3BP dynamics and control interface.}
A standard reference model for cislunar guidance is the normalized circular restricted three-body problem (CR3BP).
In the barycentric rotating frame,
\begin{equation}
\left\{
\begin{aligned}
\ddot{x}-2\dot{y}
&=\Omega_x+a_{\mathrm{ctrl},x},\\
\ddot{y}+2\dot{x}
&=\Omega_y+a_{\mathrm{ctrl},y},\\
\ddot{z}
&=\Omega_z+a_{\mathrm{ctrl},z},
\end{aligned}
\right.
\end{equation}
where $\Omega_x=\partial\Omega/\partial x$ and similarly for $\Omega_y$ and $\Omega_z$, with
\begin{equation}
\Omega(x,y,z)
=
\frac{1}{2}(x^2+y^2)
+\frac{1-\mu_{\mathrm{CR}}}{r_1}
+\frac{\mu_{\mathrm{CR}}}{r_2},
\end{equation}
and
\begin{equation}
\left\{
\begin{aligned}
r_1
&=
\sqrt{(x+\mu_{\mathrm{CR}})^2+y^2+z^2},\\
r_2
&=
\sqrt{(x-1+\mu_{\mathrm{CR}})^2+y^2+z^2}.
\end{aligned}
\right.
\end{equation}
Here,
\begin{equation}
\mu_{\mathrm{CR}}
=
\frac{M_2}{M_1+M_2}
\end{equation}
is the normalized mass parameter, $M_1$ and $M_2$ are the primary and secondary masses, the dots denote derivatives with respect to normalized time, and $\boldsymbol a_{\mathrm{ctrl}}$ is the applied control acceleration in normalized rotating-frame coordinates.
The notation $\mu_{\mathrm{CR}}$ distinguishes this mass parameter from the central-body gravitational parameter $\mu_c$ used above.
The rotating-frame Coriolis terms, coupled gravitational field, and sensitive invariant structures make cislunar guidance qualitatively different from a two-body Earth-orbit transfer.
The CR3BP is nevertheless an idealized reference model: some reviewed studies use higher-fidelity ephemeris or perturbed dynamics, whereas others employ different restricted multi-body or two-body formulations.

For Earth--Moon Lyapunov-orbit transfers, LaFarge et al. train a model-agnostic closed-loop controller that handles large initial deviations and augments traditional targeting in simulation, supporting rapid guidance without repeated onboard optimization \citep{lafarge2021autonomousClosed}.
Haq et al. use PPO to generate feasible trajectories between Earth--Moon L1 and L2 Lyapunov orbits and backward-integrate adjoint equations to initialize the indirect solver; across time-, energy-, and fuel-oriented cases, Monte Carlo tests report high solve success and shorter post-training convergence time than random-start shooting, with benchmark-level or lower-fuel solutions \citep{haq2025hybridDeep}.
Attention-augmented PPO in \citep{haq2026autonomousLow} guides Earth--Moon periodic-orbit transfers without an expert initial guess; directional cross-attention improves learning over standard PPO, and tests report robustness to sampled initial-state and thrust perturbations, while direct-optimization results vary by scenario.
For southern-halo station-keeping, Violino et al. use RL to correct position estimates from differences between observed and simulated lunar images; closed-loop simulations maintain the orbit with low-quality images and infrequent updates \citep{violino2026onboardOptical}.

For interplanetary transfer, Zavoli and Federici train PPO feedback under process and observation noise, thrust errors, and missed-thrust events; its deterministic solution is in good agreement with an indirect optimum.
Monte Carlo tests report generally high success under state and observation uncertainty, control errors, and single-step missed-thrust events, with degraded performance for multiple-step missed-thrust events; training remains computationally expensive \citep{zavoli2021reinforcementLearning}.
Federici et al. compare PPO, TD3, and SAC on a noisy, fixed-time Earth--Mars transfer; PPO best satisfies the tested terminal constraints and retains the highest final spacecraft mass among the three algorithms in this scenario \citep{federici2022comparativeAnalysis}.
Their later meta-RL study compares recurrent and feed-forward policies across dynamic uncertainty and control-execution errors; the recurrent policy uses trajectory history to adapt its feedback \citep{federici2024robustInterplanetary}.
Gao et al. instead use a sequential latent model to predict observations and infer latent information about uncertain dynamics, which then conditions the policy actions \citep{gao2026robustLow}.

These studies highlight RL's potential for closed-loop guidance in complex gravitational environments.
Direct policies can provide online decisions without repeated onboard trajectory optimization and respond to modeled navigation or dynamical uncertainty \citep{lafarge2021autonomousClosed,haq2026autonomousLow,zavoli2021reinforcementLearning}, while PPO-generated trajectories can initialize and accelerate a retained indirect solver \citep{haq2025hybridDeep}.
Recurrent and latent-variable policies extend this capability by using trajectory histories or inferred dynamical context to condition guidance under partial or uncertain information \citep{federici2024robustInterplanetary,gao2026robustLow}.

\subsection{Station-Keeping and Long-Duration Orbital Control}

Station-keeping combines orbit determination, correction timing, fuel accounting, and tolerance management; differential correction and optimization-based control remain strong comparators.
Bonasera et al. use PPO to select impulsive maneuvers near Sun--Earth $L_2$, with transfer learning to reduce training effort; in a Sun--Earth--Moon ephemeris model with solar radiation pressure, bounded motion and maneuver cost are comparable to constrained optimization, including in long-duration propagation \citep{bonasera2023designingSun}.
Suda et al. train SAC around a baseline NRHO and compare full-ephemeris Monte Carlo results with $xz$-plane-crossing control, reporting lower and more consistent station-keeping cost \citep{suda2023deepReinforcement}.
For distant retrograde orbits, RL estimates dynamics uncertainty from nominal-versus-actual trajectory deviations and predicts acceleration corrections within a double-loop predictor--corrector scheme using a robust station-keeping point \citep{huang2026robustIntelligent}.
At constellation scale, decentralized policies use relative orbital elements to select finite-pulse maneuvers.
In Monte Carlo tests, the optimal-control approach uses less fuel for both along-track and across-track corrections. For along-track correction, the optimal-control solver fails in some trials whereas RL completes the tested cases; for across-track correction, both approaches complete the tested cases. Policy evaluation is substantially faster than the tested optimal-control computation \citep{tafanidis2025reinforcementLearning}.
Related advantage actor--critic (A2C) work applies fault detection and fault-tolerant control to relative position--velocity dynamics within a Stackelberg formulation \citep{meng2024adaptiveFault}.
Together, these studies demonstrate RL's potential for efficient and adaptive orbital station-keeping: trained policies support rapid maneuver decisions and corrections under modeled dynamical uncertainty.
Across the reported scenarios, RL yields lower or more consistent station-keeping costs in some comparisons and competitive maneuver performance against the tested optimization or crossing-control methods in others \citep{bonasera2023designingSun,suda2023deepReinforcement,tafanidis2025reinforcementLearning}.

\subsection{Cross-Subdomain Findings}

In the reviewed studies, direct RL policies are used for repeated feedback decisions, whereas solver-assisted methods adapt guidance parameters or initialize retained trajectory optimizers.
For maneuver families that recur after offline training, direct policies can amortize offline training costs over repeated, low-cost policy evaluations and provide closed-loop thrust or maneuver decisions; recurrent and latent-variable architectures further condition those decisions on trajectory histories or inferred dynamics under modeled uncertainty \citep{casas2022lowThrust,zaidi2023cascadedDeep,suda2023deepReinforcement,tafanidis2025reinforcementLearning,zavoli2021reinforcementLearning,federici2024robustInterplanetary,gao2026robustLow}.
When conventional methods are limited by initialization sensitivity, RL-generated trajectories or targeter startup solutions can provide initial guesses that improve convergence of a retained solver \citep{lafarge2023adaptiveClosed,haq2025hybridDeep}.

In hybrid orbital-guidance methods, terminal accuracy and optimality also depend on the retained guidance laws and numerical solvers.
Arora and Dutta retain sequential low-level optimization, Holt et al. retain a Lyapunov-based $Q$-law and incorporate derivatives of the learned state-dependent parameters to preserve asymptotic stability, LaFarge et al. retain multiple shooting and differential correction, and Haq et al. retain indirect collocation with continuation to satisfy terminal constraints and first-order optimality conditions \citep{arora2020reinforcementLearning,holt2021optimalQ,lafarge2023adaptiveClosed,haq2025hybridDeep}.
Violino et al.'s RL navigation module corrects optical position estimates that are supplied to a separately pretrained closed-loop station-keeping policy, while Huang et al. embed learned acceleration corrections within a double-loop predictor--corrector scheme \citep{violino2026onboardOptical,huang2026robustIntelligent}.
Thus, feasibility, terminal precision, stability, or optimality in these hybrid designs cannot be attributed to RL alone: RL contributes objective weights, state-dependent guidance parameters, solver initializations, position-estimate corrections, or learned model corrections, whereas retained guidance laws, targeters, optimizers, predictor--corrector structures, and downstream controllers perform other parts of the trajectory construction, refinement, or closed-loop execution.
Where reported, direct policies provide closed-loop decisions at low post-training evaluation cost.
However, the performance claims in the reviewed direct-policy studies are supported primarily by numerical experiments rather than formal guarantees of trajectory feasibility, closed-loop stability, or global optimality \citep{casas2022lowThrust,tafanidis2025reinforcementLearning,zavoli2021reinforcementLearning}.
Solver-assisted hybrid methods preserve optimization structure while retaining an online or post-processing solve, whereas other hybrids retain guidance, navigation, or control modules without necessarily invoking an optimizer \citep{lafarge2023adaptiveClosed,haq2025hybridDeep,violino2026onboardOptical,huang2026robustIntelligent}.

Online computation, offline training, propellant use, terminal accuracy, and solve success should therefore be compared separately.
Tafanidis et al. illustrate this trade-off: optimal control uses less fuel in the tested corrections, whereas RL offers faster policy evaluation and more consistent success in the along-track tests \citep{tafanidis2025reinforcementLearning}.
The timing result concerns post-training evaluation and does not include policy-training cost, and the task-dependent fuel and success results do not identify either method as uniformly superior.

Comparisons differ in tested dynamics, uncertainty distributions, terminal tolerances, and baseline implementations.
For example, the cascaded DRL comparisons report shorter transfers than the tested automated baselines in several cases, but performance relative to traditional solvers varies by scenario; the NRHO case is slightly longer than its traditional reference and uses a mildly relaxed terminal tolerance \citep{zaidi2025automatedTrajectory}.
Zavoli and Federici demonstrate agreement with an indirect optimum in the deterministic case and robustness to modeled process, observation, thrust, and missed-thrust uncertainties, but training is computationally expensive and performance degrades for multiple-step missed-thrust events \citep{zavoli2021reinforcementLearning}.
Results obtained with simplified two-body or restricted three-body models, unmatched terminal criteria, or solver-specific failures therefore support conditional advantages---fast repeated decisions, feedback adaptation within tested uncertainty ranges, and improved initialization---rather than a general claim that RL is more accurate, fuel efficient, safe, or reliable than conventional guidance and optimization \citep{haq2025hybridDeep,zaidi2025automatedTrajectory,tafanidis2025reinforcementLearning,zavoli2021reinforcementLearning}.
When one-off fuel optimality or explicit constraint enforcement dominates and sufficient computation is available, conventional optimization remains a strong benchmark; RL is most compelling when rapid repeated decisions, feedback adaptation, or initialization robustness dominates \citep{tafanidis2025reinforcementLearning,haq2025hybridDeep,zavoli2021reinforcementLearning}.

\section{Rendezvous, Proximity Operations, and On-Orbit Servicing}
\label{sec:proximity}

Proximity operations span rendezvous and docking, planned relative-motion safety, visual inspection, and robotic capture or servicing.
Navigation uncertainty, target motion, and contact dynamics change the decisions required across mission phases.
RL has been used for direct guidance and control, learned reference generation, active inspection and sensing, and manipulator planning and control; some architectures retain conventional tracking controllers or runtime safety modules.

Here, safety concerns planned relative motion around a known target; external conjunction and debris-avoidance decisions are addressed in Section~\ref{sec:planning}.

\paragraph{Relative-motion dynamics and RL interface.}
To provide a common mathematical reference for rendezvous and proximity-operation studies, consider the Clohessy--Wiltshire (CW) model for linearized relative translation about a circular target orbit.
In a target-centered local-vertical/local-horizontal frame, with $x$, $y$, and $z$ denoting radial, along-track, and orbit-normal relative position, respectively,
\begin{equation}
\left\{
\begin{aligned}
\ddot{x}-2n\dot{y}-3n^2x &= a_x,\\
\ddot{y}+2n\dot{x} &= a_y,\\
\ddot{z}+n^2z &= a_z,
\end{aligned}
\right.
\label{eq:cw-relative-motion}
\end{equation}
where $n$ is the target-spacecraft mean motion and $\boldsymbol a_{\mathrm c}=[a_x,a_y,a_z]^{\mathrm T}$ is the commanded relative acceleration.
Defining
\begin{equation}
\boldsymbol\xi=
[x,y,z,\dot{x},\dot{y},\dot{z}]^{\mathrm T},
\qquad
\dot{\boldsymbol\xi}
=\mathbf A_{\mathrm{CW}}(n)\boldsymbol\xi+\mathbf B_{\mathrm{CW}}\boldsymbol a_{\mathrm c},
\end{equation}
a representative translational state for a CW-based RL formulation is $s_t=\boldsymbol\xi_t$; $\mathbf A_{\mathrm{CW}}(n)$ and $\mathbf B_{\mathrm{CW}}$ are the corresponding state and input matrices \citep{yang2022modelBased}.
6-DoF, rotating-target, or vision-based formulations may instead augment or replace this state with attitude, angular-rate, target-motion, image, or observation-history information.
The CW equations assume small relative separation and linearized motion about a circular reference orbit; they do not represent nonlinear relative dynamics, eccentric reference orbits, or fully coupled translational--rotational rendezvous.

\subsection{Rendezvous and Docking Guidance and Control}

Rendezvous proceeds from phasing and relative navigation to terminal approach, while docking adds alignment and contact constraints.
Far-range phasing, relative-motion guidance, MPC, convex optimization, and 6-DoF control provide phase-appropriate baselines.
RL can issue direct commands, adapt a tracker, or generate a reference for downstream control.

An early autonomous-rendezvous policy maps relative state to acceleration commands \citep{wang2020autonomousRendezvous}.
Hovell et al.'s deep-guidance policy instead sends velocity commands to a conventional controller; their follow-up extends this approach to proximity operations, and both studies report physical experiments as well as simulation \citep{hovell2020onDeep,hovell2021deepReinforcement}.
Wang et al. use actor--critic learning and fuzzy inference to tune adaptive and proportional--derivative trackers for disabled and mobile targets, reducing initial acceleration or terminal tracking error in the respective cases \citep{wang2020designOf}.
Di et al. apply PPO to far-range, high-thrust, minimum-fuel rendezvous under $J_2$ perturbation, using equal time intervals to discretize each burn \citep{di2025trajectoryOptimization}.
These studies distinguish direct state-feedback policies from RL that adapts or supplies commands to a retained tracker.

As rendezvous advances toward terminal docking, RL decisions move from relative-motion guidance toward coupled translational--rotational control under tighter alignment, contact, and actuator constraints.
This progression motivates 6-DoF policies, phase-dependent decision architectures, and hybrid guidance strategies.
For a representative state-feedback 6-DoF docking formulation, the relative translational and attitude-error states can be combined as
\begin{equation}
s_t=
\left[
\boldsymbol r_{\mathrm{rel},t}^{\mathrm T},
\boldsymbol v_{\mathrm{rel},t}^{\mathrm T},
\boldsymbol q_{e,t}^{\mathrm T},
\boldsymbol\omega_{e,t}^{\mathrm T}
\right]^{\mathrm T},
\qquad
a_t=
\left[
\boldsymbol F_t^{\mathrm T},
\boldsymbol\tau_t^{\mathrm T}
\right]^{\mathrm T},
\end{equation}
where $\boldsymbol r_{\mathrm{rel}}$ and $\boldsymbol v_{\mathrm{rel}}$ are expressed in a consistently defined target-centered frame, $\boldsymbol q_e$ and $\boldsymbol\omega_e$ are the attitude and angular-velocity errors defined consistently with Section~\ref{sec:attitude}, and $\boldsymbol F_t$ and $\boldsymbol\tau_t$ are commanded force and torque.
This wrench-level action is representative rather than universal: some policies instead output accelerations, velocity references, individual actuator commands, or phase-level decisions.
Physical translation--rotation coupling depends on the spacecraft mass distribution, actuator placement, flexible or articulated structures, and contact dynamics, while the selected coordinate frame affects how that coupling is represented mathematically; such coupling is not implied merely by a 6-DoF state or force--torque action space.

\begin{figure}[]
\centering
\includegraphics[width=\columnwidth]{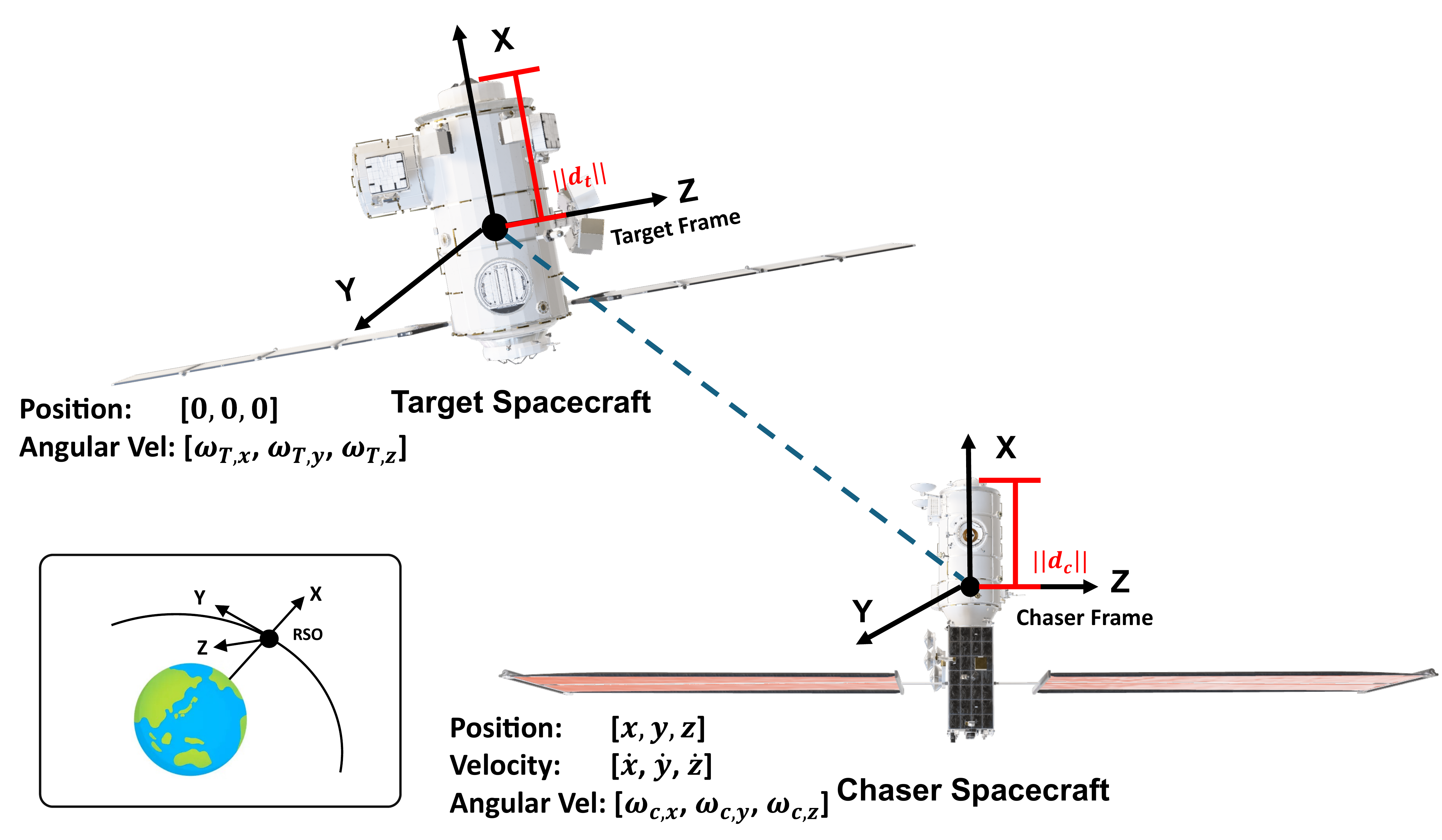}
\caption{6-DoF rendezvous-and-docking geometry, with target- and chaser-frame docking-port vectors $\boldsymbol d_t$ and $\boldsymbol d_c$.}
\label{fig:rendezvous-dynamics}
\end{figure}

Figure~\ref{fig:rendezvous-dynamics} summarizes the geometry underlying the 6-DoF relative state.
The displayed components $[x,y,z]^{\mathrm T}$ and $[\dot{x},\dot{y},\dot{z}]^{\mathrm T}$ represent $\boldsymbol r_{\mathrm{rel}}$ and $\boldsymbol v_{\mathrm{rel}}$ in the selected target-centered frame.
The uppercase $X$, $Y$, and $Z$ labels identify the illustrated target and chaser frame axes; they do not redefine the radial, along-track, and orbit-normal lowercase ordering specified for the CW model in Eq.~\eqref{eq:cw-relative-motion}.
For the 6-DoF representation, the target and chaser angular velocities $\boldsymbol\omega_T$ and $\boldsymbol\omega_C$ are first expressed in a common frame; their difference is then resolved in the chaser body frame to form $\boldsymbol\omega_e$, consistent with the definition in Section~\ref{sec:attitude}.
The attitude error $\boldsymbol q_e$ is obtained from the target and chaser attitudes using the same relative-quaternion convention.

Hu et al. use a dual-quaternion RL formulation for motion-constrained terminal control \citep{hu2021learningBased}.
Pan et al. embed actor--critic optimization in fixed-time synchronized backstepping \citep{pan2026timeSynchronized}.

Oestreich et al. first develop PPO feedback policies for 6-DoF docking, then extend them to rotating targets \citep{oestreich2020autonomousSix,oestreich2021autonomousSixDegree}.
Yang et al. formulate rotating-target rendezvous on the special Euclidean group (SE(3)) under disturbances and model uncertainty \citep{yang2025spacecraftAutonomous}.
Mohseni et al. add physics-informed temporal attention to PPO-based close-range rendezvous control \citep{mohseni2026physicsInformed}.
A vision-guided method maps monocular images directly to thrust commands, demonstrating image-based docking in simulation while retaining a performance gap from state-based policies \citep{supreeth2026visionGuided}.

For onboard use, model-based RL and neural-network compression reduce the footprint of rendezvous policies \citep{yang2022modelBased}.
Meta-RL trains finite-thrust guidance across transfer conditions, with an LSTM policy using prior-step control and reward information \citep{federici2022metaReinforcement}.
An angles-only policy includes an observability measure for angle-only navigation in its reward function \citep{yuan2022deepReinforcement}.
Hierarchical RL separates docking into phases \citep{aborizk2024multiphaseAutonomous}; a CubeSat guidance-and-attitude controller has also been evaluated in HIL testing \citep{tammam2025hierarchicalDeep}.
Du et al. incorporate Lyapunov constraints into Bayesian quadrature policy optimization and test proximity maneuvering on an air-bearing platform \citep{du2024bayesianQuadrature}.
Hovell's experiments, Du's air-bearing tests, and Tammam's HIL evaluation test different layers of guidance and control.

\subsection{Safety Assurance in Planned Proximity Operations}

Planned proximity safety covers keep-out geometry, closing speed, pointing, plume limits, and abort conditions around a known target; external conjunctions are treated in Section~\ref{sec:conjunction}.
Reward penalties, model constraints, and runtime filters encode these requirements differently (Section~\ref{sec:constraint-safety}): RL can optimize maneuver trade-offs, while filters provide an explicit check on proposed actions.

Broida and Linares include a keep-out zone in their 3-DoF PPO rendezvous simulations and report collision-free approaches in the tested cases \citep{broida2019spacecraftRendezvous}.
Qu et al.'s exploration-adaptive DDPG reduces energy consumption relative to the tested artificial-potential method in close-proximity simulations \citep{qu2022spacecraftProximity}.
Runtime assurance allows RL to propose actions while a supervisory monitor determines whether each learned command can be applied or a backup controller must take over; intervention frequency is therefore relevant to system-level comparisons \citep{dunlap2023runTime}.
Holt and Armellin show that state-dependent RL can improve time- and fuel-optimal transfers when paired with a linear-quadratic regulator (LQR); their control Lyapunov function (CLF)-based variants add a stability condition for the linear-system setting \citep{holt2025reinforcementLearning}.
Harris et al. present shielded deep RL for spacecraft command and control, an early complementary example of supervisory safety checks \citep{harris2020spacecraftCommand}.
Wijayatunga et al. use meta-RL to adapt input-constrained CBFs under thrust limits and hidden physical parameters, extending safety-filter design to changing conditions \citep{wijayatunga2026metaReinforcement}.

These studies show the value of pairing adaptive RL-based maneuver planning with explicit safety mechanisms.
Hybrid designs combine learned policies with CLF-based stability conditions, barrier-function constraints, or runtime supervision to support proximity operations while accounting for the safety properties represented in each model and its assumptions \citep{dunlap2023runTime,holt2025reinforcementLearning,wijayatunga2026metaReinforcement}.

\subsection{Vision-Based Inspection and Active Reconstruction}

Inspection couples relative motion with image quality, illumination, and the information available for later decisions.
Coverage planning and visual servoing provide useful baselines; RL can choose motion with future observations and target knowledge in view.

For shape reconstruction, DQN and A2C policies select imaging paths and improve coverage over random movement in linearized-orbit simulations \citep{brandonisio2021reinforcementLearning}.
Brandonisio et al. later test RL-based active reconstruction with noisy state inputs, including retraining with those inputs \citep{brandonisio2024deepReinforcement}.
Zhou et al. provide an open-source image-tracking benchmark and train a DQN policy for active visual tracking \citep{zhou2022spaceNoncooperative}.
Their later recurrent attention architecture treats active tracking as a partially observed problem \citep{zhou2023onDeep}.
Lighting-aware PPO selects inspection paths to improve image conditions in a 3-DoF orbital model, isolating translational path selection from attitude and camera pointing \citep{wijk2023deepReinforcement}.
Dunlap et al. use an active-set invariance filter with CBFs to enforce multiple 6-DoF inspection constraints while limiting intervention \citep{dunlap2024runTime}.
Wijk et al. use discrete CBFs for RL inspection; the formulation guarantees safety only at the discrete sampling times, not between samples \citep{wijk2024safeSpacecraft}.
Multi-agent RL inspection combines learned deputy control with runtime assurance \citep{dunlap2024deepReinforcement}.
Sample-efficient visual RL further learns proximity actions from images \citep{zhang2026sampleEfficient}.
In a direct comparison, Stephenson and Schaub find that the probabilistic roadmap method (PRM) front dominates the RL front in the tested cases, whereas RL supports closed-loop, responsive inspection with low computational requirements \citep{stephenson2026twoApproaches}.
These studies show how RL couples proximity motion with sensing: policies select paths for reconstruction coverage or favorable image conditions, while recurrent and image-based policies support feedback decisions from partial visual observations \citep{brandonisio2021reinforcementLearning,wijk2023deepReinforcement,zhou2023onDeep,zhang2026sampleEfficient}.
Safety filters and multi-agent RL extend inspection to constrained or cooperative operations \citep{dunlap2024runTime,wijk2024safeSpacecraft,dunlap2024deepReinforcement}.

\subsection{Space Robotics, Capture, and Post-Capture Control}

Space robotics and servicing span free-floating arm planning, capture contact, and post-capture control; sampling-based planning, inverse dynamics, impedance, and shared control remain useful comparators.
For a representative free-floating space manipulator, let $\boldsymbol V_b$ and $\boldsymbol V_e$ denote the base and end-effector twists and let $\boldsymbol\Phi$ collect the joint angles.
The matrices $\mathcal J_b$ and $\mathcal J_m$ are the base and manipulator Jacobians, respectively.
The end-effector twist satisfies
\begin{equation}
\boldsymbol V_e
=\mathcal J_b\boldsymbol V_b+\mathcal J_m\dot{\boldsymbol\Phi}.
\end{equation}
In the absence of external forces and torques, and assuming zero initial total momentum, linear and angular momentum conservation gives \citep{alali2024pathPlanning}
\begin{equation}
\begin{bmatrix}
\boldsymbol P\\
\boldsymbol L
\end{bmatrix}
=\mathbf H_b\boldsymbol V_b+\mathbf H_{bm}\dot{\boldsymbol\Phi}
=\boldsymbol 0,
\end{equation}
where $\boldsymbol P$ and $\boldsymbol L$ are total linear and angular momentum.
Assuming $\mathbf H_b$ is nonsingular, eliminating the base twist yields the generalized-Jacobian relation
\begin{equation}
\boldsymbol V_e
=\left(\mathcal J_m-\mathcal J_b\mathbf H_b^{-1}\mathbf H_{bm}\right)\dot{\boldsymbol\Phi}
=\mathcal J_g\dot{\boldsymbol\Phi}.
\end{equation}
The coupled dynamics can be written in block form as
\begin{equation}
\begin{bmatrix}
\mathbf H_b & \mathbf H_{bm}\\
\mathbf H_{bm}^{\mathrm T} & \mathbf H_m
\end{bmatrix}
\begin{bmatrix}
\dot{\boldsymbol V}_b\\
\ddot{\boldsymbol\Phi}
\end{bmatrix}
+
\begin{bmatrix}
\boldsymbol c_b\\
\boldsymbol c_m
\end{bmatrix}
=
\begin{bmatrix}
\boldsymbol 0\\
\boldsymbol u_j
\end{bmatrix},
\end{equation}
where $\mathbf H_b$, $\mathbf H_m$, and $\mathbf H_{bm}$ are the base, manipulator, and base--manipulator coupling inertia matrices, $\boldsymbol c_b$ and $\boldsymbol c_m$ collect velocity-dependent terms, and $\boldsymbol u_j$ is the joint-torque input.
Under the free-floating assumptions adopted by Al Ali et al., total linear and angular momentum are conserved in the absence of external forces and torques; the zero-momentum relation additionally assumes zero initial momentum \citep{alali2024pathPlanning}.
Joint torques enter as internal manipulator inputs, whereas an externally applied base wrench, disturbance, or contact requires corresponding generalized-force terms and momentum evolution \citep{alali2024pathPlanning}.
Wu et al. study RL-based dual-arm trajectory planning for a free-floating robot; simulations show that a policy trained for a fixed target position can generalize to a moving target, although success declines as target speed increases \citep{wu2020reinforcementLearning}.
Li et al. pair DDPG with artificial potential fields for constrained end-effector alignment \citep{li2022constrainedMotion}.
Dong et al. combine RL with two-axis matching and a self-guiding module to track tumbling targets.
In the reported staged simulations, the resulting trajectories remain collision-free across the tested target motions, although the authors note that no explicit collision-avoidance mechanism is included \citep{dong2024motionPlanning}.
Al Ali and Zhu use DDPG for collision- and obstacle-aware path planning under observation noise \citep{ali2024reinforcementLearning}.
Al Ali et al. also use RL to plan 6-DoF manipulator paths with pose and collision objectives \citep{alali2024pathPlanning}.
On the ISS, APIARY's actor--critic PPO policy replaced the baseline controller for 6-DoF Astrobee motion, including undocking, rotations, and translations; after the first redocking attempt triggered safe recovery, a human operator initiated a second attempt that succeeded \citep{11436341}.
In the reported $0.5$-m undocking maneuver, the RL policy undershot the commanded displacement but remained within Astrobee's pose-success criteria; its tracking was less precise than the simulated baseline reference \citep{11436341,11436888}.
Crossing describes the same ISS demonstration as APIARY, focusing on the simulation-to-ground-to-flight validation pipeline rather than a separate flight experiment \citep{11436888}.
Hu et al. introduce a continuous pose representation into RL-based trajectory planning for free-floating manipulation; comparative and disturbance simulations report higher planning success and accuracy, together with computation times meeting the paper's stated real-time requirement \citep{hu2025deepReinforcement}.
Zhuang et al. combine convolutional neural network (CNN) object detection with off-policy RL for perception-based manipulator control \citep{zhuang2025offPolicy}.
A related off-policy method coordinates heterogeneous manipulators in task space \citep{zhuang2025heterogeneousMulti}.
Yang et al. evaluate truncated quantile critics for end-effector tracking, base-drift suppression, and collision-free operation in MuJoCo simulations and Monte Carlo experiments, outperforming the tested RL baselines \citep{yang2025reinforcementLearningControl}.
Zhang et al. combine adaptive pseudospectral preplanning with RL-based postprocessing for uncertain free-floating-robot trajectories; simulations report reduced tracking errors, while air-bearing-testbed experiments verify the planning framework \citep{zhang2025learningBased}.

Xia et al. use RL to mitigate delay effects on decision-making in space teleoperation \citep{xia2025agentBased}.
Yang et al. use a PPO-based agent to adjust operator skill-training strategies \citep{yang2025skillTraining}.
Ma et al. combine sliding-mode control with actor--critic ADP for limited-input tether deployment \citep{ma2021learningBased}.
Lei et al. apply PPO to free-floating-object tracking and compare it with resolved-motion-rate control \citep{lei2022activeObject}.

Capture studies address contact and base disturbance through different RL roles.
Ai et al. combine RL with a compliant mechanism for buffer-compliance control during capture, limiting impact torque while absorbing impact energy \citep{ai2021bufferCompliance}.
Ren et al. develop RL-based rapid debris-capture trajectory optimization using relative orbital elements and an artificial-potential term to represent the positions of the chaser, obstacles, and target debris \citep{ren2025trajectoryOptimization}.
Wei et al. use TD3 for tumbling-target capture while reducing disturbance to the free-floating base \citep{wei2025trajectoryPlanning}.
Zhang et al. combine target-conditioned SAC with PD joint-velocity tracking and a CBF safety layer; MuJoCo simulations evaluate target self-spin and mass variations, and report improved collision-avoidance performance through online CBF projection \citep{zhang2026constrainedReinforcement}.

After capture, the chaser and target form a combined spacecraft with changed inertia and rotational dynamics, making attitude takeover a stabilization problem under uncertain post-capture parameters.
RL-based studies address this task through learned control of the combined system, estimate-conditioned policy inputs, or safe-reference generation for an adaptive controller.
Wu et al. formulate multi-microsatellite takeover as a differential game and use reinforcement learning with value-function approximation to solve the associated control problem \citep{wu2023spacecraftAttitude}.
Geng et al. estimate combined-spacecraft inertia from attitude histories and include it in the RL state, allowing a pre-trained policy to condition its control on the estimated dynamics \citep{geng2026inertiaParameter}.
Luo et al. use improved PPO to transform unsafe desired orbit--attitude trajectories into safe references; an RBFNN state observer estimates unavailable velocity states, while an adaptive finite-time controller tracks the references despite unknown model parameters in simulation \citep{luo2026improvedPpo}.
Together, these papers show RL contributions across planning, capture, and takeover, with classical control and safety modules often retained.
Across these separate studies, contact-model error, uncertainty in combined-spacecraft inertia, and control handover from capture to post-capture stabilization remain key integration challenges for a complete servicing sequence.

\subsection{Cross-Subdomain Findings}

Across rendezvous, proximity operations, and on-orbit servicing, RL supports learned closed-loop guidance, perception-driven decisions, and robotic manipulation under varied mission conditions \citep{oestreich2021autonomousSixDegree,zhou2023onDeep,hu2025deepReinforcement}.
Direct guidance policies and hierarchical architectures support autonomous approach and docking, while inspection-oriented RL couples spacecraft motion with active sensing, and visual RL further enables image-based proximity feedback \citep{hovell2020onDeep,aborizk2024multiphaseAutonomous,wijk2023deepReinforcement,zhang2026sampleEfficient}.
For capture and post-capture operations, RL contributes to capture-trajectory optimization, disturbance mitigation, and attitude takeover, with some architectures using estimated inertia or pairing learned references with retained tracking controllers \citep{ren2025trajectoryOptimization,wei2025trajectoryPlanning,geng2026inertiaParameter,luo2026improvedPpo}.
Hybrid architectures combine learned decisions with conventional trackers, parameter estimates, and safety filters, retaining complementary roles for feedback control and runtime supervision \citep{hovell2021deepReinforcement,geng2026inertiaParameter,dunlap2023runTime,zhang2026constrainedReinforcement}.
Air-bearing and HIL studies provide experimental evidence for selected guidance, proximity-control, and manipulation components, while the Astrobee demonstration adds an in-orbit example of RL free-flyer control \citep{hovell2021deepReinforcement,tammam2025hierarchicalDeep,du2024bayesianQuadrature,zhang2025learningBased,11436341,11436888}.
Future studies can extend these capabilities to integrated proximity-operation sequences that include autonomous approach, inspection, capture, and post-capture stabilization.

\section{Planetary Landing, Small-Body Operations, and Surface Mobility}
\label{sec:landing}

Planetary autonomy spans powered descent with integrated guidance and control, landing and hovering near small bodies, and surface mobility or hopping.
These tasks combine uncertain gravity or terrain with limited sensing and costly recovery after touchdown or contact.
This chapter compares feedback guidance, perception-driven decisions, and route or hop selection against task-appropriate control and planning methods, while keeping their distinct sensors, constraints, and test platforms clear.

\subsection{Powered Descent and Integrated Guidance, Navigation, and Control}

Powered descent must meet terminal position, velocity, attitude, and terrain constraints while limiting fuel and thrust.
Classical guidance laws and trajectory optimization provide explicit reference solutions; RL studies issue thrust commands directly or adapt an established guidance law, with integrated GNC also depending on navigation quality.
Their advantage is reusable feedback under dispersed conditions, which should be assessed alongside fuel use, constraint satisfaction, and landing success.

Gaudet and Furfaro initialize the policy through apprenticeship learning on GPOPS-generated open-loop, fuel-optimal trajectories and subsequently refine it through RL; at deployment, the policy issues acceleration commands without tracking a precomputed reference trajectory online.
Monte Carlo tests evaluate pinpoint, fuel-efficient soft landings against Apollo-like guidance under sensor and actuator uncertainty \citep{gaudet2014adaptivePinpoint}.
Gaudet et al. extend PPO to 6-DoF landing, mapping estimated state to body-frame engine commands and reporting accurate, fuel-efficient simulated trajectories across a realistic deployment ellipse, with fuel performance sensitive to reward shaping \citep{gaudet2020deepReinforcement}.

Furfaro et al. use actor--critic RL to adapt the gains and time-to-go of the classical ZEM--ZEV law, improving fuel consumption while retaining its feedback structure and accommodating path constraints \citep{furfaro2020adaptiveGeneralized}.
For atmospheric rocket landing, Carradori et al. use gated Transformer-XL meta-RL to condition guidance on observation history under aerodynamic and navigation uncertainty; perturbed Monte Carlo simulations show better performance than the tested trajectory-plus-LQR guidance-and-control baseline \citep{carradori2026transformerBased}.

Direct and hybrid RL architectures provide complementary approaches to powered-descent guidance.
Direct policies enable rapid closed-loop decisions, while RL-assisted guidance retains established control structures and supports adaptation to changing descent conditions.

Integrated RL for navigation, guidance, and control couples these decisions so that commands can respond to joint changes in estimated state and vehicle dynamics, reducing reliance on separately tuned interfaces; modular filters retain explicit intermediate estimates that may support inspection and diagnosis, although diagnostic performance has not been evaluated in these studies.

Gaudet et al. map estimated state to body-frame actuator commands in an integrated RL guidance-and-control design, reporting accurate, fuel-efficient trajectories in 6-DoF simulations with robustness to noise and parameter uncertainty \citep{gaudet2018integratedGuidance}.
Their later recurrent meta-RL policy adapts to unknown forces, mass variation, and engine failure, outperforming nonrecurrent RL under the more severe tested variations; a separate navigation filter performs better than jointly learned navigation in the reported comparison \citep{gaudet2020adaptiveGuidance}.
Federici and Furfaro examine Transformer-based meta-RL for sequential observations in space-guidance applications \citep{federici2024metaReinforcement}.
A later planetary-landing comparison evaluates recurrent meta-RL against standard RL in both guidance performance and computational complexity \citep{federici2025improvingReinforcement}.
An optical-bench HIL study runs image-based meta-RL with a camera and Jetson Nano; real-camera tests show moderate performance degradation but retain robust decisions and real-time inference, while vehicle dynamics remain simulated \citep{federici2026opticalBench}.
These studies show how RL can couple guidance to observation history and perception; modular navigation remains useful when state-estimation accuracy and explicit intermediate estimates take priority.

\subsection{Small-Body Landing, Hovering, and Proximity Operations}

Small-body missions operate with irregular gravity, uncertain shape and rotation, limited navigation aids, and weak escape margins near the surface.
Hovering, terminal descent, and close observation require simultaneous relative-state inference, thrust allocation, and collision-risk management.
Model-based estimation and robust guidance are natural baselines.
RL or meta-RL is most relevant when the policy must adapt to body properties that are unavailable before encounter, and the sensor suite must provide enough information to do so.

Gaudet and Furfaro use direct policy search to maintain hovering in unknown, nonuniform small-body gravity, testing simulated environments including Itokawa against an RL-derived optimal linear PD controller and an LQR controller \citep{gaudet2012robustSpacecraft}.
Willis et al. apply RL to small-body hovering and identify a gain problem: their motion-sensor input cannot distinguish high speed at high altitude from low speed at low altitude, although the resulting controller achieves small residual drift \citep{willis2016reinforcementLearning}.
For body-fixed hovering, Gaudet et al. use reinforcement meta-learning to train a 6-DoF policy that maps flash-based light detection and ranging (LiDAR) measurements directly to spacecraft body-frame thrust commands without position or velocity estimates, an asteroid shape model, or navigation aids; simulations test the policy on Bennu and Itokawa shape models \citep{gaudet2020sixDegree}.
Their terminal-guidance meta-RL policy uses recurrent state to adapt to randomized asteroid properties, actuator failures, center-of-mass changes, and sensor noise in 6-DoF simulations \citep{gaudet2020terminalAdaptive}.
For asteroid impact, Federici et al. use a convolutional-recurrent meta-RL policy to map camera-image sequences and mission time to thrust commands and firing times \citep{federici2022imageBased}.

Chen et al. combine recurrent meta-RL with MCTS-based policy and value learning and post-training DPW-MCTS action refinement for robust asteroid landing under uncertainty \citep{chen2025robustAdaptive}.
These studies show RL's potential for autonomous guidance and control in poorly characterized small-body environments: LiDAR-driven policies map measurements to closed-loop thrust commands without a complete pre-existing shape model, while recurrent meta-RL uses observation history to respond to randomized body properties, actuator failures, center-of-mass variation, and sensor noise \citep{gaudet2020sixDegree,gaudet2020terminalAdaptive}.
Here, shape-model-free means that control does not require a complete asteroid shape model in advance; it does not mean that the policy operates without onboard observations.

\subsection{Surface Mobility and Hopping Exploration}

Surface mobility selects routes or local actions under terrain, energy, traction, stability, and communication constraints.
For hoppers, each action additionally commits the vehicle to a ballistic landing state, so site selection and attitude recovery are coupled.
Graph search, terrain-aware planning, and model-based locomotion form the conventional foundation; RL may improve local adaptation, provided tests include terrain, contact, and resource conditions absent from training as well as nominal map following.

Using real lunar-surface terrain data in simulation, Yu et al. train a curriculum-based PPO planner with slip-aware rewards; it yields lower slip rates and angles than A* and DQN, while A* produces shorter paths \citep{yu2021learningBased}.
Hu and Zhang evaluate their hierarchical DRL planner, which separates long-range routing from local decisions, in simulation on maps of real lunar terrain; it plans faster than search and sampling baselines, though its path optimality remains below A* \citep{hu2022fastPath}.
Gao et al. use state-space DRL for full-route and stepwise rover planning, reporting faster planning than a traditional planner \citep{gao2024deepReinforcement}.
Lu et al. combine collaborative DRL with an artificial-potential-field heuristic; among the reported successful plans, the RL paths cross fewer small obstacles and have shorter calculation and response times than multi-agent A*, while the policy can still approach large obstacles in some environments \citep{lu2024lunarRover}.
Prabhu et al. use Q-learning to plan rover paths with terrain-dependent energy costs; simulated grid tests and manual path inspection illustrate energy-aware obstacle avoidance, but the study does not report a quantitative comparison with implemented A* or Dijkstra baselines \citep{prabhu2025reinforcementQ}.
Lu et al. combine action masks with Lagrangian constraint handling and reward penalties for planning under uncertain time and resource constraints; the combined method succeeds across tested task targets where the mask-only variant fails at harder settings \citep{lu2026constrainedReinforcement}.
These studies demonstrate the flexibility of RL in terrain-aware and resource-constrained surface navigation.
Learned policies incorporate terrain characteristics, slip risks, energy consumption, and mission constraints into sequential route decisions, with several studies reporting improvements in planning efficiency or task-specific performance.
Such capabilities complement conventional graph-search methods, particularly when navigation requires repeated decisions under changing environmental conditions.

Hopping vehicles face a different action model.
Jiang et al. convert mapped three-dimensional asteroid terrain into a lower-dimensional reward representation for a pretrained DRL hop planner, demonstrating route planning across the tested maps in simulation \citep{jiang2020pathPlanning}.
For quadruped hopping, Qi et al. use end-to-end MARL to train phase-specific controllers for takeoff, attitude adjustment, and soft landing; simulations test forward jumps and jumps over obstacles under weak, irregular asteroid gravity \citep{qi2023reinforcementLearning}.
Their related landing study uses the legs for integrated attitude and soft-landing control, with an automatically tuned reward outperforming handcrafted rewards in simulation \citep{qi2023integratedAttitude}.
Across separate studies, RL addresses hop-route selection and flight-phase or body control, extending surface planning beyond fixed routes; physical contact tests and transfer to unmodeled terrain remain open validation needs \citep{jiang2020pathPlanning,qi2023reinforcementLearning,qi2023integratedAttitude}.

\subsection{Cross-Subdomain Findings}

Across these applications, RL is particularly promising for repeated feedback decisions under modeled uncertainty, especially when policies can reuse learned behavior across operating conditions.
After offline training, feedforward policies can provide rapid closed-loop decisions \citep{gaudet2014adaptivePinpoint,gaudet2020deepReinforcement}, while recurrent or sequence-based meta-RL uses observation histories to adapt to modeled changes in gravity, vehicle dynamics, actuator condition, and sensing \citep{gaudet2020adaptiveGuidance,carradori2026transformerBased,gaudet2020sixDegree,gaudet2020terminalAdaptive}.
The reported comparisons remain task-specific and do not establish general superiority over conventional guidance and feedback control or trajectory optimization. Fuel efficiency should be assessed separately from closed-loop robustness across dispersed conditions, particularly because the fuel consumption of learned policies can be sensitive to reward shaping \citep{gaudet2012robustSpacecraft,chen2025robustAdaptive,gaudet2020deepReinforcement,furfaro2020adaptiveGeneralized}.
The trajectory-plus-LQR comparison should be interpreted within its tested architecture: the LQR was synthesized around a nominal trajectory without accounting for uncertainty, whereas the RL policy was trained on nonlinear dynamics with uncertainties and initial dispersions \citep{carradori2026transformerBased}.

Reported improvements must therefore be attributed to the learned and retained modules separately.
In adaptive ZEM--ZEV guidance, RL selects gains and time-to-go, whereas the established guidance law supplies the feedback structure; the reported path-constraint accommodation belongs to the combined design rather than to RL alone \citep{furfaro2020adaptiveGeneralized}.
In integrated GNC, a separate navigation filter can outperform jointly learned navigation, so estimation performance should not be credited to guidance learning alone \citep{gaudet2020adaptiveGuidance}.
In search-based asteroid landing, recurrent meta-RL supplies context-dependent policy and value information, while DPW-MCTS refines actions online at additional computational cost \citep{chen2025robustAdaptive}.
In constrained rover planning, action masks exclude encoded infeasible actions, while Lagrangian constraint handling and reward penalties address other time and resource requirements without by themselves guaranteeing trajectory-wise feasibility \citep{lu2026constrainedReinforcement}.
At deployment, feedforward, recurrent, and Transformer policies all incur inference cost; recurrent and Transformer policies process observation histories, while online search adds iterative per-decision computation to refine action selection \citep{gaudet2020deepReinforcement,gaudet2020sixDegree,carradori2026transformerBased,chen2025robustAdaptive}.

Surface-mobility results should be compared metric by metric rather than combined into a single notion of performance.
PPO reduces the reported slip rate and slip angle while A* retains the shorter path \citep{yu2021learningBased}; hierarchical DRL plans faster than the tested search and sampling methods but remains less path-optimal than A* \citep{hu2022fastPath}.
In the reported successful plans, RL paths cross fewer small obstacles and have shorter calculation and response times than multi-agent A*, while some paths still approach large obstacles \citep{lu2024lunarRover}.
Thus path length, obstacle exposure, resource use, response time, and offline training cost should remain separate measures.
Overall, the reviewed studies most directly support sensor- and history-driven feedback and repeated decisions under modeled uncertainty \citep{gaudet2020sixDegree,gaudet2020terminalAdaptive,carradori2026transformerBased}.
The reviewed studies provide limited evidence of consistent fuel-use or path-quality improvements over the tested conventional baselines, or of formal safety guarantees from learned policies alone \citep{gaudet2020deepReinforcement,yu2021learningBased,hu2022fastPath,lu2026constrainedReinforcement}.
Evidence for the superiority of weak-gravity hopping or legged landing control is also limited, and these demonstrations remain largely simulation-based \citep{jiang2020pathPlanning,qi2023reinforcementLearning,qi2023integratedAttitude}.

\section{Mission Planning, Scheduling, and Resource Management}
\label{sec:planning}

Mission planning and resource management cover Earth-observation schedules, observation and downlink, conjunction response, and multi-target servicing.
These tasks coordinate requests, target assignments, access windows, slews, processing, storage, fuel, and recovery under shared constraints.
Across the reviewed studies, RL constructs or repairs schedules, ranks requests, selects search operators, chooses conjunction responses, and sequences targets; mixed-integer linear programming (MILP), dynamic programming, and heuristics such as adaptive large-neighborhood search (ALNS) remain useful baselines or solver components.
Table~\ref{tab:rl-scheduling-taxonomy} links these RL roles to the decisions they change, the conventional or constraint-handling components retained, and useful comparison measures; its examples identify each learned component's contribution, and roles may overlap.

Miralles et al. review machine learning across Earth-observation operations, including mission planning and communications \citep{revMiralles2023}.
Wang et al. survey deep reinforcement learning for Earth-observation satellite scheduling \citep{revWang2026}.

\begin{table*}[]
\caption{Roles of RL in mission scheduling and resource management.}
\label{tab:rl-scheduling-taxonomy}
\centering
\begin{tabularx}{\textwidth}{@{}P{0.17\textwidth}YYY@{}}
\toprule
RL role & Learned decision (Section 7 examples) & Retained components or constraint-handling mechanisms & Useful comparison measures \\
\midrule
Direct schedule construction & Accept, order, or assign requests and targets; e.g., online scheduling and debris sequencing \citep{wang2019onlineScheduling,xu2023activeDebris} & Time-window and storage checks; action masking; transfer-impulse estimates and trajectory models & Schedule value or optimality gap; feasibility; online decision time \\
Request selection and multi-objective scheduling & Select requests or construct schedules for scalarized objectives \citep{he2020genericMarkov,wei2021deepReinforcement} & Constructive timing heuristics; problem-specific decoding and feasibility-state updates & Incremental benefit over the same downstream heuristic; trade-off quality \\
Search-operator control & Adapt variation-operator parameters or select/reweight search operators \citep{song2023learningAdaptive,li2026solvingThe} & Population/search updates and problem-specific schedule-construction operators & Ablations against fixed or hand-designed operator selection; total training-plus-solve cost \\
Closed-loop decisions and event response & Make closed-loop scheduling decisions, or revise plans after failures or updated conjunction data \citep{stephenson2025learningPolicies,xing2025multiSatellite,patnala2026onOrbit} & Variable-duration decisions and resource modeling; matrix-based replanning; updated conjunction-data messages and collision-risk monitoring & Recovery time; mission value retained; performance across event scenarios \\
Distributed task and network decisions & Allocate mission tasks or adapt routing decisions across satellites \citep{li2024missionPlanning,zheng2025deepReinforcement,huang2023reinforcement} & Distributed-information assumptions; action masks \citep{li2024missionPlanning}; time-window, resource, and queue constraints & Delivered mission value; information assumptions; scaling with network or team size \\
\bottomrule
\end{tabularx}
\end{table*}

Scheduling comparisons should count policy training, online inference, and any feasibility-repair step within the same solve budget.

\subsection{Agile Earth-Observation Scheduling and Dynamic Mission Planning}

Agile Earth-observation scheduling requires repeated decisions on target selection, observation sequencing, and resource allocation under time-window, maneuvering, and onboard-resource constraints.
Once trained, RL policies can support rapid online decisions; related designs use parameter transfer to reduce training costs and revise schedules after failures or as requests change \citep{herrmann2023reinforcementLearning,stephenson2025learningPolicies,wei2021deepReinforcement,xing2025multiSatellite}.
Methods range from direct schedule construction and learned priority selection to RL-guided optimization and adaptive rescheduling, with conventional feasibility checks and repair retained where needed.

A representative single-observation scheduling formulation is
\begin{equation}
\left\{
\begin{aligned}
\max_{\boldsymbol x_{\mathrm{sch}},\boldsymbol t_{\mathrm{sch}}}\quad
&\sum_{i=1}^{N_{\mathrm{sat}}}\sum_{j=1}^{N_{\mathrm{task}}}
\mathrm{profit}_{ij}x_{\mathrm{sch},ij},\\
\text{s.t.}\quad
&\sum_{i=1}^{N_{\mathrm{sat}}}x_{\mathrm{sch},ij}\leq 1,
&&j=1,\ldots,N_{\mathrm{task}},\\
&x_{\mathrm{sch},ij}=1\ \Longrightarrow\ t_{\mathrm{sch},ij}\in\mathcal W_{ij},\\
&\boldsymbol c_{\mathrm{sch}}(\boldsymbol x_{\mathrm{sch}},\boldsymbol t_{\mathrm{sch}})\leq\boldsymbol 0,\\
&x_{\mathrm{sch},ij}\in\{0,1\}.
\end{aligned}
\right.
\label{eq:representative-satellite-scheduling}
\end{equation}
Here, $N_{\mathrm{sat}}$ and $N_{\mathrm{task}}$ are the numbers of satellites and tasks, with $i$ and $j$ indexing satellites and tasks, respectively.
$x_{\mathrm{sch},ij}$ assigns task $j$ to satellite $i$, $\mathrm{profit}_{ij}$ is its mission profit, $t_{\mathrm{sch},ij}$ is the observation start time, and $\tau_{ij}$ is the task's observation duration.
For a visibility interval $[a_{ij},b_{ij}]$, $\mathcal W_{ij}=[a_{ij},b_{ij}-\tau_{ij}]$ is the feasible start-time set that ensures the full observation fits within the interval; this duration-aware interpretation is consistent with scheduling models that specify visibility windows and task durations \citep{wang2019onlineScheduling,he2020genericMarkov}.
$\boldsymbol c_{\mathrm{sch}}$ collects slew, conflict, and resource constraints.
This is a representative formulation rather than a universal model: repeated-observation and stereo-imaging missions relax or extend the single-execution constraint.

\paragraph{Direct construction.}
Wang et al. formulate online request acceptance and sequencing as a dynamic stochastic knapsack and train a DRL policy for immediate scheduling decisions \citep{wang2019onlineScheduling}.
Huang et al. apply DDPG to continuous-time observation scheduling and preprocess task clusters with a graph-based minimum-clique partition accounting for task priority and slew angle; simulations report stronger performance than tested metaheuristics, especially on large instances \citep{huang2021revising}.
Wei et al. address request-failure rate and timeliness by training DRL encoder--decoder policies for weighted-sum subproblems, then transfer parameters between neighboring policies to reduce training expense and approximate a Pareto front.
Across randomly generated tests with 50--200 requests, comparisons with the Non-dominated Sorting Genetic Algorithm II (NSGA-II), the Multiobjective Evolutionary Algorithm based on Decomposition (MOEA/D), and the Indicator-Based Evolutionary Algorithm (IBEA) report competitive solution quality, spread, and computation time \citep{wei2021deepReinforcement}.
Herrmann et al. train neural value approximators on MCTS data; the resulting policies approximate MCTS at much lower execution time, while their advantage over a genetic algorithm weakens at target densities outside training \citep{herrmann2023reinforcementLearning}.
Together, these studies show how RL can shift part of repeated combinatorial search to policy training and support rapid scheduling decisions online \citep{wang2019onlineScheduling,wei2021deepReinforcement,herrmann2023reinforcementLearning}.

\paragraph{Priority selection with heuristic feasibility.}
He et al. pair Q-learning request selection with heuristic timing, assuming that collected imaging data are returned to the ground center within each scheduling period \citep{he2020genericMarkov}.
For variable-duration imaging, Wang et al. train two variable-duration policies to choose observation start times and durations; one favors solution quality, while the other runs faster \citep{wang2025deepReinforcement}.
Chen et al. train a DRL priority network and use backward-slack and top-insert scheduling to place requests while retaining constraint checks \citep{chen2025neuralPriority}.
Their reconstruction-mechanism model revises suboptimal request choices as the neural policy constructs an agile-satellite schedule \citep{chen2025rcmA}.
Li et al. use DQN for multi-satellite task allocation and ensemble heuristics for the subsequent single-satellite schedules \citep{li2025optimizingTime}.
Zhou et al.'s two-phase hybrid optimization learns task sequencing and uses a maximum-residual-capacity heuristic for resource allocation \citep{zhou2025deepReinforcement}.
Liu et al. combine pointer-network sequencing with a second DRL stage for observation scheduling, using local selection and pruning to improve solution quality and computation \citep{liu2025twoStage}.
These results reflect the complete RL--heuristic pipelines: learning selects priorities or sequences, while conventional components enforce timing and resource rules.

\paragraph{RL-controlled search operators.}
Song et al. use a gated recurrent unit (GRU) to adjust variation-operator parameters, with policy-gradient RL updating the GRU for electro-magnetic satellite scheduling \citep{song2023learningAdaptive}.
Adapting their notation to distinguish genetic-algorithm generations from RL time steps, the learned action, policy distribution, and relative fitness-change reward are
\begin{equation}
\left\{
\begin{aligned}
\boldsymbol a_k^{\mathrm{GA}}
&=
\begin{bmatrix}
\mathrm{CR}_k & \mathrm{MR}_k
\end{bmatrix}^{\mathrm T},\\
\pi_\theta\!\left(\boldsymbol a_k^{\mathrm{GA}}\mid s_k^{\mathrm{GA}}\right)
&=
\mathcal N\!\left(
\boldsymbol a_k^{\mathrm{GA}}\mid
\operatorname{GRU}(s_k^{\mathrm{GA}};\mathbf W_{\mathrm{GRU}}),
\boldsymbol\Sigma
\right),\\
r_k^{\mathrm{GA}}
&=
\frac{F_k^{\mathrm{best}}-F_{k-1}^{\mathrm{best}}}
     {F_{k-1}^{\mathrm{best}}},
\end{aligned}
\right.
\label{eq:song-rl-guided-ga}
\end{equation}
where $s_k^{\mathrm{GA}}$ is the search state at generation $k$, $\mathbf W_{\mathrm{GRU}}$ denotes the GRU parameters, and $\mathrm{CR}_k$ and $\mathrm{MR}_k$ are the crossover and mutation rates.
$\operatorname{GRU}$ supplies the Gaussian mean, and $\boldsymbol\Sigma$ is the action covariance matrix.
$F_k^{\mathrm{best}}$ and $F_{k-1}^{\mathrm{best}}$ are consecutive best fitness values; the final expression is their relative change and requires $F_{k-1}^{\mathrm{best}}\ne0$.
Thus RL adapts search parameters, while selection, crossover, mutation, elite retention, and local search remain within the genetic algorithm.

\begin{figure}[]
\centering
\includegraphics[width=\columnwidth]{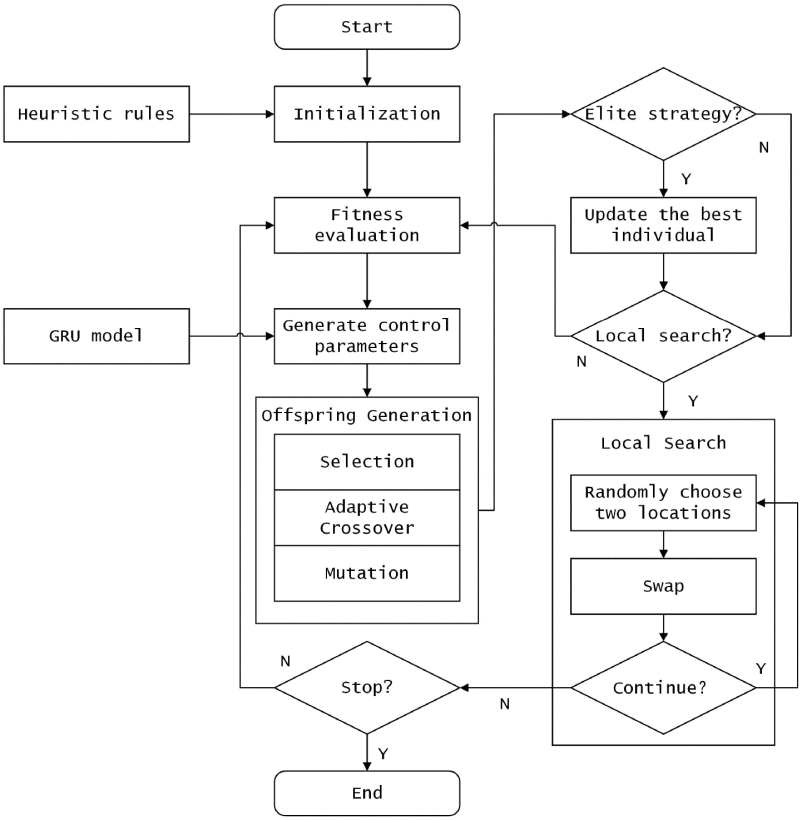}
\caption{GRU-guided adaptation of genetic search operators \citep{song2023learningAdaptive}.}
\label{fig:rl-controlled-search-operators}
\end{figure}

Figure~\ref{fig:rl-controlled-search-operators} illustrates this division between the learned parameter controller and the retained genetic search.

Huang et al. use Q-learning to adjust dung-beetle operator proportions; tests report gains in optimization performance, convergence speed, and scheduling effectiveness \citep{huang2025reinforcementLearning}.
The RL controller learns which neighborhoods to select, while problem-specific insertion operators construct schedules; ablations compare the learned controller with the hand-designed selector \citep{li2026solvingThe}.
Zhu et al. use DQN to select genetic operators and combine them with local intensification \citep{zhu2026dqnGuided}.
Here RL directs search effort rather than replacing the optimizer, allowing learned operator choices to work with established schedule-construction and repair procedures.

\paragraph{Event-driven replanning.}
Stephenson et al. formulate target sequencing as a partially observable semi-Markov decision process (POsMDP) and compare variable-interval closed-loop RL with a discretized MILP benchmark in high-fidelity simulation.
In uniform-request tests at high density, the agents outperform the one-hour MILP incumbent with about 10~ms per decision and collect about 80\% of the solver upper bound \citep{stephenson2025learningPolicies}.
Xing et al. pair DQN task selection with matrix-sorting replanning to respond to satellite failures without retraining the network \citep{xing2025multiSatellite}.
Yao et al. use meta-RL to adapt scheduling policies across mission distributions with limited retraining \citep{yao2026metaReinforcement}.
\paragraph{Related onboard planning and cooperative scheduling.}
Harris et al. formulate orbit-insertion and science-planning decisions as POMDPs to explore DRL for high-level spacecraft autonomy \citep{harris2019spacecraftDecision}.
Their later procedure-generation study applies DRL to spacecraft operations procedures and examines policy behavior under model and hyperparameter variation \citep{harris2022generationOf}.
Piccinin et al. use DRL to select imaging actions for mapping small celestial bodies \citep{piccinin2022deepReinforcement}.
Qi et al. integrate mission planning and data-transmission scheduling for deep-space detectors; the PPO-based method outperforms the tested actor--critic (AC), DQN, genetic-algorithm, and greedy baselines in reward, especially under mission congestion \citep{qi2024missionPlanning}.
Lu et al. use RL-generated attitude torques to track time-sensitive moving targets \citep{lu2024reinforcementLearning}.
Cong et al. divide moving-target observations into subtasks and report improved task assignment and computation against the tested baselines \citep{cong2025autonomousCollaborative}.
Luo et al. add attention, hybrid dynamic conflict resolution, and incremental replanning to multi-agent deep deterministic policy gradient (MADDPG) for cooperative observation in dynamic environments \citep{luo2026researchOn}.
Cao et al. combine a double deep Q-network for variable-number targets (DDQN-VNT) with heuristic rules to assign pairs of satellites to time-sensitive targets while accounting for target-loss risk \citep{cao2026multiSatellite}.
Across these applications, RL supports rapid choices for changing mission requests, while the tested benefits depend on the scheduling or coordination system around the policy.

\subsection{Observation, Data Delivery, and Communication Resource Management}

Joint observation and downlink scheduling couples target selection and imaging with processing, storage, and transmission before access windows close.
RL can coordinate these decisions as queues and links change.
Mission-level evaluation should account for useful data delivered, not only images selected.

Li et al.'s attention-based distributed DRL lets satellites plan observations and downlinks asynchronously, reporting higher returns and shorter computation in the tested scenarios; its decoder masks target points that violate observation constraints \citep{li2024missionPlanning}.
Zheng et al. use multiagent PPO to schedule precedence-dependent observation, on-orbit computation, and downlink operations within limited time windows \citep{zheng2025deepReinforcement}.
Cheng et al. integrate observation and data transmission in a hierarchical DRL scheduler, reporting stronger profit across the tested problem scales while maintaining computational efficiency \citep{cheng2026uhtsDrl}.
\paragraph{Communication-network routing and resource allocation.}
Beyond observation and downlink scheduling, RL supports adaptive communication-network management through routing, beam allocation, and satellite handover.
These policies adapt communication resources to changing network conditions, influencing data-delivery efficiency and reliability.
Huang et al. propose a Q-learning-based dynamic distributed routing scheme for LEO satellite networks, in which satellites update Q-tables from neighboring-node information to adapt routing to topology changes while optimizing end-to-end delay and traffic overhead.
Simulations show decreasing delay during maintenance and better overhead balance at high loads; the centralized shortest-path baseline retains lower delay at low load and early in some maintenance runs.
Although initial Q-table training takes longer, the authors' complexity analysis gives the distributed scheme lower per-step maintenance complexity than the centralized shortest-path baseline \citep{huang2023reinforcement}.
For multi-beam satellites, Huang et al. use hybrid-action DQN to select beam patterns and transmitter power; simulations report 40\%--60\% lower power consumption than conventional DQN under higher traffic demands, with a small trade-off in transmission delay and packet loss \citep{huang2023sequential}.
Xia et al. combine centrally initialized Q-tables with distributed multiagent routing; simulations report lower latency and packet loss than the tested centralized and distributed baselines \citep{xia2025centralizedDistributed}.
Zhao et al. separate PPO handover timing from DQN target-satellite selection, reporting fewer handovers, higher handover success, transmission rate, and load balance than traditional and non-hierarchical RL methods \citep{zhao2026learningWhen}.
These studies highlight RL's potential for adaptive satellite communication management and motivate mission-level evaluations that connect network performance with scientific data delivery and observation-task completion.

\subsection{Conjunction Response and Debris Avoidance}
\label{sec:conjunction}

Conjunction response selects whether, when, and how to maneuver as encounter geometry, risk estimates, and propellant limits change.
RL can add learned maneuver selection to the conventional risk-screening and constrained-optimization process.

Mu et al. formulate avoidance with a variable number of debris as a CMDP and use an LSTM network with penalized proximal policy optimization (P3O) to account for collision-probability and energy costs under operational constraints \citep{mu2024autonomousSpacecraft}.
Han et al. use a GRU-augmented actor to select avoidance maneuvers; simulations report better safety, fuel economy, and training stability than DDPG \citep{han2025spacecraftAutonomous}.
Patnala et al. develop an RL-enabled servicing vehicle that monitors updated conjunction data messages (CDMs), identifies high-risk encounters, and plans collision-avoidance support maneuvers \citep{patnala2026onOrbit}.
Collision-risk information enters the decision process through CMDP costs, recurrent policy state, or updated conjunction data that prompt maneuver planning \citep{mu2024autonomousSpacecraft,han2025spacecraftAutonomous,patnala2026onOrbit}.
Together, these architectures link risk assessment to adaptive maneuver selection while accounting for modeled fuel and operational constraints.

\subsection{Multi-Target Servicing and Active Debris-Removal Planning}
\label{sec:multi-target-servicing}

Multi-target servicing and active debris removal select target order, transfer windows, refueling or removal actions, and time allocation; Section~\ref{sec:proximity} treats each target's approach and contact.
RL may rank targets, assist trajectory optimization, or support emergency replanning.
Transfer models, optimization, action masks, and repair procedures handle constraints in the simulated planning problems.

Xu et al. combine a pointer-network actor--critic with neural transfer-impulse estimates to choose debris order and rendezvous times, outperforming greedy and ant-colony baselines in the tested scenarios \citep{xu2023activeDebris}.
Poupon et al. use RL for risk-aware target sequencing under fuel and time limits and adapt the debris sequence as new collision-risk information becomes available \citep{poupon2024aiDriven}.
Qi et al. embed RL in a multiobjective snake optimizer to adapt its search strategy, while a repair strategy addresses boundary-violating solutions in cooperative spacecraft planning \citep{qi2024cooperativeMission}.
Zhou et al. combine improved RL with online optimization for emergency debris-removal planning and report complete removal in the tested simulation setup \citep{zhou2025improvedReinforcement}.

In a co-elliptic transfer and refueling study, masked PPO visits more debris than the greedy baseline and plans faster than MCTS in randomized simulations \citep{bandyopadhyay2025optimalMulti}.
A related study adds adaptive collision avoidance and emergency detours to masked PPO and reports the highest average debris coverage among the tested RL--greedy configurations across randomized scenarios \citep{bandyopadhyay2026optimizingMission}.
Both studies are simulation-only and do not test physical debris removal.

These studies highlight RL's potential for sequential servicing decisions under coupled transfer, fuel, time, and collision-risk constraints.
Learned target-selection policies and hybrid optimization frameworks support efficient mission sequencing and adaptive replanning, offering promising approaches to increasingly complex multi-target servicing and debris-removal missions \citep{poupon2024aiDriven,qi2024cooperativeMission,zhou2025improvedReinforcement,bandyopadhyay2025optimalMulti,bandyopadhyay2026optimizingMission}.

\subsection{Cross-Subdomain Findings}

Across mission planning, scheduling, and resource management, RL offers potential online efficiency when trained policies can be reused across related scheduling instances, reducing repeated online search \citep{wang2019onlineScheduling,herrmann2023reinforcementLearning,stephenson2025learningPolicies}.
In Stephenson et al.'s uniform-request evaluation, the tested RL agents outperform the MILP solver's one-hour open-loop incumbent at high request densities with about 10~ms of computation per closed-loop decision.
Their reported cumulative reward is about 80\% of the solver's conservative upper bound in this setting.
These are distinct comparisons: the incumbent is a feasible time-limited solution, whereas the upper bound is not an executable schedule or evidence that the learned policies are near-optimal.
Performance also depends on the request distribution; under high-density city-distributed requests, the best tested agent reaches 68\% of the upper bound \citep{stephenson2025learningPolicies}.

Scheduling performance depends on both the learned task-selection policy and the retained schedule-construction procedures.
Direct policies learn reusable state-to-action rules offline and can reduce online request- or target-selection time.
In hybrid designs, however, learned priorities, targets, or search operators remain coupled to schedule construction and feasibility checks \citep{chen2025neuralPriority}, transfer-impulse estimates and trajectory models \citep{xu2023activeDebris}, action masks \citep{li2024missionPlanning,bandyopadhyay2025optimalMulti}, or search and repair procedures \citep{li2026solvingThe,qi2024cooperativeMission}; solution quality and feasibility consequently belong to the combined pipeline rather than to RL alone.
Similarly, failure response in Xing et al. combines DQN task selection with matrix-based replanning, while the parameter-transfer mechanism of Wei et al. reduces retraining across related scheduling objectives rather than per-decision inference time \citep{xing2025multiSatellite,wei2021deepReinforcement}.

Direct policies favor low post-training decision latency but incur offline training cost and may lose performance outside the training distribution \citep{wang2019onlineScheduling,herrmann2023reinforcementLearning,stephenson2025learningPolicies}.
They still depend on problem-specific constraint handling, such as storage and time-window checks, decoder or action masks, and transfer-impulse estimates and trajectory models \citep{wang2019onlineScheduling,xu2023activeDebris,li2024missionPlanning,bandyopadhyay2025optimalMulti}.
Learned guidance of heuristic or optimization search retains schedule-construction and repair structure and therefore includes the associated online search or repair time, while making the learned component's incremental contribution harder to isolate \citep{chen2025neuralPriority,li2026solvingThe,qi2024cooperativeMission}.
Parameter-transfer and meta-RL methods trade additional source-task or meta-training effort for faster adaptation, with benefits that depend on similarity between training and deployment tasks \citep{wei2021deepReinforcement,yao2026metaReinforcement}.
Distributed and multi-agent schedulers offer potential scalability benefits, subject to communication, information-sharing, and coordination assumptions \citep{li2024missionPlanning,zheng2025deepReinforcement,huang2023reinforcement}.

Current evidence therefore supports RL as a reusable decision mechanism for recurring and time-critical scheduling problems, not as a generally superior replacement for MILP, ALNS, or other well-tuned optimization methods \citep{wei2021deepReinforcement,herrmann2023reinforcementLearning,stephenson2025learningPolicies,li2026solvingThe}.
Fair comparisons should report offline training cost, policy inference time, optimizer and repair time, feasibility rates, request distributions, solver time limits, incumbent objective values, and solver bounds or optimality gaps separately.
Because the studies use different request distributions, simulation environments, solver budgets, and baseline configurations, cross-paper rankings of solution quality or computational advantage remain tentative \citep{wei2021deepReinforcement,herrmann2023reinforcementLearning,stephenson2025learningPolicies}.

\section{Multi-Spacecraft Autonomy and Coordination}
\label{sec:multi}

Multi-spacecraft autonomy includes formation control and sensing, reconfiguration, and orbital pursuit--evasion.
These tasks require each vehicle to account for other agents when controlling relative geometry, safety, information gathering, or game objectives.
RL can provide local control, formation references, reconfiguration decisions, or adversarial strategies, alongside conventional consensus, distributed-control, optimization, and differential-game methods.
A central distinction among multi-spacecraft RL architectures is the information available during training and execution: methods may use CTDE, fully decentralized learning, hierarchical coordination, or explicit modeling of other agents' behavior.
These choices affect communication requirements, coordination capability, and adaptation to changing multi-spacecraft configurations.

\paragraph{Multi-spacecraft dynamics and policy organization.}
Let $\boldsymbol x_i$ and $\boldsymbol u_i$ denote the physical state and control input of spacecraft $i$, respectively.
The joint decision state and action for a system of $N$ spacecraft can be written as
\begin{equation}
\boldsymbol s_t
=
\left(s_t^1,\ldots,s_t^N\right),
\qquad
\boldsymbol a_t
=
\left(a_t^1,\ldots,a_t^N\right).
\end{equation}
The decision state $s_t^i$ may contain $\boldsymbol x_i(t)$ together with relative states, mission variables, or information received from other spacecraft.

A representative interacting multi-spacecraft model is
\begin{equation}
\dot{\boldsymbol x}_i
=
f_i\!\left(
\boldsymbol x_i,
\boldsymbol u_i,
\boldsymbol x_{\mathcal N_i(t)},
\boldsymbol d_i
\right),
\qquad
\boldsymbol x_{\mathcal N_i(t)}
=
\operatorname{col}_{j\in\mathcal N_i(t)}
\boldsymbol x_j,
\end{equation}
where $\mathcal N_i(t)$ is the set of spacecraft whose states are physically, operationally, or informationally relevant to spacecraft $i$, and $\boldsymbol d_i$ represents disturbances or model uncertainty.
When spacecraft have no direct physical interaction, the individual dynamics need not depend on $\boldsymbol x_{\mathcal N_i(t)}$; coupling may instead enter through relative observations, shared objectives, communication, collision constraints, resource allocation, or adversarial rewards.

A centralized policy can treat the multi-spacecraft system as one decision maker and generate a joint action according to
\begin{equation}
\boldsymbol a_t
\sim
\pi_\theta\!\left(\cdot\mid\boldsymbol s_t\right).
\end{equation}
By contrast, decentralized execution uses local policies,
\begin{equation}
a_t^i
\sim
\pi_{\theta_i}^i\!\left(\cdot\mid o_t^i\right),
\qquad i=1,\ldots,N,
\end{equation}
where $o_t^i$ denotes the information locally available to spacecraft $i$.
Policies may have distinct parameters or share parameters across homogeneous spacecraft.
Under CTDE, a critic or training objective may use the joint state and actions, while each deployed actor remains conditioned on its local information.

Accordingly, a multi-spacecraft mission need not use MARL: a centralized policy operating on the joint state and action can still constitute a single-agent formulation.
Conversely, MARL assigns decisions to multiple agents, whose policies may differ in their local information, parameter sharing, communication, and training architecture.
The learned action $a_t^i$ may directly specify $\boldsymbol u_i$ or may instead provide a reference, maneuver, or high-level command that is subsequently processed by a retained controller, optimizer, or safety mechanism.

\subsection{Formation Flying, Cooperative Orbital Control, and Distributed Sensing}

Existing RL-based formation-control approaches can be broadly distinguished by their decision roles: augmenting distributed feedback and consensus controllers, generating formation references or maneuver schedules, and learning decentralized coordination policies from local observations.
Formation control maintains relative position and attitude while coordinating motion, separation, and, in some missions, sensing or connectivity.
Consensus, distributed control, and model-predictive control remain useful references because they make coordination structure explicit.
Chen et al. combine RL with a distributed fixed-time observer and controller for attitude consensus under model uncertainty \citep{chen2023reinforcementLearning}.
Kankashvar et al. use transfer-based multi-agent Q-learning to learn discrete-time formation-reconfiguration controllers online without system-dynamics knowledge, and demonstrate collision avoidance in simulation \citep{kankashvar2023multiAgent}.
Sun et al. combine adaptive distributed RL with cooperative control and soft and hard constraints for spacecraft-swarm reconfiguration around obstacles \citep{sun2023distributedCooperative}.
Meng et al. pair RL with adaptive control for spacecraft-cluster consensus under disturbances and collision-avoidance requirements \citep{meng2024obstacleAvoidance}.
Qu et al. use an RL optimal-control strategy to add low-thrust actions to Lorentz-force-driven charged-spacecraft reconfiguration, since Lorentz force alone cannot accomplish the required formation changes \citep{qu2024learningBased}.
Sun et al. combine prescribed-performance control and reinforcement learning to develop a dynamic event-triggered approximate optimal tracking controller for spacecraft formations near an asteroid \citep{sun2025nearAsteroid}.
Compared with static event triggering, the dynamic scheme reduces control updates while introducing trade-offs in velocity accuracy and control cost.
Zhang et al. use distributed inverse Q-learning to infer a cost function and a consistent formation-control policy from expert demonstrations, without requiring the learner to know the system matrices \citep{zhang2025inverseReinforcement}.
Ren et al. integrate RL with decentralized MPC, using a learned full-horizon maneuver schedule to guide impulsive control without a predefined reference trajectory \citep{ren2025learningBased}.
Jin et al. train local-observation multi-agent proximal policy optimization (MAPPO) policies to produce desired states and velocities on $SE(3)$, then use an improved Lloyd algorithm to refine the formation along a target curve \citep{jin2026deepReinforcement}.
Their eight air-bearing simulators track precomputed trajectories, demonstrating the downstream formation-control setup rather than online execution of the learned multi-agent policy.
Yu et al. combine improved PPO with a stability-constrained robust-control framework for rapid spacecraft-swarm rounding-up and target tracking in elliptical orbits \citep{yu2026improvedLearning}.

Cooperative sensing and inspection add observation quality to the formation objective, but the cited studies assign RL different roles.
Dunlap et al. pair multi-agent RL for scalable spacecraft inspection with a runtime assurance layer that filters control outputs against translational and rotational constraints \citep{dunlap2024deepReinforcement}.
Lu et al. pair parallel-island differential evolutionary algorithm (PIDEA)-based multisatellite stereo-observation scheduling with the RL attitude-tracking controller described in Section~\ref{sec:attitude} \citep{lu2025multisatelliteScheduling}.
A related study combines HIDE for constellation-level task allocation with an SAC policy for single-spacecraft attitude tracking of moving targets under modeled actuator faults \citep{lu2026constellation}.
Wang et al. use a capability-aggregated MARL framework with decentralized policies trained using convolutional-augmented MAPPO (CA-MAPPO) for cooperative tracking of maneuvering targets; the policies use local observations and three-axis control-torque actions in multi-scenario simulations \citep{wang2026cooperativeSatellite}.
Huang et al. formulate collaborative exploration by a swarm of space robots as a multi-agent RL problem, extending the reviewed mission-system applications to planetary mobility \citep{huang2020multi}.
Together, these studies illustrate complementary approaches to cooperative observation: multi-agent policies coordinate inspection or tracking, while hybrid scheduling-and-control systems combine constellation-level task allocation with RL-based attitude tracking \citep{dunlap2024deepReinforcement,lu2025multisatelliteScheduling,lu2026constellation,wang2026cooperativeSatellite}.

\subsection{Formation Reconfiguration and Self-Reconfigurable Space Systems}

Reconfiguration changes formation geometry or module arrangement, and may require RL to select maneuvers, assign roles, or coordinate high- and low-level decisions.
Graph search and assignment optimization remain interpretable alternatives, while shields and validity checks can preserve explicit feasibility rules in learning-based systems.
Zhang et al. use RL task planning to select cell-by-cell moves for constructing a desired cellular-satellite structure \citep{zhang2021reinforcementLearning}.
Ye et al. formulate partially observable self-reconfiguration path planning for a single module; recurrent Q-learning uses observation history, and preemptive action shielding removes moves that violate safety specifications \citep{ye2024poSrpp}.
Tests on an electromagnetic self-reconfigurable-satellite rig support the feasibility of this local planner under the tested setup.
Liu et al. combine multi-agent SAC with sequential convex programming (SCP) expert trajectories in a physics-based environment; mixed expert replay accelerates convergence over conventional RL, while the resulting method reduces energy use relative to SCP \citep{liu2025efficientMulti}.
Jiang et al. represent heterogeneous cluster topology with a graph-attention network and divide RL decisions between a manager and low-level workers \citep{jiang2026gnnGuided}.
Their state-aware action bias and inference-time validity check encode energy and collision rules.
Simulations report strong completion under modeled fault scenarios and transfer to larger clusters, with comparisons against MAPPO and greedy baselines reported for selected scenarios.
These studies demonstrate RL's flexibility in coordinating discrete reconfiguration choices, multi-agent motion, and hierarchical task execution \citep{zhang2021reinforcementLearning,ye2024poSrpp,liu2025efficientMulti,jiang2026gnnGuided}.
Expert-guided learning can accelerate policy training, while graph-based representations and action shielding provide structured ways to represent changing configurations and enforce encoded constraints in reconfigurable spacecraft tasks \citep{liu2025efficientMulti,ye2024poSrpp,jiang2026gnnGuided}.

\subsection{Orbital Pursuit--Evasion and Adversarial Interaction}

Pursuit--evasion requires maneuver decisions against an opponent whose actions or objectives may be only partly known.
RL-based pursuit--evasion research addresses three closely related challenges: decision-making under uncertain opponent behavior, temporally structured maneuver planning, and strategic coordination among multiple cooperating or competing spacecraft.
Differential games provide a model-based reference, while RL offers learned feedback, opponent estimation, and multi-agent strategy generation.
Yang et al. separate impulsive pursuit into far-range rendezvous optimization and a close-range DDPG pursuit policy under $J_2$ perturbation; Monte Carlo tests vary initial states and evasion strategies \citep{yang2021twoStage}.
Zheng et al. study linear spacecraft pursuit--evasion with complete and incomplete information, using Q-learning to obtain the pursuer's control gain without system-matrix knowledge in the latter setting \citep{zheng2023nonzeroSum}.
Wei et al. address action delay in continuous-time pursuit decisions through scheduled action execution and use hierarchical RL subgoals for long-range pursuit; simulations cover two tasks \citep{wei2025continuousTime}.
Yu et al. train distributed RL policies with hybrid discrete--continuous maneuver actions and a population-based procedure to produce diverse strategies in one-on-one GEO games \citep{yu2025diverseStrategy}.
Jia et al. apply counterfactual-baseline multi-agent proximal policy optimization (CFB-MAPPO) to impulsive swarm-to-swarm pursuit--evasion, using counterfactual baselines to address credit assignment across spacecraft and reward functions for multiple constraints \citep{jia2026swarmTo}.
Zhang et al. combine online estimation of unknown evader control weights with pre-trained approximate dynamic programming under actuator saturation \citep{zhang2026spacecraftPursuit}.
In the reported comparison, the offline pre-training and online iterative approximate dynamic programming (PIADP) method achieves the shortest capture time and lowest $\Delta v$ among the tested methods; the Riccati comparator, however, is evaluated with a relaxed saturation limit of $3\times10^{-3}\,\mathrm{km/s^2}$, compared with $2.5\times10^{-3}\,\mathrm{km/s^2}$ for the standard setup.
Guo et al. combine team--opponent cross-attention with variable-size multi-agent TD3, producing interpretable cooperative maneuvers in simulation; reported robustness covers team-size ratios similar to those used in training \citep{guo2026pursuitEvasion}.
In these studies, RL offers a practical route to feedback and strategy generation when analytical game solutions or repeated numerical optimization are difficult, while transfer claims remain specific to the opponent models, information assumptions, and team-size ranges tested.

\subsection{Cross-Subdomain Findings}

Across multi-spacecraft autonomy and coordination, RL can provide value by learning reusable decisions from local observations, adapting to configuration changes represented during training, and conditioning pursuit--evasion actions on observed or estimated opponent behavior.
These advantages remain conditional on the training distribution, observation design, and available inter-spacecraft information \citep{wang2026cooperativeSatellite,ye2024poSrpp,guo2026pursuitEvasion}.

The reviewed multi-agent architectures differ in how information is centralized during training and execution.
CTDE methods use global information or a centralized critic during training but decentralized actors during execution \citep{wang2026cooperativeSatellite,jia2026swarmTo}.
For partially observable self-reconfiguration, Ye et al.'s locally executed recurrent policy uses observation history to inform module-level movement decisions \citep{ye2024poSrpp}.
Yu et al.'s population-based distributed training parallelizes strategy learning, while the game formulation assumes shared orbital information \citep{yu2025diverseStrategy}.
Opponent-aware attention and online parameter estimation add a separate capability by conditioning decisions on teammate--opponent relations or inferred adversary behavior \citep{guo2026pursuitEvasion,zhang2026spacecraftPursuit}.

RL supplies local actions, reference states, maneuver schedules, or high-level decisions, while conventional controllers, MPC, Lloyd refinement, SCP expert trajectories, event-triggering mechanisms, and safety filters provide complementary functions in feedback control, geometric refinement, expert guidance, update scheduling, and constraint handling \citep{chen2023reinforcementLearning,ren2025learningBased,jin2026deepReinforcement,liu2025efficientMulti,sun2025nearAsteroid,dunlap2024deepReinforcement,jiang2026gnnGuided}.
In particular, Jin et al.'s air-bearing simulators track precomputed trajectories rather than executing the learned multi-agent policy online \citep{jin2026deepReinforcement}.

Comparisons should separate formation error, propellant use, communication load, post-training decision time, offline training cost, and constraint violations.
The reported trade-offs vary by architecture: event-triggering reduces control-update frequency while affecting tracking and control-cost metrics, whereas graph-based, recurrent, and attention-based methods address scalability, partial observability, or coordination under their respective tested conditions \citep{sun2025nearAsteroid,jiang2026gnnGuided,ye2024poSrpp,guo2026pursuitEvasion}.
Reported advantages remain specific to the tested dynamics, communication and observation assumptions, team-size ranges, opponent models or strategies, terminal criteria, and baseline configurations \citep{dunlap2024deepReinforcement,jin2026deepReinforcement,ren2025learningBased,yu2025diverseStrategy,guo2026pursuitEvasion,zhang2026spacecraftPursuit}.
Constraint satisfaction or safety properties supplied by preemptive action shielding, runtime-assurance filters, constrained MPC, or inference-time action-validity checks should be attributed to the complete hybrid architecture rather than to the learned policy alone \citep{ye2024poSrpp,dunlap2024deepReinforcement,ren2025learningBased,jiang2026gnnGuided}.
These mechanisms cover the constraints encoded in their respective models and operate under stated information, feasibility, and dynamics assumptions; they do not establish unrestricted mission-level safety.

\section{Research Gaps and Future Directions for Dependable Spacecraft Autonomy}
\label{sec:future}

Dependable spacecraft autonomy requires reliable task completion under expected conditions, safe operation within mission constraints, robustness to modeling and environmental uncertainty, tolerance of component faults, and behavior that supports systematic verification and validation.
This chapter examines how decision models, data-efficient adaptation, integrated architectures, and progressively realistic tests can advance these objectives.
Established spacecraft applications are distinguished from methods and uses that remain prospective.

\subsection{Decision-Making under Partial Observability, Uncertainty, and Constraints}
\label{sec:mdp-variants}

Section~\ref{sec:foundations} introduced these MDP variants.
Here, we examine their use in selected spacecraft applications and identify open evaluation questions.

Selected sensor-limited studies explicitly formulate POMDPs for active imaging, visual proximity guidance, and mission planning, whereas Yuan et al. formulate angles-only rendezvous as an MDP with an observability-aware reward \citep{brandonisio2021reinforcementLearning,brandonisio2024deepReinforcement,supreeth2026visionGuided,zhang2026sampleEfficient,harris2019spacecraftDecision,harris2022generationOf,yuan2022deepReinforcement}.
Their implementations address partial observability differently, including belief-based state representations, image-conditioned policies, and temporal frame stacking; a POMDP formulation does not, by itself, imply exact Bayesian belief filtering.
Comparisons should report the estimates and observation histories available to each policy and test sensitivity to image quality, missed observations, and viewing geometry.

Constraint-aware studies place mission limits at different points in learning and action selection: Mu et al. formulate debris avoidance as a CMDP \citep{mu2024autonomousSpacecraft}; W. Lu et al. formulate constraint-aware attitude control for agile target tracking as a CMDP and combine worst-case SAC with data-driven actuator-fault identification \citep{lu2025safeReinforcement}; and S. Lu et al. formulate constrained lunar-rover planning as a CMDP, combining action masks, Lagrangian constraint handling, and reward penalties for uncertain time and resource use \citep{lu2026constrainedReinforcement}.
Evaluation should distinguish task performance, expected costs, per-run violations, and action-level feasibility.

Across transfer, landing, and control tasks, recurrent meta-RL and latent-variable policies use observation histories to adapt to changing or hidden conditions \citep{gaudet2020adaptiveGuidance,federici2024robustInterplanetary,chen2025robustAdaptive,gao2026robustLow}.
Explicit robust-MDP formulations, including distributionally robust variants, and Bayes-adaptive MDP formulations are not represented in the selected application studies.
Matched comparisons with recurrent adaptation could assess worst-case performance under specified uncertainty sets and the value of explicit posterior-based decisions under hidden dynamics.

Variable-duration scheduling uses a POsMDP, while cooperative observation and on-orbit scheduling use multi-agent POMDP formulations \citep{stephenson2025learningPolicies,luo2026researchOn,zheng2025deepReinforcement}.
Local observations in inspection and formation studies do not alone establish a Dec-POMDP formulation \citep{dunlap2024deepReinforcement,jin2026deepReinforcement}; comparisons should report decision intervals, communication assumptions, team size, and differences between training and execution.

Table~\ref{tab:spacecraft-mdp-applications} summarizes reviewed applications and open evaluation questions.

\begin{table*}[]
\caption{Reviewed applications and open evaluation questions for spacecraft decision models.}
\label{tab:spacecraft-mdp-applications}
\centering
\begin{tabularx}{\textwidth}{@{}P{0.22\textwidth}Y@{}}
\toprule
Model family & Reviewed applications and open evaluation questions \\
\midrule
MDP & Angles-only rendezvous uses an observability-aware reward; report the state and sensor information available to the policy \citep{yuan2022deepReinforcement}. \\
POMDP & Explicit formulations in selected imaging, visual proximity-guidance, and mission-planning studies; distinguish belief-based state representations from image-conditioned or frame-stacked policies, and assess sensitivity to observation quality, estimation error, and missing data \citep{brandonisio2021reinforcementLearning,brandonisio2024deepReinforcement,supreeth2026visionGuided,zhang2026sampleEfficient,harris2019spacecraftDecision,harris2022generationOf}. \\
CMDP & Debris avoidance, constraint-aware attitude control, and constrained lunar-rover planning; report expected costs alongside per-run violations and action-level feasibility \citep{mu2024autonomousSpacecraft,lu2025safeReinforcement,lu2026constrainedReinforcement}. \\
Robust / distributionally robust MDP & Explicit formulations are not represented in the selected application studies; compare uncertainty-set policies with recurrent adaptation under matched disturbances and report worst-case outcomes. \\
Bayes-adaptive MDP & Explicit formulations are not represented in the selected application studies; compare posterior-based decisions with recurrent or meta-RL under the same hidden parameters. \\
POsMDP / Semi-MDP & Variable-duration Earth-observation scheduling; report how partial observations and elapsed time enter decisions, rewards, discounting, and online decision latency \citep{stephenson2025learningPolicies}. \\
Dec-POMDP-related models & Cooperative observation and scheduling studies use multi-agent POMDP formulations, with CTDE specified in some studies; inspection and formation studies use local observations without explicitly specifying complete Dec-POMDP models.
Assess communication, team size, and training--execution assumptions \citep{luo2026researchOn,zheng2025deepReinforcement,dunlap2024deepReinforcement,jin2026deepReinforcement}. \\
\bottomrule
\end{tabularx}
\end{table*}
\subsection{Data-Efficient Learning, Transfer, and Policy Adaptation}

Spacecraft RL studies address two related challenges: reducing training costs across related problems and adapting decisions to changing dynamics or incomplete observations.
Wei et al. decompose the multi-objective agile Earth-observation scheduling problem into scalarized subproblems and transfer network parameters between neighboring subproblems to reduce training expense \citep{wei2021deepReinforcement}.
For powered descent, Gaudet et al. use recurrent meta-RL to adapt guidance from observation histories; their tests include Mars landing with random engine failure and asteroid landing under unknown environmental dynamics \citep{gaudet2020adaptiveGuidance}.
For interplanetary low-thrust transfer, Federici and Zavoli compare recurrent and feed-forward policies under dynamic uncertainty and control-execution errors; the recurrent network forms an internal representation of the environment distribution to condition control decisions \citep{federici2024robustInterplanetary}.
For asteroid landing, Chen et al. combine recurrent meta-RL with MCTS-based training and post-training DPW-MCTS action refinement under uncertainty \citep{chen2025robustAdaptive}.
Gao et al. use a sequential latent-variable model to represent uncertain environments and predict future observations, then use the stochastic latent PPO (SLPPO) scheme to derive policy actions from learned stochastic latent variables for low-thrust trajectory design \citep{gao2026robustLow}.
Together, these studies examine complementary mechanisms: parameter transfer reduces training expense across related scheduling problems, while recurrent or latent representations and search-based action refinement support decisions under modeled guidance uncertainties.
These studies report transfer or adaptation within their evaluated settings; continual online policy-weight updates remain a distinct research direction.

Further improvements in data efficiency may come from model-based and offline learning.
World-model methods such as Dreamer use trajectories imagined in a learned latent dynamics model for policy learning \citep{hafner2019dreamTo}, whereas offline RL learns from previously collected data without additional online data collection \citep{levine2020offlineReinforcement}.
Return-conditioned sequence models such as Decision Transformer provide another approach by predicting actions from desired returns and trajectory histories \citep{chen2021decisionTransformer}.
Dreamer-style latent imagination, offline RL, and return-conditioned sequence modeling offer potential ways to exploit simulation archives and mission data, but their spacecraft-specific benefits remain insufficiently established in the reviewed studies.
Their applicability to spacecraft tasks therefore requires further evaluation under distribution shift and limited dataset coverage \citep{levine2020offlineReinforcement}; for spacecraft, coverage of safety-critical states is especially important.

Continual adaptation presents additional challenges beyond initial policy training.
Policy updates for changing dynamics or mission requirements should be assessed for loss of previously validated capabilities and violations of established constraints.
Future research should therefore examine data-efficient adaptation together with catastrophic forgetting, onboard computational limits, and mechanisms for monitoring, validating, and, when necessary, reverting policy updates.
An important objective is to determine when a learned policy can be safely adapted onboard and when adaptation should instead be restricted to a validated parameter or control architecture.

\subsection{Integrated Spacecraft Autonomy and Cross-Module Coordination}

Integrated spacecraft autonomy requires coordinated decisions across perception, navigation, guidance, control, fault management, and mission planning.
Existing studies illustrate integration at two complementary levels.
At the control level, investigated interfaces include image-based navigation coupled to a separately trained station-keeping controller, learned guidance followed by conventional tracking, hierarchical rendezvous guidance and attitude control, and fault or health-state estimation coupled to fault-tolerant control \citep{violino2026onboardOptical,hovell2020onDeep,tammam2025hierarchicalDeep,lu2025neuralNetwork,he2026mambaKan}.
At the mission level, Zheng et al. coordinate observation, on-orbit computation, and downlink scheduling under communication and resource constraints, while Qi et al. integrate deep-space mission planning with data-transmission scheduling \citep{zheng2025deepReinforcement,qi2024missionPlanning}.
Validation ranges from numerical simulation to planar air-bearing and hardware-in-the-loop experiments, while the reviewed mission-level scheduling studies remain simulation-based.
These results demonstrate bounded subsystem integration rather than end-to-end autonomy spanning perception, GNC, fault management, and mission planning.

A central research challenge is how estimation errors, uncertainty, delays, and resource constraints propagate across interconnected modules.
For example, inaccurate navigation estimates may affect learned guidance decisions, while attitude maneuvers can influence observation quality, communication availability, and subsequent mission scheduling.
Future architectures should also clarify the allocation of decision authority between RL policies and conventional components.
Rather than replacing the entire autonomy stack, RL may generate high-level decisions or control references while established estimators, feedback controllers, optimizers, and safety supervisors retain their respective functions; learned guidance followed by conventional tracking provides one example of such an interface \citep{hovell2020onDeep}.

Future research should investigate tighter coordination between active perception and guidance, fault diagnosis and recovery planning, and mission scheduling and resource-aware control.
Modular, hierarchical, and end-to-end architectures should be compared under matched sensing, computational, and operational conditions, with explicit assessment of decision authority, error propagation, and module-level contributions.
Beyond individual control accuracy or scheduling reward, evaluations should consider mission completion, constraint violations, fault recovery, communication and computation overhead, and the behavior of the integrated system when individual modules degrade or fail.

\subsection{Testing, Verification, and Deployment Readiness}
\label{sec:testing-readiness}

\begin{figure}[]
\centering
\includegraphics[width=\columnwidth]{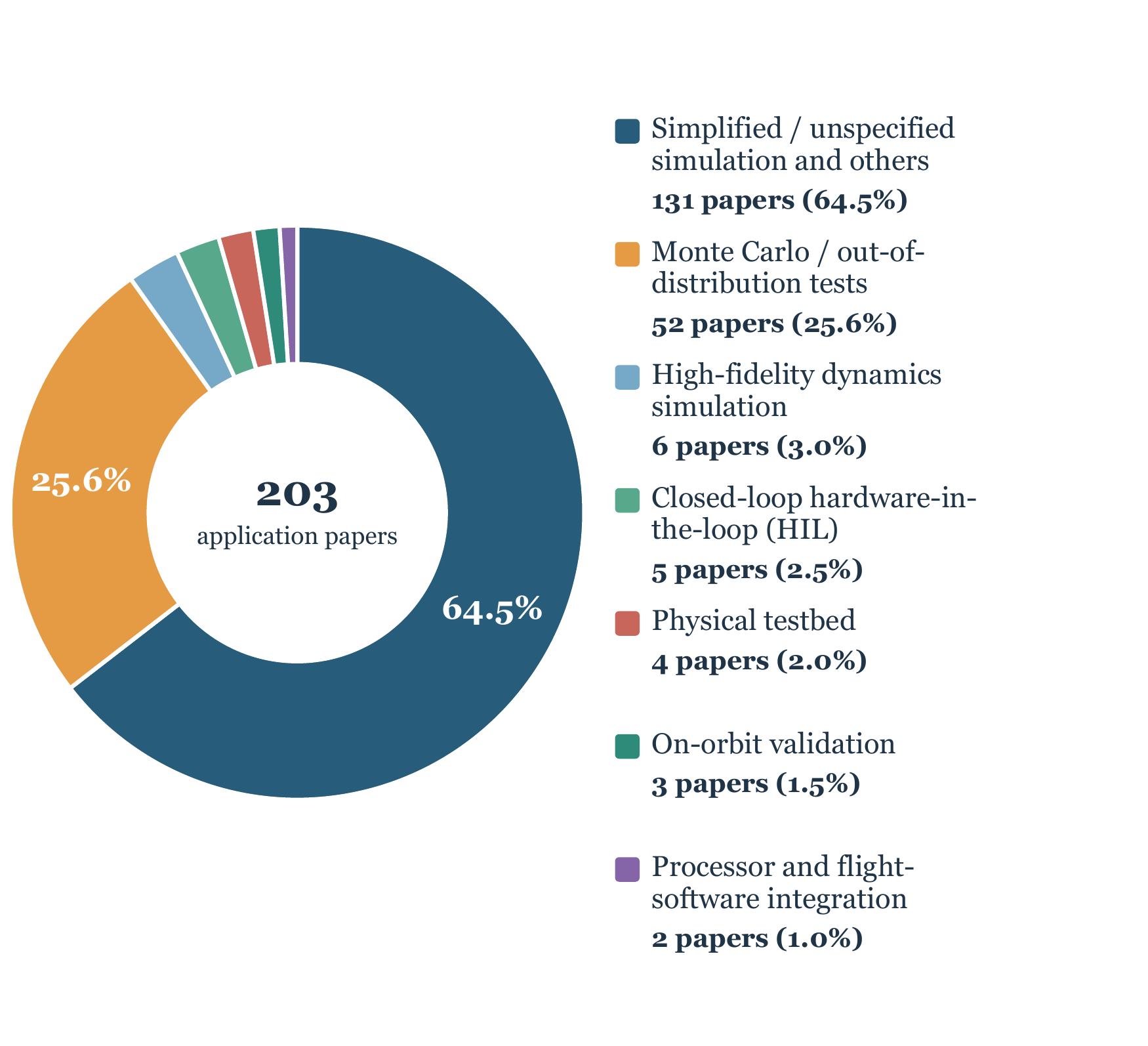}
\caption{Primary validation categories assigned to the 203 spacecraft application papers according to the stated counting rule.}
\label{fig:validation-evidence-pie}
\end{figure}

Deployment readiness requires evidence of computational feasibility, closed-loop performance, robustness, and operational safety under clearly specified assumptions.
The representative studies in Table~\ref{tab:domain-evidence-overview} differ in policy roles, retained system components, and validation settings.
Figure~\ref{fig:validation-evidence-pie} summarizes the validation settings of 203 application papers reviewed in Sections~3--8.
Each application paper is assigned to one category using the following priority: on-orbit validation, physical testbed, closed-loop HIL, processor and flight-software integration, Monte Carlo or out-of-distribution testing, high-fidelity dynamics simulation, and simplified or unspecified simulation and other cases.
Processor and flight-software integration includes either learned-policy execution on embedded hardware, such as processor-in-the-loop testing, or flight-model software and interface validation; membership in this category does not imply that the RL policy itself executed onboard or achieved closed-loop spacecraft control.
For example, Barrenechea et al.'s OPS-SAT work is included because it tested SEPP software and flight-model interfaces, although the in-orbit runs used the Algorithm Driver without the HOPAS AI algorithm \citep{barrenechea2023enabling}.
For this review, on-orbit validation includes RL policies executed onboard orbiting spacecraft or spaceborne robotic platforms, using actual flight measurements to affect closed-loop control or mission decisions.
The resulting counts are 3 on-orbit application papers (representing two distinct flight demonstrations), 4 physical-testbed papers, 5 closed-loop HIL papers, 2 processor and flight-software integration papers, 52 Monte Carlo or out-of-distribution papers, 6 high-fidelity simulation papers, and 131 papers assigned to simplified or unspecified simulation and other cases.
These mutually exclusive categories follow a counting rule rather than a hierarchy of validation maturity.
The reported counts describe primary validation categories under the stated assignment rule, rather than the total prevalence of each validation characteristic across the corpus.
The pie-chart center shows the total number of application papers, and its legend reports the count and percentage for each setting.

Ground-based studies provide complementary but distinct forms of validation.
Reaction-wheel HIL and triaxial-turntable HIL experiments evaluate selected attitude-control functions, while the hierarchical CubeSat GNC platform examines rendezvous and attitude-control behavior \citep{sarjoughian2026hardwareValidated,dong2021learningBased,tammam2025hierarchicalDeep}.
In Jin et al.'s formation study, eight air-bearing spacecraft simulators track precomputed trajectories; the experiment does not execute the learned RL policy live on hardware \citep{jin2026deepReinforcement}.
Muñoz et al. assess attitude-policy execution on an embedded processor with a simulated satellite environment, whereas Yang et al. investigate policy compression and formal neural-network verification for rendezvous guidance on a workstation \citep{munoz2025feasibilityOf,yang2022modelBased}.
An optical-bench experiment processes real camera images using onboard-like computing but retains simulated spacecraft motion \citep{federici2026opticalBench}.
Other studies examine robustness under randomized dynamics, CLF/QP-based safety enforcement in simulation, autonomous operations-procedure generation, and runtime assurance \citep{fabisch2026deep,chen2026safeReinforcement,harris2022generationOf,dunlap2023runTime,dunlap2024runTime}.
These studies validate different properties and operating conditions.
Processor-in-the-loop testing can demonstrate learned-policy execution in a closed loop with simulated spacecraft dynamics, whereas flight-software interface testing does not necessarily establish execution of the RL policy itself.
Neither constitutes in-orbit closed-loop RL validation, and formal verification and physical testbeds provide evidence only within their specified properties and experimental conditions.
Formal verification should distinguish properties of the policy network from closed-loop properties of the complete spacecraft system.
Future work should examine how verified operating regions, model uncertainty, sensor errors, and runtime safety mechanisms can contribute to system-level assurance arguments.

\begin{figure*}[]
\centering
\begin{subfigure}[t]{0.31\textwidth}
  \centering
  \includegraphics[width=\linewidth,height=0.12\textheight,keepaspectratio]{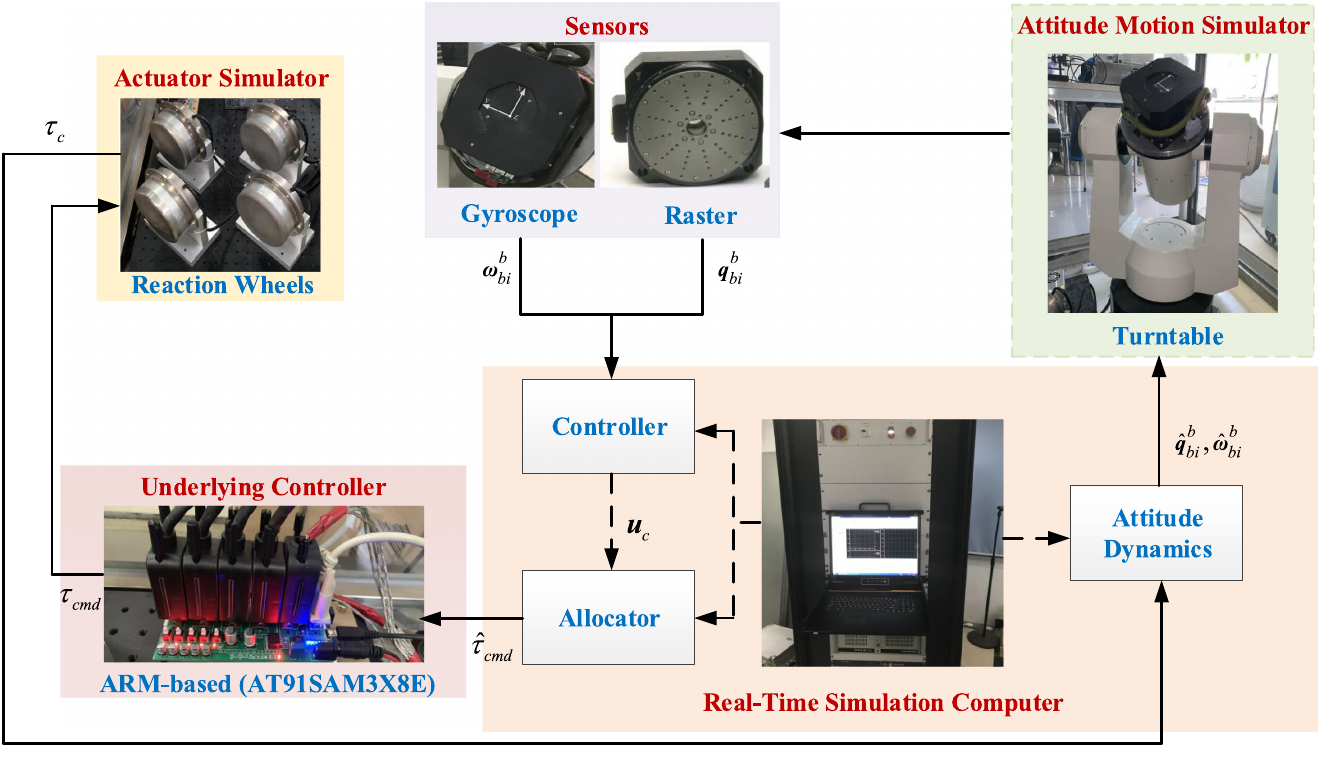}
  \caption{Attitude-tracking HIL testbed \citep{dong2021learningBased}.}
\end{subfigure}\hfill
\begin{subfigure}[t]{0.31\textwidth}
  \centering
  \includegraphics[width=\linewidth,height=0.12\textheight,keepaspectratio]{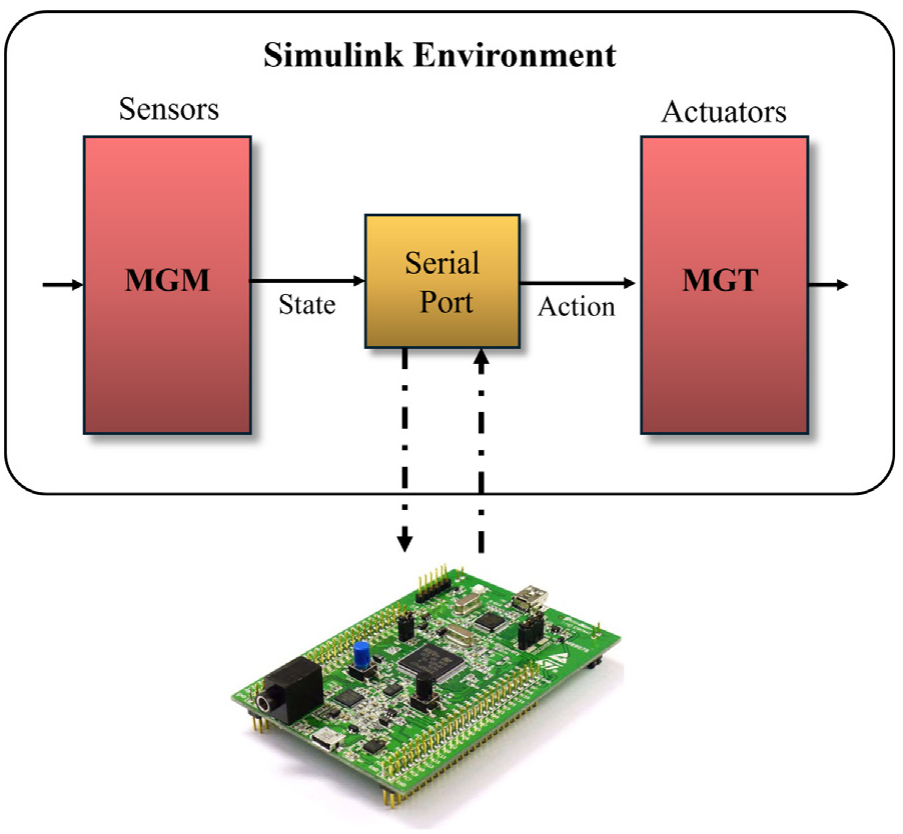}
  \caption{UPMSat-2 processor-in-the-loop setup \citep{munoz2025feasibilityOf}.}
\end{subfigure}\hfill
\begin{subfigure}[t]{0.31\textwidth}
  \centering
  \includegraphics[width=\linewidth,height=0.12\textheight,keepaspectratio]{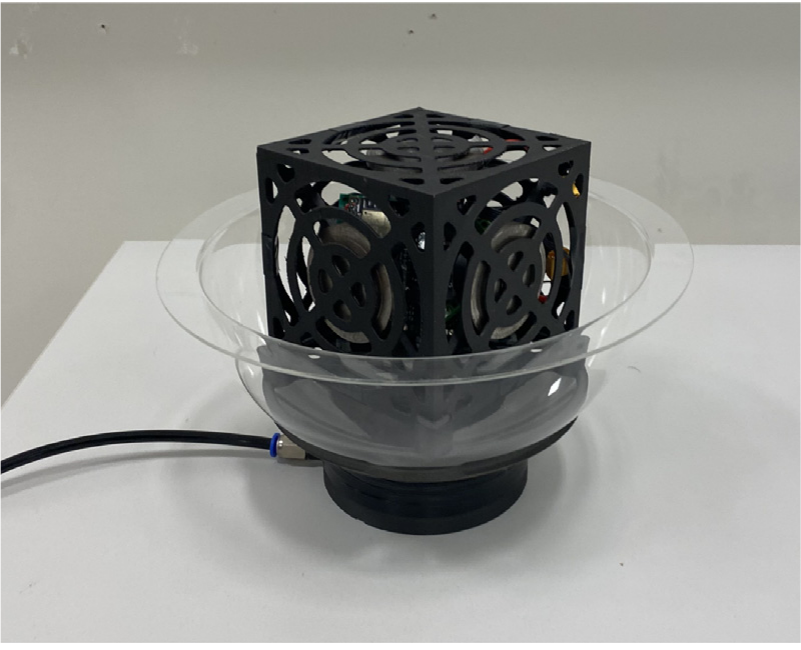}
  \caption{CubeSat attitude-dynamics test platform \citep{tammam2025hierarchicalDeep}.}
\end{subfigure}

\vspace{4pt}
\begin{subfigure}[t]{0.31\textwidth}
  \centering
  \includegraphics[width=\linewidth,height=0.12\textheight,keepaspectratio]{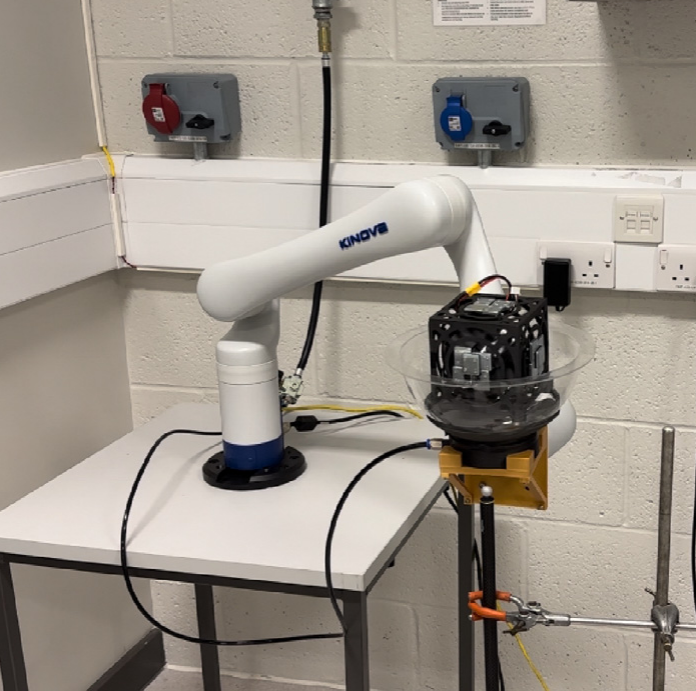}
  \caption{Combined attitude-dynamics and rendezvous testing platform \citep{tammam2025hierarchicalDeep}.}
\end{subfigure}\hfill
\begin{subfigure}[t]{0.31\textwidth}
  \centering
  \includegraphics[width=\linewidth,height=0.12\textheight,keepaspectratio]{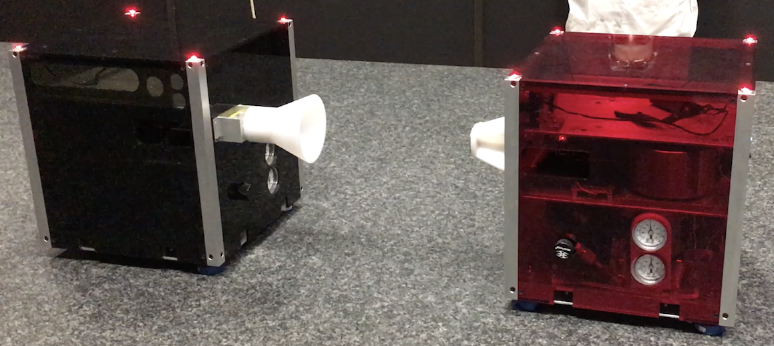}
  \caption{SPOT air-bearing spacecraft platforms \citep{hovell2020onDeep}.}
\end{subfigure}\hfill
\begin{subfigure}[t]{0.31\textwidth}
  \centering
  \includegraphics[width=\linewidth,height=0.12\textheight,keepaspectratio]{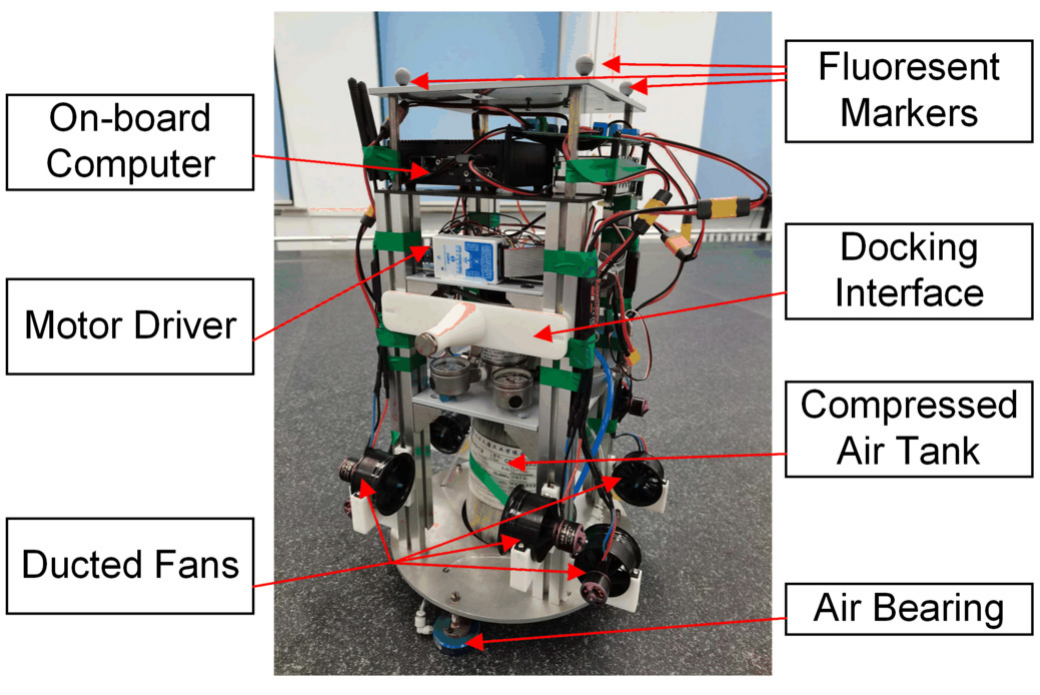}
  \caption{Air-bearing chaser simulator \citep{du2024bayesianQuadrature}.}
\end{subfigure}

\vspace{4pt}
\begin{subfigure}[t]{0.31\textwidth}
  \centering
  \includegraphics[width=\linewidth,height=0.12\textheight,keepaspectratio]{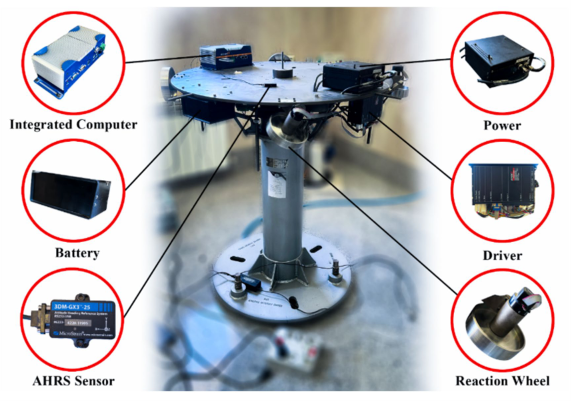}
  \caption{3-DoF reaction-wheel simulator \citep{sarjoughian2026hardwareValidated}.}
\end{subfigure}\hfill
\begin{subfigure}[t]{0.31\textwidth}
  \centering
  \includegraphics[width=\linewidth,height=0.12\textheight,keepaspectratio]{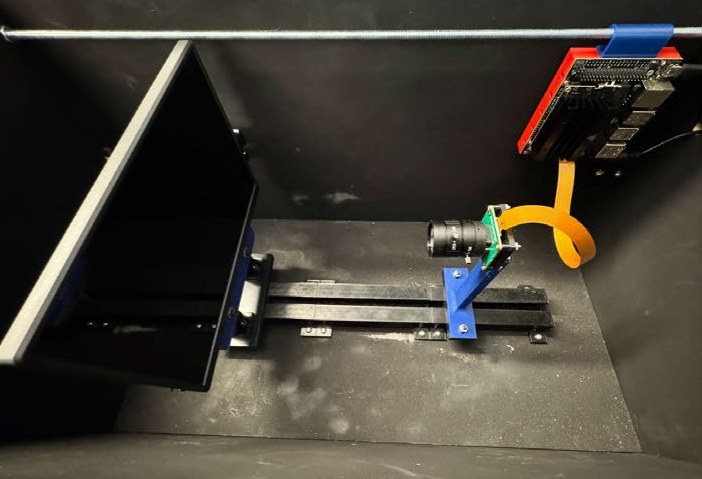}
  \caption{Image-based spacecraft-GNC optical bench \citep{federici2026opticalBench}.}
\end{subfigure}\hfill
\begin{subfigure}[t]{0.31\textwidth}
  \centering
  \includegraphics[width=\linewidth,height=0.12\textheight,keepaspectratio]{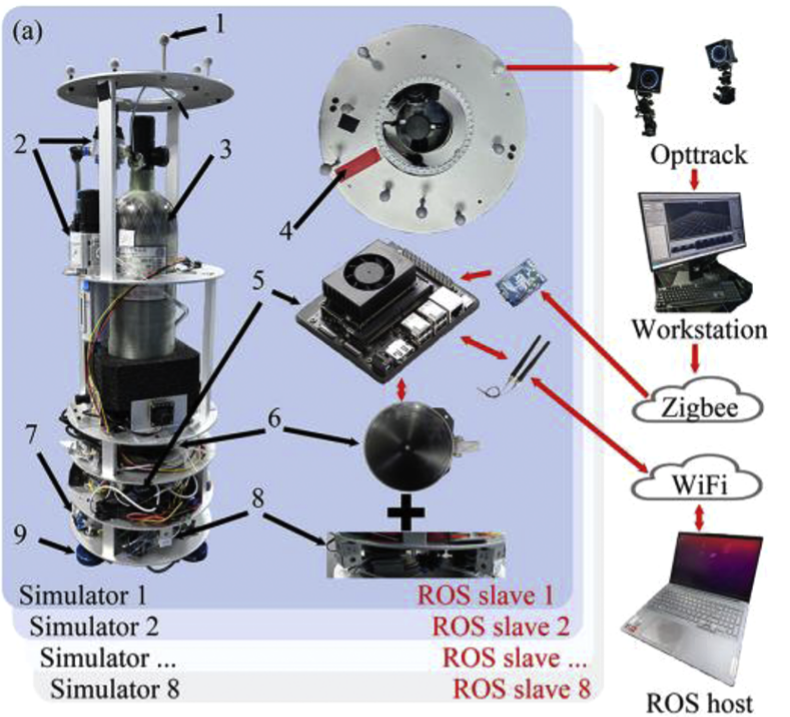}
  \caption{Air-bearing simulators tracking precomputed trajectories \citep{jin2026deepReinforcement}.}
\end{subfigure}
\caption{Representative ground platforms used to evaluate spacecraft RL methods and related hardware-based tests.}
\label{fig:ground-validation-platforms}
\end{figure*}

Figure~\ref{fig:ground-validation-platforms} shows representative ground-based platforms spanning closed-loop HIL, processor-in-the-loop execution, physical testbeds, and optical-bench evaluation.
The set includes two CubeSat test setups reported by Tammam et al. \citep{tammam2025hierarchicalDeep} and an image-based optical bench that processes real camera images while retaining simulated spacecraft motion \citep{federici2026opticalBench}.
Additional examples include Hovell et al.'s SPOT planar testbed with air-bearing spacecraft \citep{hovell2020onDeep}, Du et al.'s air-bearing chaser simulator for proximity maneuvers \citep{du2024bayesianQuadrature}, and Sarjoughian et al.'s 3-DoF reaction-wheel attitude simulator \citep{sarjoughian2026hardwareValidated}.
Jin et al.'s air-bearing simulators tracked precomputed trajectories rather than live RL outputs, illustrating platform-based trajectory tracking rather than hardware execution of the learned policy \citep{jin2026deepReinforcement}.

Recent flight experiments provide more direct evidence of RL deployment while revealing important differences in decision authority and validation scope.
On ESA's OPS-SAT, Barrenechea et al. tested onboard software and telemetry/command interfaces, including timing and open-loop wheel-speed commands, but did not achieve RL-based closed-loop control \citep{barrenechea2023enabling}.
In contrast, LeLaR deployed a simulation-trained attitude policy onboard the InnoCube 3U satellite for reaction-wheel pointing maneuvers.
All seven repeated RL-controlled trials met the reported $1^\circ$ per-axis criterion, compared with three of seven trials for the onboard PD controller, which served as a contextual reference \citep{11457578}.
APIARY and Crossing describe the same ISS Astrobee demonstration rather than independent flight experiments.
The first redocking attempt triggered safe recovery, while the successful second attempt required operator initiation; an RL-controlled $0.5$-m undocking maneuver undershot the command but satisfied the reported pose-success criteria \citep{11436341,11436888}.
These cases distinguish successful closed-loop flight control from preliminary onboard integration and show why operator intervention, safety supervision, and the authority granted to RL must be reported explicitly.

\begin{figure*}[]
\centering
\begin{subfigure}[t]{0.46\textwidth}
  \centering
  \includegraphics[width=\linewidth,height=0.22\textheight,keepaspectratio]{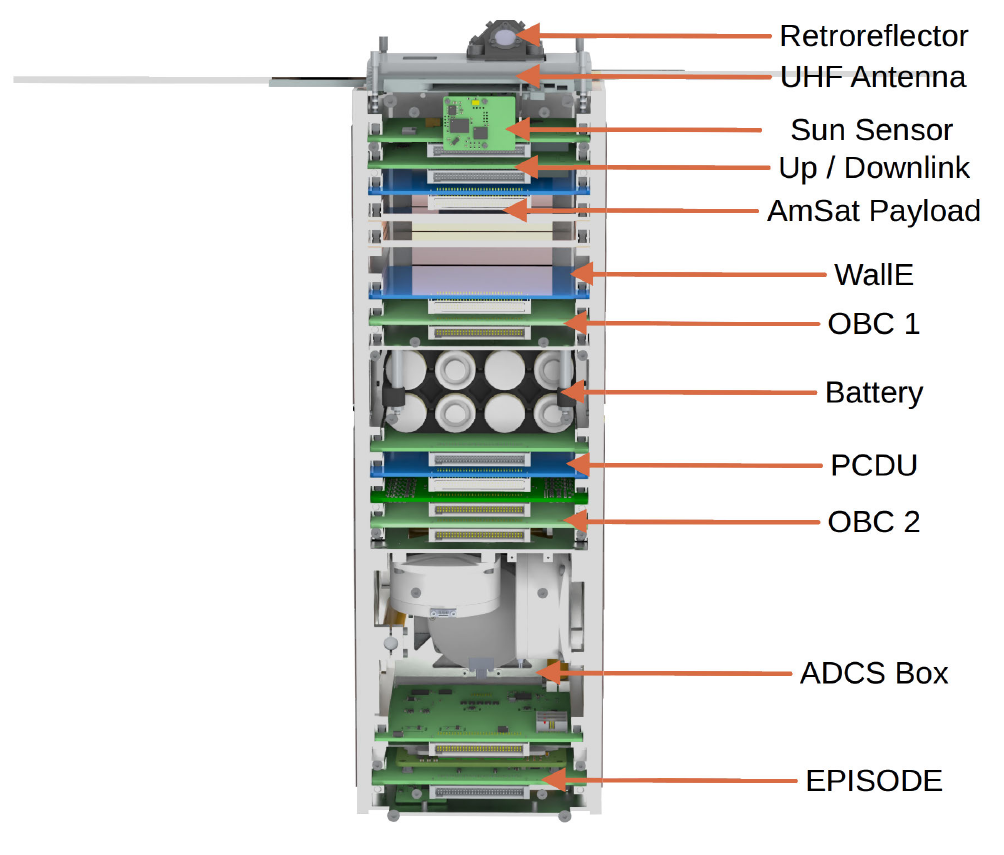}
  \caption{InnoCube 3U satellite configuration \citep{11457578}.}
\end{subfigure}\hfill
\begin{subfigure}[t]{0.46\textwidth}
  \centering
  \includegraphics[width=\linewidth,height=0.22\textheight,keepaspectratio]{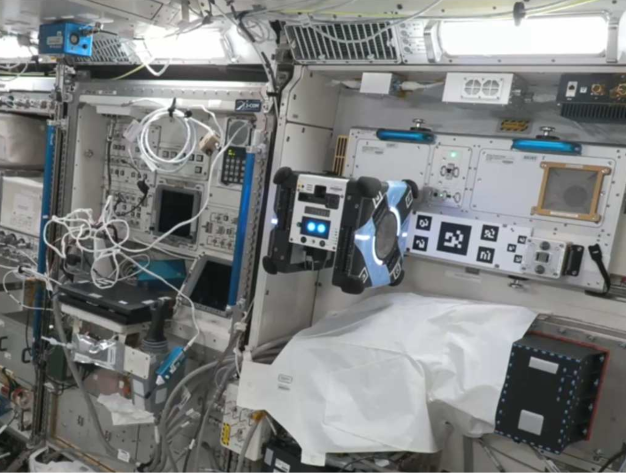}
  \caption{Astrobee during ISS flight testing \citep{11436341,11436888}.}
\end{subfigure}
\caption{Spaceborne platforms used in the reviewed in-orbit RL demonstrations.}
\label{fig:flight-platforms}
\end{figure*}

Figure~\ref{fig:flight-platforms} shows the InnoCube satellite and the ISS Astrobee platform associated with the two reviewed flight demonstrations; the Astrobee images come from the single campaign reported by APIARY and Crossing \citep{11457578,11436341,11436888}.

Simulation environments and mission-level benchmarks provide additional evidence but should not be confused with flight validation.
Stephenson et al. compare shielded RL with probabilistic-roadmap inspection planning in a realistic Basilisk/BSK-RL simulation environment, while OrbitZoo provides high-fidelity orbital simulation and several RL mission examples \citep{stephenson2026twoApproaches,oliveira2025orbitZoo}.
OrbitZoo's comparison with Starlink ephemerides evaluates orbital propagation accuracy rather than the flight performance of an RL policy \citep{oliveira2025orbitZoo}.
Future research should progress toward reproducible, integrated mission-level validation covering nominal operations, uncertainty, sensing degradation, actuator faults, safety interventions, and replanning.
Evaluation reports should identify the functions actually tested, the physical and simulated components, the assumptions and uncertainty coverage, and the decision authority assigned to learned policies.
Consistent reporting of mission completion, constraint violations, fallback behavior, inference latency, memory use, and computational cost will be essential for assessing deployment readiness.

\section{Conclusion}
\label{sec:conclusion}

This review has examined RL for spacecraft autonomy across six application domains: attitude control, orbital guidance, proximity operations, planetary exploration, mission planning and scheduling, and multi-spacecraft coordination.
The reviewed studies demonstrate that RL serves diverse roles, ranging from direct feedback control and adaptive guidance to optimizer assistance, mission-level decision-making, and multi-agent coordination.
RL's contribution depends on its decision authority and integration with conventional spacecraft subsystems.
In hybrid architectures, performance improvements should be attributed to the complete system rather than the learned policy alone.
Across application domains, RL's potential value is particularly evident in repeated decision-making, feedback adaptation under modeled uncertainties, and optimizer assistance.
Conventional methods remain important benchmarks because of their established feedback structures, explicit optimization formulations, and constraint-handling mechanisms.
Reported RL advantages vary with baseline selection, operating conditions, and evaluation metrics.

Safety assurance, generalization beyond training conditions, and onboard implementation remain unevenly validated across the reviewed studies.
The reviewed approaches address different safety, feasibility, and operational constraints through mechanisms such as reward penalties, constrained optimization, action masking, safety filters, and runtime assurance.
Their enforcement scope depends on the constrained quantity, intervention point, and implementation assumptions.
Most of the 203 reviewed application papers report simulation-based evaluation; three report on-orbit validation, corresponding to two distinct flight demonstrations.
These demonstrations are important steps toward deployment, but provide limited evidence of integrated mission-level autonomy under realistic uncertainty, failures, and operational constraints.

Future studies should evaluate RL within integrated spacecraft systems that include navigation, guidance, control, fault management, and mission planning.
These systems also require decision models for partial observability, uncertain dynamics, and multi-timescale interactions, together with data-efficient adaptation to changing mission conditions.
Evaluation should use rigorous control and optimization baselines, report safety interventions and computational requirements, and provide reproducible evidence across progressively realistic test settings.
Operational adoption will require demonstrated mission performance, reliability, constraint handling under specified conditions, and compatibility with onboard hardware and established subsystems.

% Place all remaining tables and figures before the reference list.
\clearpage

% CAS model 2: alphabetic author--year references.
\bibliographystyle{unsrtnat}
\bibliography{cas-refs}

\end{document}